\documentclass[journal,onecolumn]{IEEEtran}

\usepackage{setspace}

\usepackage{ifpdf}
\usepackage{setspace}
\usepackage{cite}
\usepackage{amsthm}

\usepackage{cancel}

\newtheorem{theorem}{Theorem}
\newtheorem{example}{Example}
\newtheorem{lemma}{Lemma}
\newtheorem{corollary}{Corollary}

\ifCLASSINFOpdf
  \usepackage[pdftex]{graphicx}
  \DeclareGraphicsExtensions{.pdf,.jpeg,.png,.jpg}
\else
  \usepackage[dvips]{graphicx}
  \graphicspath{{figures/}}
  \DeclareGraphicsExtensions{.eps}
\fi

\ifCLASSOPTIONcaptionsoff
  \usepackage[nomarkers]{endfloat}
 \let\MYoriglatexcaption\caption
 \renewcommand{\caption}[2][\relax]{\MYoriglatexcaption[#2]{#2}}
\fi

\usepackage[cmex10]{amsmath}
\usepackage{amssymb}
\usepackage{amscd}
\usepackage{dsfont}

\usepackage{algorithm,algorithmicx,algpseudocode}

\usepackage{array}

\usepackage{mdwmath}
\usepackage{mdwtab}

\usepackage{eqparbox}

\usepackage[caption=false]{caption}
\usepackage[font=footnotesize]{subfig}
\usepackage{stfloats}

\newcommand{\xqedhere}[2]{%
  \rlap{\hbox to#1{\hfil\llap{\ensuremath{#2}}}}}

\usepackage{mathtools}
\mathtoolsset{showonlyrefs}

\usepackage{stackrel}
\usepackage{multirow}

\newcommand{\hf}{\frac{1}{2}}

\renewcommand{\a}{\alpha}
\renewcommand{\b}{\beta}
\newcommand{\g}{\gamma}
\newcommand{\s}{\sigma}
\renewcommand{\r}{\rho}

\newcommand{\flr}[1]{\left\lfloor #1 \right\rfloor}
\newcommand{\ceil}[1]{\left\lceil #1 \right\rceil}
\newcommand{\flrfrac}[2]{\left\lfloor\! \frac{#1}{#2}\! \right\rfloor}
\newcommand{\deltafrac}[2]{\delta\!\left(\! \frac{#1}{#2}\! \right)}
\newcommand{\abs}[1]{\left| #1 \right|}
\newcommand{\Ex}[1]{\text{E}\!\left[  #1 \right]}

\newcommand{\saw}[1]{\left(\!\left(#1\right)\!\right)}
\newcommand{\saww}[1]{\left(\!\!\left(#1\right)\!\!\right)}
\newcommand{\sawww}[1]{\left(\!\!\!\left(#1\right)\!\!\!\right)}

\newcommand{\sawfrac}[2]{\left(\!\!\!\left(\!\frac{#1}{#2}\!\right)\!\!\!\right)}
\newcommand{\sawfracinline}[2]{\left(\!\!\left(\!\frac{#1}{#2}\!\right)\!\!\right)}

\newcommand{\sawfracprodinline}[4]{\sawfracinline{#1}{#2} \!\!\! \sawfracinline{#3}{#4}}

\newcommand{\dedsuminline}[6]{\sum_{#1}^{#2}\!\sawfracprodinline{#3}{#4}{#5}{#6}}

\def\urltilde{\kern -.15em\lower .7ex\hbox{\~{}}\kern .04em}
\def\urldot{\kern -.10em.\kern -.10em}
\def\urlhttp{http\kern -.10em\lower -.1ex\hbox{:}\kern -.12em\lower 0ex\hbox{/}\kern -.18em\lower 0ex\hbox{/}}

\newcommand{\cpi}{\stackbin{\circ}{\pi}}
\newcommand{\ppi}[1]{\stackbin{\circ}{\pi}\!\left( #1 \right)}
\newcommand{\ppin}[2]{\stackbin{\circ}{\pi}_{#1}\!\left( {#2} \right)}

\makeatletter
\def\old@comma{,}
\catcode`\,=13
\def,{%
  \ifmmode%
    \old@comma\discretionary{}{}{}%
  \else%
    \old@comma%
  \fi%
} \makeatother

\begin{document}
%
\title{Pruned Bit-Reversal Permutations: Mathematical\\[-.5em] Characterization, Fast Algorithms and Architectures}

\author{Mohammad~M.~Mansour,~\IEEEmembership{Senior Member,~IEEE}
\thanks{Mohammad M. Mansour is with the Department of Electrical and Computer Engineering at the American University of Beirut, Lebanon. E-mail: mmansour@aub.edu.lb.}}



%


%
\maketitle
\vspace{-0.3in}

%

\begin{abstract}
A mathematical characterization of serially-pruned permutations (SPPs) employed in variable-length permuters and their associated fast pruning algorithms and architectures are proposed. Permuters are used in many signal processing systems for shuffling data and in communication systems as an adjunct to coding for error correction. Typically only a small set of discrete permuter lengths are supported. Serial pruning is a simple technique to alter the length of a permutation to support a wider range of lengths, but results in a serial processing bottleneck. In this paper, parallelizing SPPs is formulated in terms of recursively computing sums involving integer floor and related functions using integer operations, in a fashion analogous to evaluating Dedekind sums. A mathematical treatment for bit-reversal permutations (BRPs) is presented, and closed-form expressions for BRP statistics including descents/ascents, major index, excedances/descedances, inversions, and serial correlations are derived. It is shown that BRP sequences have weak correlation properties. Moreover, a new statistic called permutation inliers that characterizes the pruning gap of pruned interleavers is proposed. Using this statistic, a recursive algorithm that computes the minimum inliers count of a pruned BR interleaver (PBRI) in logarithmic time complexity is presented. This algorithm enables parallelizing a serial PBRI algorithm by any desired parallelism factor by computing the pruning gap in lookahead rather than a serial fashion, resulting in significant reduction in interleaving latency and memory overhead. Extensions to 2-D block and stream interleavers, as well as applications to pruned fast Fourier transforms and LTE turbo interleavers, are also presented. Moreover, hardware-efficient architectures for the proposed algorithms are developed. Simulation results of interleavers employed in modern communication standards demonstrate 3 to 4 orders of magnitude improvement in interleaving time compared to existing approaches.
\end{abstract}

\begin{IEEEkeywords}\vspace{-0.1in}
Bit-reversal permutations, pruned interleavers, turbo interleavers, permutation polynomials, permutation statistics.
\end{IEEEkeywords}

%
\IEEEpeerreviewmaketitle

\vspace{-0.1in}

%
\section{Introduction and Motivation}
\label{s:intro}
\IEEEPARstart{P}{ermuters} are devices that reorder a sequence of symbols according to some permutation~\cite{Ramsey_1970}. They have a variety of applications in communication systems, signal processing, networking, and cryptography. In communication systems, permuters are used as an adjunct to coding for error correction~\cite{Ramsey_1970,Forney_1971} and are more commonly known as \emph{interleavers}. Interleavers are a subclass of permuters with carefully chosen permutations to break certain patterns in the input sequence, and strategically reposition symbols according to their relevance in protecting the overall sequence against errors. Examples include interleavers in turbo codes~\cite{1993_Berrou_turbo_codes}, edge permuters in Tanner graphs~\cite{tanner} for low-density-parity check (LDPC) codes~\cite{gallager}, channel interleavers in bit-interleaved coded modulation schemes~\cite{Zehavi_1992}, and carrier interleaving for diversity gain in multi-carrier wireless systems with frequency-selective fading and multiple-access interference~\cite{1990_Bingham}.

In signal processing, permuters are used to shuffle streaming data~\cite{Parsons_2009} into a particular order such as in signal transform (e.g., fast Fourier transform (FFT)~\cite{CooleyTukey_1965,Burrus_1988}, discrete cosine transform~\cite{Skodras_1991}, Hartley transform~\cite{1987_Evans}), matrix transposition~\cite{1999_Kim,1999_Portnoff}, and matrix decomposition algorithms~\cite{1991_Verbauwhede}. In networking, permuters are widely used as interconnection and sorting networks for switching and routing~\cite{1999_Chang}. In cryptography, permuters are commonly used in cipher algorithms for encryption~\cite{2004_Rivest}.

The theory of interleavers has been established in the classic papers~\cite{Ramsey_1970,Forney_1971} and more recently in~\cite{2001_Garello}. Interleavers can be implemented using hard-wired connections, reconfigurable interconnection networks, or memory buffers with address generators depending on the desired throughput, reconfigurability, and resource requirements. A class of computationally efficient interleavers with simple address generation are \emph{block} interleavers~\cite{2001_Garello} of power-of-2 length $k\!=\!2^n$. They are expressed in closed-form by $\rho\!: \mathds{Z}_k \!\rightarrow\! \mathds{Z}_k, \r(j) \!=\! k_1\!\cdot\!\pi_1(j\!\bmod\! k_1) \!+\! \pi_2\left(\!\flrfrac{j}{k_1}\!\right)$, where $\pi_1\!:\! \mathds{Z}_{k_1} \!\rightarrow\! \mathds{Z}_{k_1}$ and $\pi_2\!:\! \mathds{Z}_{k_2} \rightarrow \mathds{Z}_{k_2}$ are basic permutations of lengths $k_1\!=\!2^{n_1}$ and $k_2\!=\!2^{n_2}$, respectively, and $k\!=\!k_1k_2$. Here the $k$ symbols are written row-wise into a $k_2\!\times \!k_1$ array and read column-wise after permuting the rows by $\pi_1$ and the columns by $\pi_2$. Example permutations proposed in the literature or adopted in modern communications standards~\cite{LTE_phy_layer,IEEE_802.20_3GPP2,IEEE_802.16e} include the bit-reversal permutation (BRP) $\pi(j) \!= \! \text{BRP}(j_{\text{(2)}})$~\cite{IEEE_802.20_3GPP2} which reverses the order of bits in $j_{(2)}$, and polynomial-based permutations $\pi(j) \!=\! f_p(j)\bmod k$ where $f_p(j)$ is a degree-$p$ permutation polynomial (PP) over the ring $\mathbb{Z}_k$~\cite{Takeshita_2005}. Commonly used polynomials include circular shift by a constant $f_1(j)\! =\! j\! +\! c$ (e.g.,~\cite{2001_Crozier}, the parity and column twist (PCT) interleaver~\cite{DVB-T2_ETSI_302_755}), linear PPs $f_1(j) \!=\! jh \!+ \!c$ (e.g.,~\cite{Knuth_Seminumerical,IEEE_802.20_3GPP2}, almost regular permutations (ARP)~\cite{2004_Berrou}, dithered relative prime (DRP) interleavers~\cite{2001_Crozier}), and quadratic PPs (QPPs) $f_2(j) \!=\! jh \!+\! j^2b\!+\!c$, where $c,h,b$ are appropriately chosen integers (e.g.~\cite{Takeshita_2005,Nimbalker_2008,LTE_phy_layer}).

Many practical interleavers are limited to a small set of discrete lengths. Pruning is a technique used to support more flexible block lengths $k$~\cite{1999_Eroz_Hammongs_prunable_interleavers,2002_ferrari,2005_Dinoi_Benedetto_S_random}. Communication standards~\cite{LTE_phy_layer,IEEE_802.20_3GPP2,IEEE_802.16e} typically vary $k$ depending on the input data rate requirements and channel conditions. To support any length $\b$, interleaving is done using a mother interleaver with smallest $k \!>\! \b$ such that outlier interleaved addresses $\geq\b$ are excluded. However, pruning alters the spread characteristics of the mother interleaver, and creates a serial bottleneck since interleaved indices become address-dependent. Hence permuting streaming data in parallel on the fly is no longer practically feasible~\cite{Parsons_2009}. Expensive buffering of the data is required to maintain a desired system throughput.
Hence it is essential to characterize the pruned permutation structure to study its spread characteristics, and to parallelize the pruning operation to reduce latency and memory overhead by interleaving an address without interleaving all its predecessors.

Alternatively, pruning can also be employed to design more efficient FFTs by eliminating redundant or vacuous computations when the input vector has many zeros and/or when the required outputs may be very sparse compared to the transform length.

Pruning interleavers has motivated the following problem. Given a set of integers $[k]=\{0,1,\cdots,k-1\}$ and a permutation $\pi$ on $[k]$, determine how many of the first $\a\leq k$ integers in $[k]$ are mapped to indices less than some $\b<k$ in the permuted sequence. For example, for the permutation $\pi =\left( \begin{smallmatrix} 0& 1& 2& 3& 4& 5& 6& 7& 8& 9\\ 9& 1& 7& 2& 5& 8& 6& 4& 0& 3 \end{smallmatrix}\right)$, and $\a=5,\b=6$, out of the first five integers only three $\{1,3,4\}$ map to positions less than six. Surprisingly, this problem has largely been unattempted before in the literature. In~\cite{2009_Mansour_LCSI}, a solution for linear permutation polynomials based on Dedekind sums~\cite{Knuth_Seminumerical,1971_Dieter} was proposed.

In this paper, we propose a mathematical formulation of this problem for general permutations using sums involving integer floor and the so-called ``saw-tooth'' functions (Section~\ref{s:prob_definition}), analogous to Dedekind sums. The arithmetic properties of these sums are analyzed in Section~\ref{s:rec_relations_eval_perm_stats}, and a set of mathematical identities used to solve the problem recursively are derived. We specialize to BRPs and give a mathematical characterization of these permutations, which have been mainly treated using numerical techniques in the literature to speed up radix-2 FFT computations and related transforms (e.g., see~\cite{1974_Polge,1987_Evans,1988_Rodriguez,1989_Elster,1991_Biswas,1991_Yong,1992_Orchard,1992_Jeong,1995_Rius,2001_Drouiche,2004_Prado,2007_Pei}). In~\cite{2009_Mansour_PBRI} a combinatorial solution based on bit manipulations was proposed. Here we derive in Section~\ref{s:perm_statistics} closed-form expressions for BRP statistics including descents/ascents, major index, excedances/descedances, inversions, serial correlations, and show that BRP sequences have weak correlation properties (i.e., a permuted index $\pi(j)$ strongly depends on the unpermuted index $j$). We propose a new statistic called permutation inliers, and prove that it characterizes the pruning gap of pruned interleavers. Using this statistic, we derive a recursive algorithm in Section~\ref{s:pruned_interl_min_inliers} to compute the minimum inliers count in a pruned BR interleaver (PBRI) in logarithmic time complexity, and apply it to parallelize a serial PBRI and reduce its latency and memory overhead. In Section~\ref{s:ext_composite_interleavers} we extend the discussion to block and stream interleavers that are composed of two or more permutations. In Section~\ref{s:applications}, we apply the inliers problem to design parallel BRPs for pruned FFTs, as well as parallel
pruned interleavers for LTE turbo codes. In Section~\ref{s:practical examples_impl_aspects}, we consider implementation aspects of the proposed algorithms and present hardware-efficient architectures. We perform simulations using several practical examples to demonstrate the advantages of the proposed algorithms. Finally, Section~\ref{s:conclusion} provides concluding remarks. Proofs of all theorems and lemmas are included in the Appendix.

%
\section{Preliminaries and Problem Formulation}\label{s:prob_definition}
Consider the set of integers $[k]\!\triangleq\!\{0,1,\cdots, k\!-\!1\}$, and let $\pi$ be a permutation on $[k]$. Denote by $(j_{n-1} \cdots j_1j_0)_2$ the $n$-bit binary representation of $j\!\in\![k]$, where $k\!=\!2^n$ and $j_i\!\in\!\{0,1\}$ for $i\!=\!0,\cdots,n\!-\!1$. The bit-reversal of $j$ is defined as $\pi_n(j)\!\triangleq\!(j_0j_1 \cdots j_{n-1})_2\!=\!\sum_{i=0}^{n-1}j_i2^{n-i}$. Note that $\pi_n\!(\pi_n(j))\!=\!j$ and hence $\pi_{n}^{-1}\!=\!\pi_n$. The goal is to characterize the so-called \emph{permutation statistics} of $\pi$ when $\pi$ is the bit-reversal permutation. The subject of permutation statistics dates back to Euler~\cite{1997_Clarke_euler-mahonian}, but was formally established as a discipline of mathematics by MacMahon in~\cite{1960_MacMahon_books,1913_MacMahon}. We start with some definitions.

A \emph{fixed point} of $\pi$ is an integer $i\in[k]$ such that $\pi(i)=i$. An \emph{excedance}~\cite{1960_MacMahon_books} of $\pi$ is an integer $i$ such that $\pi(i)>i$. Denote by $\text{FP}(\pi)$ and $\text{EXC}(\pi)$ the sets consisting of all fixed points and all excedances of $\pi$, respectively, and by $\#\text{FP}(\pi)$ and $\#\text{EXC}(\pi)$ the number of fixed points and excedances of $\pi$ (sometimes called \emph{excedance number}). An element of a permutation that is neither a fixed point nor an excedance is called a \emph{descedance}. For example, the permutation $\pi =\left( \begin{smallmatrix} 0& 1& 2& 3& 4& 5& 6& 7& 8& 9\\ 3& 1& 7& 2& 5& 8& 6& 4& 0& 9 \end{smallmatrix}\right)$ has the fixed points $\text{FP}(\pi)=\{1,6,9\}$ and the excedances $\text{EXC}(\pi)=\{0,2,4,5\}$, and hence $\#\text{FP}(\pi)=3$ and $\#\text{EXC}(\pi)=4$.

We say that $i\leq k-2$ is a \emph{descent}~\cite{1960_MacMahon_books} of $\pi$ if $\pi(i) > \pi(i+1)$. Similarly, $i\leq k-2$ is an \emph{ascent} of $\pi$ if $\pi(i) < \pi(i+1)$. Denote by $\text{DES}(\pi)$ and $\text{ASC}(\pi)$ the set of descents and the set of ascents of $\pi$, respectively, and by $\#\text{DES}(\pi)$ and $\#\text{ASC}(\pi)$ denote the number of descents and ascents of $\pi$. The \emph{major index}~\cite{1960_MacMahon_books} of $\pi$, $\text{maj}(\pi)$, is the sum of the descents, i.e. $\text{maj}(\pi)=\sum_{i\in \text{DES}(\pi)}i$. In the previous example, the descents are $\text{DES}(\pi)=\{0,2,5,6,7\}$ and hence $\#\text{DES}(\pi)=5$, the ascents are $\text{ASC}(\pi)=\{1,3,4,8\}$ and hence $\#\text{ASC}(\pi)=4$, and the major index is $\text{maj}(\pi)=0+2+5+6+7=20$.

A pair $(\pi(i),\pi(j))$ is called an \emph{inversion}~\cite{1960_MacMahon_books} of $\pi$ if $i < j$ and $\pi(i) > \pi(j)$. The set consisting of all inversions of $\pi$ is denoted by $\text{INV}(\pi)$ and its size by $\#\text{INV}(\pi)$. Continuing our example, the inversions are $\text{INV}(\pi)=\{(0,1),(0,3),(0,8),(1,8),(2,3),(2,4),(2,6),(2,7),(2,8),(3,8),(4,7),(4,8),(5,6),(5,7),(5,8),(6,7),(6,8),(7,8)\}$, and $\#\text{INV}(\pi)=18$.

The \emph{spread} of entries $i,j$ with span $\abs{i\!-\!j}\!<\! \a$ of $\pi$ measures how far $i,j$ are spread apart after permuting. The minimum spread~\cite{1995_Divsalar} of all distinct entries of $\pi$ with a span $<\a$ is defined as $\text{SP}_{\a}(\pi)\!=\!\min_{i,j\in[k]}\abs{\pi(i)\!-\!\pi(j)}\!+\!\abs{i\!-\!j},i\!\neq \!j$. For our example, no 2 consecutive entries map into consecutive entries, but entries 0, 1 map to $\abs{\pi(0)\!-\!\pi(1)}\!=\!2$, hence $\text{SP}_{2}(\pi)\!=\!3$.

\begin{figure}[hbtp]
\centering
\includegraphics[scale=0.80]{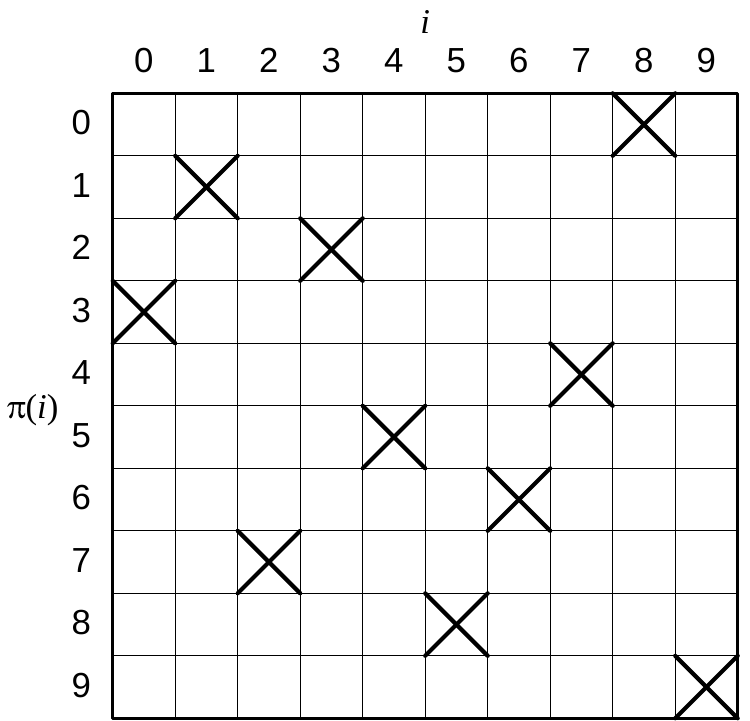}\vspace{-0.1in}
\caption{Array representation of the permutation $\pi =(3, 1, 7, 2, 5, 8, 6, 4, 0, 9)$.}
\label{f:permutation_array}
\end{figure}

 Often it is convenient to represent a permutation on $[k]$ by a $k\times k$ array with a cross in each of the squares $(i,\pi(i))$. Fig.~\ref{f:permutation_array} shows the array representation of the permutation in the previous example. Fixed points correspond to crosses on the main diagonal, excedances to crosses to the right of this diagonal, while descedances are represented by crosses on the left.\vspace{-0.1in}
\vspace{-0.1in}

In this paper we introduce a new permutation statistic useful for analyzing pruned interleavers called \emph{permutation inliers}. An integer $i\in [k]$ is called an $(\a,\b)$-\emph{inlier} of $\pi$ if $i<\a$ and $\pi(i)<\b$. Let $\text{INL}_{\a,\b}(\pi)$ denote the set of all $(\a,\b)$-inliers,\vspace{-0.1in}
\begin{equation}
\label{e:inliers_set_definition}
\text{INL}_{\a,\b}(\pi)\triangleq\{j\in [k] ~|~ j < \a, \pi(j) < \b\}, ~0<\a,\b \leq k,\\[-0.7em]
\end{equation}
and $\#\text{INL}_{\a,\b}(\pi)$ the number of $(\a,\b)$-inliers of $\pi$. We call determining $\text{INL}_{\a,\b}(\pi)$ for arbitrary $\pi$ the \emph{permutation inliers problem}. Similarly, an integer $i\in [k]$ is called an $(\a,\b)$-\emph{outlier} if $i<\a$ and $\pi(i)\geq\b$. $\text{OUL}_{\a,\b}(\pi)$ denotes the set of all $(\a,\b)$-outliers, and $\#\text{OUL}_{\a,\b}(\pi)$ their number: $\text{OUL}_{\a,\b}(\pi) \triangleq\{j\in [k] ~|~ j < \a, \pi(j) \geq \b\}, ~0<\a,\b \leq k$, or equivalently\vspace{-0.06in}
\begin{align}
\text{OUL}_{\a,\b}(\pi) &= [\a] - \text{INL}_{\a,\b}(\pi),\label{e:outliers_set_definition}\\[-3.1em]
\end{align}
where `$-$' is the set-difference operator. Referring to the $k\times k$ array diagram of $\pi$ in Fig.~\ref{f:permutation_array}, the $(\a,\b)$-inliers correspond to the crosses included in the rectangle with diagonal vertices $(0,0)$ and $(\a-1,\b-1)$. In the previous example, the $(5,7)$-inliers are $\text{INL}_{5,7}(\pi)=\{0,1,3,4\}$, while the outliers are the complement set $\text{OUL}_{5,7}(\pi)=[5]-\text{INL}_{5,7}(\pi)=\{2\}$.

The more general case of counting inliers in a bounded region $\a_1\!\leq\! j\! <\!\a_2$ and $\b_1\!\leq\! \pi(j)\! <\! \b_2$, $\text{INL}_{\a_1,\b_1,\a_2,\b_2}(\pi) \!=\! \{j\!\in\! [k] ~|~ \a_1\!\leq\! j \!<\!\a_2,~\b_1\!\leq\! \pi(j) \!<\! \b_2\}, ~0\!\leq\! \a_1 \!<\!\a_2 \!\leq\! k, ~0\!\leq\! \b_1\! <\!\b_2 \!\leq\! k$,
reduces to the original problem in~\eqref{e:inliers_set_definition} by observing that $\text{INL}_{\a_1,\b_1,\a_2,\b_2}(\pi) = \left\{\text{INL}_{\a_2,\b_2}(\pi) - \text{INL}_{\a_2,\b_1}(\pi)\right\} -
              \left\{\text{INL}_{\a_1,\b_2}(\pi) - \text{INL}_{\a_1,\b_1}(\pi)\right\}$.
Hence without loss of generality, we focus on~\eqref{e:inliers_set_definition} in the remainder of this paper.


There are no known techniques in number theory to analyze the structure of $\text{INL}$ for arbitrary permutations $\pi$ in the form presented above. However, with the help of the following lemma, we can recast the problem into one of evaluating a summation that involves integer floors, a device which is well-studied in number theory.\vspace{-0.1in}
\begin{lemma}
\label{lem:INL_definition}
The number of $(\a,\b)$-inliers of $\pi$ is given by\vspace{-0.1in}
\begin{align}
\label{e:INL_definition}
\#\text{INL}_{\a,\b}(\pi) = \sum_{j=0}^{k-1}\flrfrac{j-\a}{k}\!\!\flrfrac{\pi(j)-\b}{k} \leq \min{(\a,\b)}\\[-3.3em]
\end{align}
\IEEEproof
The floor function $\flr{x}$ is the largest integer less than or equal to the real number $x$. The first floor function in~\eqref{e:INL_definition} evaluates to $-1$ for $0\leq j<\a$ and 0 otherwise, while the second evaluates to $-1$ for $0\leq \pi(j)<\b$ and 0 otherwise. Hence the sum of their product counts the number of elements in $\text{INL}_{\a,\b}$. The number of outliers in the complement set is simply\vspace{-0.1in}
\begin{align}
\label{e:OUL_definition}
\#\text{OUL}_{\a,\b}(\pi) = \a - \sum_{j=0}^{k-1}\flrfrac{j-\a}{k}\!\!\flrfrac{\pi(j)-\b}{k}\\[-3.1em]
\end{align}
\end{lemma}
Moreover, if $\pi$ is an \emph{involution} (i.e., $\pi=\pi^{-1}$), then $\#\text{INL}_{\a,\b}(\pi)$ is symmetric in $\a$ and $\b$. Also, if $\pi$ is flipped into $\s:\s(j)\!=\!k\!-\!1\!-\!\pi(j)$, then the $({\a,\b})$-inliers of $\s$ are the $({\a,k\!-\!\b})$-outliers of $\pi$.
\begin{lemma}[Properties of $\#\text{INL}_{\a,\b}$]
\label{lem:properties of INL function}
If $\pi=\pi^{-1}$, then $\#\text{INL}_{\a,\b}=\#\text{INL}_{\b,\a}$ and hence\vspace{-0.07in}
\begin{align}
\label{e:INL_involution_symmetry}
\sum_{j=0}^{k-1}\flrfrac{j-\a}{k}\!\!\flrfrac{\pi(j)-\b}{k} = \sum_{j=0}^{k-1}\flrfrac{j-\b}{k}\!\!\flrfrac{\pi(j)-\a}{k}\\[-3.1em]
\end{align}
Moreover, if $\s(j)\!=\!k\!-\!1\!-\!\pi(j)$, then $\#\text{INL}_{\a,\b}(\s)\!=\!\a\!-\!\#\text{INL}_{\a,k\!-\!\b}(\pi)$ for $0<\a,\b<k$.
\IEEEproof Since $\pi\!=\!\pi^{-1}$, if $j_1\!<\!\a$ maps to $j_2\!=\!\pi(j_1)\! <\!\b$, then $j_2 \!<\!\b$ maps to $\pi(j_2)\!=\!j_1\!<\!\a$. Hence the two sums in~\eqref{e:INL_involution_symmetry} count the same elements. To prove $\#\text{INL}_{\a,\b}(\s)\!=\!\a\!-\!\#\text{INL}_{\a,k\!-\!\b}(\pi)$, substitute $\s(j)$ in~\eqref{e:INL_definition} and use $\flrfrac{-m}{n}\!=\!-\!\flrfrac{m-1}{n}\!-\!1$.~\IEEEQED
\end{lemma}

Similarly, we can recast the inversions problem into floor summations using the following lemma. First observe that inversions are the union of the outlier sets $\text{OUL}_{1,\pi(1)},\cdots,\text{OUL}_{k-1,\pi(k-1)}$, where the elements of each set $\text{OUL}_{\a,\pi(\a)}$ are paired with $\a$:\vspace{-0.1in}
\begin{equation}
\label{e:INV_outliers_relation}
\text{INV}(\pi) = \bigcup_{\a=1}^{k-1}\left(\a,\text{OUL}_{\a,\pi(\a)}(\pi)\right).\\[-0.6em]
\end{equation}
The notation $\left(\a,\text{OUL}_{\a,\pi(\a)}(\pi)\right)$ is the set of pairs $\{(\a,j) ~|~ j\in \text{OUL}_{\a,\pi(\a)}(\pi)\}$, $\a=1,\cdots,k-1$.
\begin{lemma}[Inversions]
\label{lem:Inversions_definition}
The number of inversions is given by\vspace{-0.1in}
\begin{align}
\label{e:I_definition}
\#\text{INV}(\pi)  = \frac{k(k-1)}{2} - \sum_{\a=0}^{k-1}\sum_{j=0}^{k-1}\flrfrac{j-\a}{k}\!\!\flrfrac{\pi(j)-\pi(\a)}{k}\\[-3.1em]
\end{align}
\IEEEproof From~\eqref{e:INV_outliers_relation}, it follows that $\#\text{INV}(\pi)$ is the sum of $\#\text{OUL}_{\a,\pi(\a)}$ for $\a=1,\cdots,k-1$. Also, $\#\text{OUL}_{0,\pi(0)}=0$ when $\a=0$. Summing~\eqref{e:OUL_definition} with $\b=\pi(\a)$ for $\a=0,\cdots,k-1$, the result follows.~\IEEEQED
\end{lemma}

For certain permutations such as the circular shift permutation $\pi(j)\!=\!j+c\!\pmod{k}$, $0\leq c<k$, it is possible to obtain closed form expressions for~\eqref{e:INL_definition} and~\eqref{e:I_definition}. First we need the following lemma.

%
\begin{lemma}
\label{lemma:prod_floors}
For any integers $p,q$, we have\vspace{-0.05in}
\begin{align}
\label{e:floorsum_pq}
\sum_{j=0}^{k-1}\flrfrac{j-p}{k}\!\! \flrfrac{j-q}{k}
        = \min(p\bmod k,q\bmod k) + \flrfrac{p}{k}\!\!\flrfrac{q}{k}k
                    +\flrfrac{p}{k}(q\bmod k) +\flrfrac{q}{k}(p\bmod k)\\[-3.2em]
\end{align}
\IEEEproof Write $p\!=\!\flrfrac{p}{k}k\!+\!p\!\pmod{k}$ and $q\!=\!\flrfrac{q}{k}k\!+\!q\!\pmod{k}$, then substitute in summation~\eqref{e:floorsum_pq}.~\IEEEQED
\end{lemma}
Now applying~\eqref{e:floorsum_pq} in~\eqref{e:INL_definition} for $\pi(j) \!=\!j+c\! \pmod{k}$, we have
\begin{align}
\#\text{INL}_{\a,\b} &= \sum_{j=0}^{k-1}\flrfrac{j-\a}{k}\!\!\flrfrac{(j+c)\bmod k-\b}{k}
              = \sum_{j=0}^{k-1}\flrfrac{j-\a}{k}\!\!\left(\flrfrac{j+c-\b}{k}-\flrfrac{j+c}{k}\right)\notag\\
              &= \min(\a,(\b-c)\bmod k) + \a\flrfrac{\b-c}{k} - \min(\a,k-c) + \a\label{e:P_circular_perm}\\[-3.1em]
\end{align}
since $0\leq\a,\b,c<k$. For example, for $k=32,c=7,\a=15,\b=19$, the number of $(15,19)$-inliers is $\#\text{INL}_{15,19}=12$.

To count the inversions, we substitute~\eqref{e:P_circular_perm} with $\b\!=\!\pi(\a)\!=\!(\a\!+\!c)\bmod k$ in~\eqref{e:OUL_definition}, then~\eqref{e:I_definition}. It is easy to verify that $\min(\a,(\left(\a\!+\!c\right)\bmod k\!-\!c)\bmod k)\!=\!\a$, and that $\a\flrfrac{\left(\a+c\right)\bmod k-c}{k}=-\a$ when $\a+c\geq k$, and 0 otherwise. Then
from~\eqref{e:I_definition} and~\eqref{e:OUL_definition}\vspace{-0.07in}
\begin{align}
\#\text{INV} &= \sum_{\a=0}^{k-1} \left(
                    - \min(\a,(\left(\a+c\right)\bmod k-c)\bmod k)
                    - \a\flrfrac{\left(\a+c\right)\bmod k-c}{k}
                    + \min(\a,k-c)\right)\notag\\
              &= - \sum_{\a=0}^{k-1}\a
                 + \sum_{\a=k-c}^{k-1}\a
                 + \sum_{\a=0}^{k-c-1}\a
                 + \sum_{\a=k-c}^{k-1}\!\!\left(k-c\right)
              = c(k-c)\label{e:I_circular_perm}
\end{align}
Equation~\eqref{e:I_circular_perm} agrees with the intuitive result because integers $c$ to $k-1$ occupy the first $k-c$ entries in ascending order in the permuted sequence, while integers $0$ to $c-1$ occupy the remaining $c$ entries. Hence the product $c(k-c)$ gives $\#\text{INV}$.

For other types of permutations such as polynomial-based permutations or BRPs, finding a closed form expression for sum~\eqref{e:INL_definition} is not as straightforward due to the presence of the floor functions. Fortunately, such summations can be more conveniently manipulated by replacing the floor function with the ``saw-tooth" function\vspace{-0.1in}
\begin{align}
\label{e:saw_function}
\saw{x}\triangleq x -\flr{x} - \frac{1}{2}+\frac{1}{2}\delta(x),\\[-3.2em]
\end{align}
where $\delta(x)=1$ if $x$ is an integer, and 0 otherwise.

It will be shown in this paper that for any permutation $\pi$ that fixes the zero element (i.e., $\pi(0)=0$),\vspace{-0.05in}
\begin{align}
\label{eq:INL_saw_formula_K}
\#\text{INL}_{\a,\b} = \frac{\a\b}{k} + \sum_{j=0}^{k-1}
            \left[\!\sawfrac{j-\a}{k}\!-\!\sawfrac{j}{k}\!\right]\!\!
            \left[\!\sawfrac{\pi(j)-\b}{k}\!-\!\sawfrac{\pi(j)}{k}\!\right] + K_{\text{INL}},\\[-3.2em]
\end{align}
where $K_{\text{INL}}$ is a constant. Hence in the remainder of this paper, we focus on evaluating summations of the form\vspace{-0.05in}
\begin{align}
\label{eq:sum_product_saw_fractions}
\sum_{j=0}^{k-1}\!\sawfrac{j-\a}{k}\!\!\!\sawfrac{\pi{(j)}-\b}{k}, \quad \a,\b\in[k],\\[-3.2em]
\end{align}
when $\pi$ is the BRP, which are reminiscent of Dedekind sums~\cite{Knuth_Seminumerical}. Evaluating~\eqref{eq:sum_product_saw_fractions} for arbitrary permutations is still an open research problem. For BRPs, we show that these summations can be evaluated recursively in $\log_2k-1$ steps using only integer addition and shift operations. This result is extended to evaluate summations~\eqref{e:INL_definition} and~\eqref{e:I_definition} using simple mathematical recursions.

Moreover, for the purposes of characterizing the randomness of pseudo-random numbers generated by BRPs, we study the serial correlations between an entry in the bit-reversed sequence and all its successors. We show that these serial correlations require evaluating related sums of the form $\sum_{j=0}^{k-1}\!\saww{\!\frac{\pi(j)}{k}\!}\!\!\!\saww{\!\frac{\pi(j+p)}{k}\!}$ for all $0\leq p < k$ successors. We propose a simple recursive integer algorithm to evaluate these sums in logarithmic time-complexity.


%
\section{Recursive relations for Evaluating Permutation Statistics}\label{s:rec_relations_eval_perm_stats}
In this section, we derive recursive expressions for summations involving the saw-tooth function that are useful for computing permutation statistics. We start with the following basic properties which immediately follow from the definition in~\eqref{e:saw_function}: $\saw{\hf}=0$, $\saw{n}=0$, $\saw{-x} =-\saw{x}$,
$\saw{\frac{n+\theta}{k}} = \saw{\frac{\vphantom{\theta}n}{k}} + \frac{\theta}{k} -\hf\deltafrac{\vphantom{\theta}n}{k}$ for integers $n,k$, real $\theta$, $0<\theta<1$, and \vspace{-0.25in}
\begin{singlespace}
\begin{align}
  \sawfrac{\vphantom{1}n}{2}    &=0,\quad\text{integer $n$},\label{e:saw_property_n_over_2}\\
  \saw{x+n}         &=\saw{x},\quad\text{integer $n$},\label{e:saw_property_x_plus_integer}\\
  \sawww{x\pm\hf}      &= \saw{2x} - \saw{x}, \label{e:saw_property_x_plus_half}
\end{align}
\end{singlespace}

Next, consider the sum of product of the $m^{\text{th}}$-power integers $0^m,\cdots, (k\!-\!1)^m$ and the bit-reversed integers $\pi_{\!n}(0),\cdots, \pi_{\!n}(k\!-\!1)$:\vspace{-0.05in}
\begin{align}
\label{e:sum_j^m_pi_j}
   J_m(k) \triangleq \sum_{j=0}^{k-1}j^m\pi_n{(j)},~ m \geq 0\\[-3.1em]
\end{align}

%
\begin{theorem} $J_m$ can be evaluated using the following recurrence:\vspace{-0.05in}
\label{th:sum_j^m_pi_j_formula}
\begin{align}
\label{e:sum_j^m_pi_j_formula}
   J_m(k) = 2J_m(k/2)+(k/2)^m + \sum_{r=0}^{m}\! \binom{m}{r}\frac{k^r}{2^r}
            \left[2J_{m-r}(k/2)
                +\frac{(k/2-1)^{m-r+1}}{m-r+1}\sum_{s=0}^{m-r}(1-k/2)^{-s}B_s\binom{m-r+1}{s}
            \right]\\[-3.1em]
\end{align}
with initial conditions $J_m(1)=0$ and $m\geq 0$, where $B_s$ are the Bernoulli numbers. Also, since in~\eqref{e:sum_j^m_pi_j} the order in which integers are summed is irrelevant and $\pi_n=\pi_n^{-1}$, we have $\sum_{\pi_n(j):j=0}^{k-1}(\pi_n\!(j))^m j = \sum_{j=0}^{k-1}j^m\pi_n{(j)} = J_m(k)$.~\IEEEQED
\end{theorem}

%
%
\begin{corollary} \label{cor:sum_j12_pi_j_formula}
For $m=0$ we have $J_0(k)=4J_0(k/2)+k/2=k(k-1)/2$. For $m=1,2$, we have
\begin{align}
    J_1(k) &= \sum_{j=0}^{k-1}j\pi_n{(j)} = 4J_1(k/2) + \frac{k(k-2)(k+1)}{8} + \frac{k^2}{4}
           = \frac{k^3}{4} + \frac{(n-4)k^2}{8} + \frac{k}{4}\label{e:sum_j1_pi_j_formula}\\
    J_2(k) &= \sum_{j=0}^{k-1}j^2\pi_n{(j)} = 4J_2(k/2) + \frac{k^4}{8}+ \frac{(3n-7)k^3}{48} - \frac{k^2}{8} +
                \frac{k}{12}
           = \frac{k^4}{6} - \frac{(20-6n)k^3}{48} + \frac{(8-3n)k^2}{24} - \frac{k}{12}\notag\qquad~\IEEEQED
\end{align}
\end{corollary}

Moreover, the function $\saw{x}$ possesses many interesting properties when $x$ is a rational number $j/k$, specifically when $\saw{j/k}$ is summed over a complete residue system modulo $k$. The following lemma summarizes some of these identities:\vspace{-0.05in}
%
\begin{lemma}[Sum of saw-fractions over a complete residue system]
\label{lem:saw_sums_over_residue_system}
\begin{singlespace}
\begin{align}
  \sum_{j=0}^{k-1}\!\sawfrac{j}{k}       &= 0 \label{e:property_sawsum_j}\\[-0.4em]
  \sum_{j=0}^{k-1}\!\sawfrac{j+w}{k}     &= \saw{w};~~\text{$w$ any real}\label{e:property_sawsum_j_plus_w}\\[-0.4em]
  \sum_{j=0}^{k-1}\!\sawfrac{\pi(j)}{k}  &= 0;~\pi~\text{any permutation on $[k]$}\label{e:property_sawsum_pi_j}\\[-0.4em]
  \sum_{j=0}^{k-1}\!\sawfrac{jh}{k}
        &= 0;~~\text{$h$ and $k$ not necessarily co-prime}\label{e:property_sawsum_jh}\\[-0.4em]
  \sum_{j=0}^{k-1}\!\sawfrac{\pi(j)+w}{k} &= \saw{w};~\pi~\text{any permutation on $[k]$; $w$ any real.}\label{e:property_sawsum_pi_j_plus_w}~\IEEEQED\\[-2.5em]
\end{align}
\end{singlespace}
\end{lemma}

Further properties are derived when $\saww{\!\frac{\vphantom{\pi_n(j)}j-b}{k}\!}$ or
$\saww{\!\frac{\pi_n(j)-b}{k}\!}$ for $\pi_n$ are summed over \emph{half} a residue system for shift values $b$.
\begin{lemma}[Sum of saw-fractions over half a residue system]
\label{lem:saw_sums_over_half_residue_system} Let $b$ be a non-negative integer, then\vspace{-0.05in}
\begin{align}
    4\sum_{j=0}^{k/2-1}\!\!\sawfrac{j-b}{k} &{=}
        \left\{\!\!\!
          \begin{array}{ll}
            \!\hphantom{-} 2(b\bmod k) - k/2 + 1, & \hbox{$0\leq b\bmod k<k/2$;} \vspace{-0.1in}\\[-0.2em]
            \! -2(b\bmod k) + 3k/2 - 1, & \hbox{$b\bmod k \geq k/2$.}
          \end{array}
        \right.\label{e:saw_sum_k+b_2}\\[-0.4em]
    \sum_{j=0}^{k/2-1}\!\!\sawfrac{\pi_n(j)\pm b}{k} \!&=\! 0 \label{e:saw_sum_pi_+b_2}\\[-3.3em]
\end{align}
In particular, when $b=0$, $4\sum_{j=0}^{k/2-1}\!\!\saww{\frac{j}{k}} =1-k/2$.~\IEEEQED
\end{lemma}
\vspace{-0.05in}

Summations of saw-fractions $\saww{j^2/k}$ and floor-fractions $\flr{j^2/k}$ involving squared integers have never been attempted before in the literature. Below we derive an interesting identity for these sums over a complete residue system.

%
\begin{lemma}[Sum of saw and floor fractions involving squared integers over a complete residue system]\vspace{-0.07in}
\label{lem:saw_sum_j2_over_residue_system}
\begin{align}\\[-3.1em]
  \sum_{j=0}^{k-1}\!\sawfrac{j^2}{k}  &= \left\{
                                           \begin{array}{ll}
                                            \!\!\!-\sqrt{k}+ 3/2, & \hbox{$\log_2{k}$ even;}\vspace{-0.1in} \\
                                            \!\!\!-3\sqrt{k/8}+ 3/2, & \hbox{$\log_2{k}$ odd.}
                                           \end{array}
                                         \right.\label{e:saw_sum_j2}\vspace{-0.2in}\\
    \sum_{j=0}^{k-1}\!\flrfrac{j^2}{k}  &= \left\{
                                           \begin{array}{ll}
                                            \!\!\!\frac{k^2}{3}-k+\frac{3\sqrt{k}}{2}-\frac{4}{3}, &
                                                \hbox{$\log_2{k}$ even;} \vspace{-0.1in}\\
                                            \!\!\!\frac{k^2}{3}-k+\sqrt{2k}-\frac{4}{3}, & \hbox{$\log_2{k}$ odd.}
                                           \end{array}
                                         \right.\label{e:floor_sum_j2}\\[-3.1em]
\end{align}
\end{lemma}
\vspace{-0.05in}

Moreover, for the arithmetic analysis of BRPs $\pi_n$, summations that involve products of saw-fractions of the form $\saww{\!\frac{\vphantom{\pi_n{(j)}}j}{k}\!}\!\!\!\saww{\!\frac{\pi_n{(j)}}{k}\!}$ and their variations are of particular interest.

%
\begin{lemma}[Sum of products of saw-fractions]\vspace{-0.05in}
\label{lem:prod_saw_fractions}
\begin{align}\\[-3.1em]
        R(k) \triangleq 4k\sum_{j=0}^{k-1}\!\sawfrac{j}{k}\!\!\!\sawfrac{\pi_n{(j)}}{k}
             = \frac{k\log_2(k)}{2} - k + 1 \label{e:R_k_definition}~\IEEEQED\\[-3.1em]
\end{align}
%
\end{lemma}
\vspace{-0.05in}

More generally, sums of products of the form $\saww{\!\frac{\vphantom{\pi_n(j)}j-b}{k}\!}\!\!\!\saww{\!\frac{\pi_n(j)-c}{k}\!}$ for shift integers $b,c$ can also be evaluated efficiently.\vspace{-0.05in}

\begin{lemma}[Sum of products of saw-fractions with a shift] Let $c^*\!=\!\pi_{\!n\!-\!1}\!\left( (c\!-\!1)/2 \right)$ if $c$ is odd, and $c^*\!=\!\pi_{\!n\!-\!1}\!\left( c/2 \right)$ if $c$ is even.\vspace{-0.25in}
\label{lemma:S_sum}
\begin{align}\\[-3.1em]
\label{e:S_sum}
    S(k,b,c) &\triangleq 4k\sum_{j=0}^{k-1}\!\sawfrac{j-b}{k}\!\!\!\sawfrac{\pi_n(j)-c}{k}
             \!=\! \left\{\!
                  \begin{array}{ll}
                    \!\! 2S(k/2,b,(c-1)/2) + 2k\!\saww{\!\frac{c^*-b}{k}\!} + K_{S}, & \hbox{$c$ odd;} \vspace{-0.05in}\\
                    \!\! 2S(k/2,b,c/2) -2k\!\saww{\!\frac{c^*-b}{k}+\hf\!} + K_{S}, & \hbox{$c$ even.}
                  \end{array}
                \right.\\[-0.2em]
    K_{S} &=   \left\{\!
                        \begin{array}{ll}
                          \!\!\! -2b + k/2 - 1 , & \hbox{$0\leq b<k/2$;}\vspace{-0.1in}\\
                          \!\!\! \hphantom{-}2b -3k/2 + 1 , & \hbox{$k/2\leq b<k$.}
                        \end{array}
                      \right.\label{w:K_S}\\[-3.1em]
\end{align}
\end{lemma}

Furthermore, we investigate summations that involve products of differences of saw-functions similar to those in~\eqref{eq:INL_saw_formula_K}:\vspace{-0.05in}
\begin{align}
\label{e:T_definition}
T(k,b,c) \triangleq 4k \sum_{j=0}^{k-1}\!\left[\!\sawfrac{j-b}{k}\!\!-\!\!\sawfrac{j}{k}\!\right]
            \!\!\!\left[\!\sawfrac{\pi_{n}(j)-c}{k}\!\!-\!\!\sawfrac{\pi_{n}(j)}{k}\!\right]\\[-3.1em]
\end{align}

\begin{lemma} If $c\!=\!0$ or $\!b=\!0$, then $T(k,b,c)\!=\!0$. Else, let $c^*\!=\!\pi_{\!n\!-\!1}\!\left(\! (c\!-\!1)/2 \!\right)$ if $c$ is odd, $c^*\!=\!\pi_{\!n\!-\!1}\!\left( c/2 \right)$ if $c$ is even. Then\vspace{-0.08in}
\label{lemma:T_sum_minus}
\begin{align}
\label{e:T_sum_minus}\\[-3.0em]
    T(k,b,c) \!=\! \left\{\!
                  \begin{array}{ll}
                    \!\! 2T(k/2,b,(c-1)/2) \!-\! 4b \!-\! k \!\left(\!
                        2\!\flrfrac{c^*-b}{k} \!-\! 2\!\flrfrac{2b}{k} \!-\! \deltafrac{c^*-b}{k}
                        \!+\!\deltafrac{c^*}{k} \!+\! \deltafrac{2b}{k}\!\right), & \hbox{$c$ odd;} \vspace{-0.05in}\\
                    \!\! 2T(k/2,b,c/2) + k\!\left(2\!\flr{\frac{c^*-b}{k}+\hf} \!+\!2\!\flr{\frac{b}{k}\!+\!\hf}
                    \!-\! \delta\!\left(\frac{c^*-b}{k}+\hf\right) \!-\! \delta\!\left(\frac{b}{k}+\hf\right)
                    \!\right), & \hbox{$c$ even.}
                  \end{array}
                \right.\\[-3.2em]
\end{align}
This recursion (and~\eqref{e:S_sum}) can be evaluated using integer arithmetic in at most $n-1$ steps since $T(2,b,c)=0$.~\IEEEQED
\end{lemma}

Note that the recursive solution in~\eqref{e:T_sum_minus} is similar to that for linear permutation polynomials involving Dedekind sums~\cite{2009_Mansour_LCSI}.

Specifically, when $c=\pi_n{(b)}$ in~\eqref{e:T_definition}, a closed form expression for the sum $\sum_{b=0}^{k-1}T(k,b,\pi_n(b))$ can be derived.
These sums appear in equations similar to~\eqref{e:I_definition} for counting inversions.
%
\begin{lemma}
\label{lemma:U_sum_minus}\vspace{-0.06in}
\begin{align}
    U(k) &\triangleq \sum_{b=0}^{k-1}\sum_{j=0}^{k-1}\!
         \left[\!\sawfrac{j-b}{k}\!\!-\!\!\sawfrac{j}{k}\!\right]\!\!\!
         \left[\!\sawfrac{\pi_n(j)-\pi_n(b)}{k}\!\!-\!\!\sawfrac{\pi_n(j)}{k}\!\right]
         = \frac{k(\log_2{k}-2)}{8}+\frac{1}{4}\label{e:U_sum_minus}\\[-3.1em]
\end{align}
\end{lemma}

We next consider sums of products of saw-fractions involving $\frac{\pi_n(j)}{k}$ and their $p^\text{th}$ successors $\frac{\pi_n(j+p)}{k}$. These sums are used in studying the serial correlation properties of BRPs.

%
\begin{lemma}[Sum of products of saw-fractions and their $p^\text{th}$-successors] Let $0\leq p < k$, then\vspace{-0.05in}
\label{lemma:C_k_p_sum}
\begin{align}
\label{e:C_k_p_sum}\\[-3.1em]
    C(k,p) &\triangleq k^2\sum_{j=0}^{k-1}\!\sawfrac{\pi_n(j)}{k}\!\!\!\sawfrac{\pi_n(j+p)}{k}
             \!=\! \left\{\!
                  \begin{array}{ll}
                    \!\! k(k-1)(k-2)/12, & \hbox{$p=0$;} \vspace{-0.1in}\\
                    \!\! k(k-2)(k-4)/12, & \hbox{$p=k/2$;}\vspace{-0.1in}\\
                    \!\! 8C(k/2,p) +\left(1 - \frac{3}{2^{v+1}}\right)\frac{k^2}{2} + \max(p,k-p) & \hbox{otherwise;}
                  \end{array}
                \right.\\[-3.1em]
\end{align}
where $0\leq v<n$ is the position of the least-significant one-bit in the binary representation of $p$ (starting from 0).~\IEEEQED
\end{lemma}
For example, when $n=6$, $k=2^n=64$ and $p=1=00000\underline{\mathbf{1}}_{(2)}$, we have $v=0$ and\vspace{-0.05in}
\begin{align}
\label{e:ser_corr_C_k_1}
    C(k,1) = 8C(k/2,1) - \frac{k^2}{4} + k - 1 = \frac{(k-1)(k-2)(-5k+6)}{84},\\[-3.1em]
\end{align}
and when $p=2=0000\underline{\mathbf{1}}0_{(2)}$, we have $v=1$ and $C(k,2) = 8C(k/2,2) + \frac{k^2}{8} + k - 2 = \frac{(k-4)(8k^2+11k-12)}{168}$.
A simple algorithm for computing $C(k,p)$ using integer operations is shown below. Note that $k^2((2k-12)u^2 + 18u - 5k)/24u^2$ is an integer since $12u^2$, $18u$, and $2u^2-5$ are divisible by 3 since $u$ is a power of 2 (easily proved by induction).
\begin{algorithm}
\centering
\caption{Integer algorithm to compute $C(k,p)$. $k\!=\!2^n$ and $u\!=\!2^v$.}\label{alg:C}
\begin{algorithmic}[0]
\State $C \gets 0, ~k'\gets k$
\For{$j=0$ to $n-v-2$}
    \State $C \gets C + 8^j\max{(p,k'-p)}$
    \State $k' \gets k'/2$
    \State $p \gets p \bmod k'$
\EndFor
\State $C \gets C + k^2((2k-12)u^2 + 18u - 5k)/24u^2$ \Comment Accumulation of the terms $\left(1 - \frac{3}{2^{v+1}}\right)\frac{k^2}{2}$ in~\eqref{e:C_k_p_sum}
\end{algorithmic}
\end{algorithm}

Another related sum is one involving \emph{shifted} saw-fractions $\frac{\pi_n(j)-a}{k}$ and their first-successors $\frac{\pi_n(j+1)-b}{k}$, for shift values $a,b$. These sums are used in studying the probability of consecutive BRP terms falling within specific intervals. Let\vspace{-0.05in}
\begin{align}
\label{eq:V_k_a_b_def}
    V(k,a,b)&\triangleq k^2\sum_{j=0}^{k-1}\!\sawfrac{\pi_{n}(j)\!-\!a}{k}\!\!\!\sawfrac{\pi_n(j\!+\!1)\!-\!b}{k}\\[-3.4em]
\end{align}

%
\begin{lemma}[Sum of products of shifted saw-fractions and their shifted successors]
\label{lem:V_k_a_b}
\begin{align}
\label{e:V_k_a_b}\\[-3.1em]\hspace{-0.25in}
    V(k,a,b) &= 8V(k/2,a'/2,b'/2)
        \!+\! k^2\!\left[\!\!\sawfrac{a\!+\!1}{k}\!\!-\!\! \sawfrac{a\!+\!2}{k} \!\!\right]\!\!\!
                   \left[\!\!\sawfrac{b}{k} \!\!-\!\! \sawfrac{b\!-\!1}{k} \!\!\right]
        \!-\! \frac{k^2}{2}\!\!\sawfrac{a''}{k}
        \!-\! \frac{k^2}{2}\!\!\sawfrac{b''}{k}
        \!+\! \left(\! \frac{k^2}{4}\deltafrac{b''}{k} \!-\! \frac{k}{2} \!\right)\!\cdot\! e\\[-3.1em]
\end{align}
where $a'=a$ if $a$ is even and $a'\!=\!a-1$ if $a$ is odd, $b'=b$ if $b$ is even and $b'\!=\!b-1$ if $b$ is odd, $a'' = 2\pi_{n-1}(\pi_{n-1}(a'/2)+1) - b'$ if $a$ is even and $a'' = -2\pi_{n-1}(\pi_{n-1}(a'/2)+1) + b'$ if $a$ is odd, $b'' = 2\pi_{n-1}(\pi_{n-1}(b'/2)-1) - a'$ if $b$ is even and $b'' = -2\pi_{n-1}(\pi_{n-1}(b'/2)-1) + a'$ if $b$ is odd, and $e=1$ if both $a,b$ are odd or both even and $e=0$ otherwise.~\IEEEQED
\end{lemma}

Finally, generalizing~\eqref{eq:V_k_a_b_def} into products of differences we have the following lemma:\vspace{-0.05in}
\begin{align}
\label{e:W_k_a_b_def}\\[-3.1em]
W(k,a,b)&\triangleq k^2\sum_{j=0}^{k-1}\!
            \left[\!\sawfrac{\pi_{n}(j)\!-\!a}{k}\!\!-\!\!\sawfrac{\pi_{n}(j)}{k}\!\right]\!\!\!
            \left[\!\sawfrac{\pi_n(j\!+\!1)\!-\!b}{k}\!\!-\!\!\sawfrac{\pi_n(j\!+\!1)}{k}\!\right],\\[-3.3em]
\end{align}

%
%
\begin{lemma}[Sum of products of differences of shifted saw-fractions and their shifted successors]\vspace{-0.05in} \label{lem:W_k_a_b}
\begin{align}
\label{e:W_k_a_b_iterative_formula}\\[-3.1em]
    W(k,a,b) &= 8W(k/2,a'/2,b'/2) +
            (2e_b\!-\!1)\frac{k^2}{2}\!\!\left[\!\sawfrac{b''\!-\!a'}{k}\!\!-\!\!\sawfrac{b''}{k}\!\right] \!+\!
            (2e_a\!-\!1)\frac{k^2}{2}\!\!\left[\!\sawfrac{a''\!-\!b'}{k}\!\!-\!\!\sawfrac{a''}{k}\!\right]\\
            &\qquad +\frac{k^2}{4}\!\left[2\!\sawfrac{k/2\!-\!b'}{k} \!-\! (1\!-\!e_b) \deltafrac{k/2\!-\!b'}{k}
                    \!+\!e\deltafrac{a''\!-\!b'}{k}
                    \!-\! \deltafrac{a'\!+\!2}{k}\!\!\!
                    \left(\!\deltafrac{b'}{k} \!-\! 1 \!-\! e_a\! \right)\!\right]
                    \!-\! \frac{a'k}{2} \!-\!e_a e_b k \notag\\[-3.1em]
\end{align}
where $e_a\!=\!0$ if $a$ is even and 1 if $a$ is odd, $e_b\!=\!0$ if $b$ is even and 1 if $b$ is odd, $e\!=\!e_a e_b\!+\!(1\!-\!e_a)(1\!-\!e_b)$, $a'\!=\!a\!-\!e_a$, $b'\!=\!a\!-\!e_b$, $a''\!=\! 2\pi_{n\!-\!1}\!\left(\pi_{n\!-\!1}\left(a'/2\right)\!+\!1\right)$, and $b'' \!=\! 2\pi_{n\!-\!1}\!\left(\pi_{n\!-\!1}\left(b'/2\right)\!-\!1\right)$.~\IEEEQED
\end{lemma}
\vspace{-0.05in}

%
\section{Permutation Statistics of Bit-Reversal Permutations}\label{s:perm_statistics}
In this section we derive permutation statistics for BRPs and present a solution for the permutation inliers problem using the results from Section~\ref{s:rec_relations_eval_perm_stats}. Let $\{X_j\}$ be the sequence $X_j \!=\! \pi_n(j), j\!=\!0,1,\cdots,2^{n}\!-\!1$ generated by the BRP on $n$ bits.\vspace{-0.06in}

%
\subsection{Descents and Major Index}
We start by determining the number of descents induced by a BRP.\vspace{-0.05in}

%
\begin{lemma}[A priori law and number of descents]
\label{lem:theorem_apriori_law}
The probability that $X_j > X_{j+1}$ is $\hf$ and the number of descents is $\#\text{DES}(\pi_n)=k/2$. More generally, the probability that $X_j > X_{j+1} > \cdots > X_{j+t}$ is 0 for $t\geq 2$.
\IEEEproof Consider the $n$-bit binary representation of $j,~j=0,\cdots,2^n-1$. We count the number of occurrences of $X_j > X_{j+1}$ in the $\{X_j\}$, with subscripts taken $\bmod~k$. Obviously, even integers have a $0$ in their least-significant bit position, while odd integers have a $1$. Then $X_j \!>\! X_{j+1}$ and $X_j\! >\! X_{j-1}$ when $j$ is odd ($j+1$ is even) since $\pi_n(j) > \pi_n(j+1)$ and $\pi_n(j) > \pi_n(j-1)$. Hence $X_{j} > X_{j+1}$ exactly $k/2$ times, which equals the number of descents. When $t\geq 2$, then $X_j > X_{j+1} > \cdots > X_{j+t}$ cannot occur because the superscript $j+m$ is even at least for one $0\leq m \leq t-1$ and hence $X_{j+m} < X_{j+m+1}$.~\IEEEQED
\end{lemma}\vspace{-0.05in}

Obviously, the number of ascents is $\#\text{ASC}(\pi_n) = k - \#\text{DES}(\pi_n)=k/2$. The \emph{major index} of $\pi_n$ is the sum of the indices of the first number in each pair that is a descent.\vspace{-0.05in}

%
\begin{lemma}[Major index]
\label{lem:majro index}
The major index of a BRP is $\text{maj}(\pi_n)=k^2/4$.
\IEEEproof From Lemma~\ref{lem:theorem_apriori_law}, descents occur at odd indices, hence the major index is $\sum_{j=0,~\mathrm{odd}}^{k-1}j=k^2/4$.~\IEEEQED
\end{lemma}
\vspace{-0.1in}

%
\subsection{Fixed Points, Excedances, and Descedances}\vspace{-0.05in}
For a BRP, the number of fixed points is the number of palindromes (when $\pi_n(i)=i$):\vspace{-0.1in}
\begin{align}\\[-3.1em]
\label{eq:num_fixed_points}
    \#\text{FP}(\pi_n)= 2^{\lceil \log_2(k)/2 \rceil}\\[-3.3em]
\end{align}
The sum of all fixed points, as well as their squares, can be evaluated using the following lemma:
%
\begin{lemma}[Sum of Fixed Points]\vspace{-0.15in}
\label{lem:sum_of_fixed_points}
\begin{align}
    F_1(k) &\triangleq \sum_{i \in \text{FP}(\pi_n)}i =\left\{\!\!
                      \begin{array}{ll}
                        \!\!\sqrt{k}(k-1)/2, & \hbox{$n$ even;} \vspace{-0.1in}\\
                        \!\!\sqrt{k}(k-1)/\sqrt{2}, & \hbox{$n$ odd.}
                      \end{array}
                    \right.\label{eq:sum_of_fixed_points}\\[-0.2em]
    F_2(k) &\triangleq \sum_{i \in \text{FP}(\pi_n)}i^2 =\left\{\!\!
                      \begin{array}{ll}
                        \!\!k^2\sqrt{k}/3+k\sqrt{k}(n-4)/8 +\sqrt{k}/6, & \hbox{$n$ even;} \vspace{-0.1in}\\
                        \!\!2k^2\sqrt{k}/3\sqrt{2}+k\sqrt{k}(n-5)/4\sqrt{2} +\sqrt{k}/3\sqrt{2}, & \hbox{$n$ odd.}
                      \end{array}  
                    \right.\label{eq:sum_of_squared_fixed_points}\\[-3.1em]
\end{align}
\end{lemma}

An \emph{excedance} of $\pi_n$ is an integer $j\!\in\![k]$ such that $\pi_n(j)\! >\! j$.
\vspace{-0.05in}

%
\begin{lemma}[Excedance Number and Probability $X_j > j$]
\label{lem:prob_j_greater_Xj}
The excedance number of a BRP is $\#\text{EXC}(\pi_n)=(k-2^{\lceil \log_2(k)/2 \rceil})/2$
and the probability that $X_j > j$ is $\#\text{EXC}(\pi_n)/k$.
\IEEEproof Consider the $n$-bit binary representation of $j,~j=0,\cdots,2^n-1$. These representations can be partitioned such that $j=\pi_n(j)$, $j<\pi_n(j)$, or $j>\pi_n(j)$. From~\eqref{eq:num_fixed_points}, the number of palindromes is $2^{n/2}$ when $n$ is even or $2^{(n+1)/2}$ when $n$ is odd. There are equal number of remaining representations corresponding to $j<\pi_n(j)$ and $j>\pi_n(j)$. Hence the number of times $j > X_{j}$ or $X_{j} > j$ is $(2^n-2^{\lceil n/2 \rceil})/2$, and the probability that $X_j> j$ is $(2^n-2^{\lceil n/2 \rceil})/2k$.~\IEEEQED
\end{lemma}

%
\begin{corollary}\vspace{-0.05in}
    \label{cor:flrsum_pij-j}
    \begin{align}\\[-3.5em]
        \label{e:flrsum_pij-j}
            -\sum_{j=0}^{k-1}\flrfrac{j-\pi_n(j)}{k} =  \frac{k-2^{\lceil \log_2(k)/2 \rceil}}{2}\\[-3.1em]
    \end{align}
\IEEEproof The floor functions $-\!\flrfrac{j-\pi_n(j)}{k}$ evaluate to $+1$ when $\pi_n(j) > j$, hence their sum is the excedance number.~\IEEEQED
\end{corollary}\vspace{-0.05in}

Next we consider the sum of all excedances of $\pi_n$, $E_1(k)\triangleq\sum_{j\in\text{EXC}(\pi_n)}j$.\vspace{-0.05in}

%
\begin{lemma}[Sum of Excedances]\vspace{-0.05in}
\label{lem:sum of excedances}
\begin{align}\\[-3.3em]
        \label{e:sum of excedances}
            E_1(k) = -\sum_{j=0}^{k-1}j\flrfrac{j-\pi_n(j)}{k} \!= \!
                \left\{\!\!\!\!
                         \begin{array}{ll}
                              \frac{k^2}{6}-\frac{k}{6}-\frac{\sqrt{k}}{4}(k-1)+\frac{k\log_2(k)}{16},
                                & \hbox{$n$ even;} \vspace{-0.08in}\\
                              \frac{k^2}{6}-\frac{k}{12}-\frac{\sqrt{k}}{2\sqrt{2}}(k-1)+\frac{k(\log_2(k)-1)}{16},
                                & \hbox{$n$ odd.}
                         \end{array}
                  \right.\\[-3.1em]
    \end{align}
\end{lemma}

%
\begin{corollary}[Sum of Descedances]\vspace{-0.05in}
\label{cor:sum of descedances}
\begin{align}
        \label{e:sum of descedances}\\[-3.4em]
            -\sum_{j=0}^{k-1}\pi_n(j)\flrfrac{j-\pi_n(j)}{k} \!= \!
                \left\{\!\!\!\!
                         \begin{array}{ll}
                              \frac{k^2}{3}-\frac{k}{3}-\frac{\sqrt{k}}{4}(k-1)-\frac{k\log_2(k)}{16},
                                & \hbox{$n$ even;} \vspace{-0.08in}\\
                              \frac{k^2}{3}-\frac{5k}{12}-\frac{\sqrt{k}}{2\sqrt{2}}(k-1)-\frac{k(\log_2(k)-1)}{16},
                                & \hbox{$n$ odd.}
                         \end{array}
                  \right.\\[-3.1em]
    \end{align}
\IEEEproof Note that $\sum_{j=0}^{k-1}\pi_n(j)\flrfrac{j-\pi_n(j)}{k} = \sum_{j=0}^{k-1}j\flrfrac{\pi_n(j)-j}{k}$ which sums all $-j$ such that $\pi_n(j)<j$. Hence~\eqref{e:sum of descedances} follows since $-\sum_{j=0}^{k-1}j\flrfrac{\pi_n(j)-j}{k}+E_1(k)+F_1(k)=k(k-1)/2$.~\IEEEQED
\end{corollary}

In fact the sum of the squares of all excedances $E_2(k)$ can be similarly evaluated.\vspace{-0.05in}

%
\begin{lemma}[Sum of Squares of Excedances]\vspace{-0.05in}
\label{lem:sum of square_excedances}
\begin{align}
        \label{e:sum of square excedances}\\[-3.4em]
            E_2(k) = -\sum_{j=0}^{k-1}j^2\flrfrac{j-\pi_n(j)}{k} \!= \!
                \left\{\!\!\!\!
                         \begin{array}{ll}
                              \frac{k^3}{12} - \frac{k^2\sqrt{k}}{6} + \frac{k^2(3n-4)}{48}
                                - \frac{k\sqrt{k}(n-4)}{16} - \frac{kn}{16} - \frac{\sqrt{k}}{12},
                                & \hbox{$n$ even;} \vspace{-0.08in}\\
                              \frac{k^3}{12} - \frac{k^2\sqrt{k}}{3\sqrt{2}} + \frac{k^2(n-1)}{16}
                               -\frac{k\sqrt{k}(n-5)}{8\sqrt{2}} - \frac{k(3n+1)}{48} - \frac{\sqrt{k}}{6\sqrt{2}},
                                & \hbox{$n$ odd.}
                         \end{array}
                  \right.\\[-3.1em]
    \end{align}

\IEEEproof The proof is similar to Theorem~\ref{lem:sum of excedances}. When $n$ is even, we have for $P_{01}$, $\sum_{01}=\sum_{i=0}^{n/2-1}(2i+1)^2$, for $P_{00}$, $\sum_{00}=4E_2(k/4)$, and for $P_{11}$, $\sum_{11}=\sum_{i\in\text{EXC}(\pi_{n-2})}(k/2+2i+1)^2=(k/2+1)^2\times\#\text{EXC}(\pi_{n-2})+4E_2(k/4) + 4(k/2+1)E_1(k/4)$. Summing terms, we obtain the recursion $E_2(k) = 8E_2(k/4) + \frac{k(k+2)\log_2(k)}{32}-\frac{\sqrt{k}(k^2+k-2)}{8}
            +\frac{7k^3}{96}+\frac{k^2}{48}-\frac{k}{4}$ when $k\geq 4$. Similarly, when $n$ is odd, we obtain $E_2(k) = 8E_2(k/4) + \frac{k(k+2)\log_2(k)}{32}-\frac{\sqrt{k}(k^2+k-2)}{4\sqrt{2}}
            +\frac{7k^3}{96}+\frac{k^2}{32}-\frac{11k}{48}$ when $k\geq 8$.~\IEEEQED

\end{lemma}

%
\begin{corollary}[Sum of Squares of Descedances]\vspace{-0.05in}
\label{cor:sum of squares_descedances}
\begin{align}
        \label{e:sum of squares_descedances}\\[-3.2em]
            -\sum_{j=0}^{k-1}\pi_n^2(j)\flrfrac{j-\pi_n(j)}{k} \!= \!
                \left\{\!\!\!\!
                         \begin{array}{ll}
                              \frac{k^3}{4} - \frac{k^2\sqrt{k}}{6} - \frac{k^2(3n+20)}{48}
                                - \frac{k\sqrt{k}(n-4)}{16} + \frac{k(3n+8)}{48} - \frac{\sqrt{k}}{12},
                                & \hbox{$n$ even;} \vspace{-0.08in}\\
                              \frac{k^3}{12} - \frac{k^2\sqrt{k}}{3\sqrt{2}} + \frac{k^2(n+7)}{16}
                               -\frac{k\sqrt{k}(n-5)}{8\sqrt{2}} + \frac{k(n+3)}{16} - \frac{\sqrt{k}}{6\sqrt{2}},
                                & \hbox{$n$ odd.}
                         \end{array}
                  \right.\\[-3.2em]
    \end{align}
\IEEEproof The proof is similar to Corollary~\eqref{cor:sum of descedances}.~\IEEEQED
\end{corollary}
\vspace{-0.15in}

%
\subsection{Minimum Spread}\vspace{-0.05in}
\begin{lemma}
\label{lemma:min_spread}
The minimum spread $\text{SP}_{\a}(\pi_n)\!=\!\min_{i,j\in[k]}\abs{\pi_n(i)\!-\!\pi_n(j)}\!+\!\abs{i\!-\!j},\abs{i\!-\!j}\!<\!\a,i\!\neq \!j$, of a BRP with $k\geq 8$ is\vspace{-0.07in}
\begin{align*}
\text{SP}_{\a} = \left\{
                   \begin{array}{ll}
                     k/4+1, & \hbox{$\a=2$;} \\[-0.75em]
                     k/8+2, & \hbox{$\a=3$;} \\[-0.75em]
                     \min(6,k/8+2), & \hbox{$\a\geq 4$.}
                   \end{array}
                 \right.\\[-3.3em]
\end{align*}
\IEEEproof For $\a\!=\!2$, $\abs{\pi_n(i\!+\!1)\!-\!\pi_n(i)}\!=\!k/2$ if $i$ is even. For $i$ odd, the minimum occurs when $i\!=\!1\!\pmod{4}$, in which case $\abs{\pi_n(i\!+\!1)\!-\!\pi_n(i)}\!=\!k/2\!-\!k/4\!=\!k/4$. For $\a\!=\!3$, $\min_{i\in[k]}\abs{\pi_n(i\!+\!2)\!-\!\pi_n(i)}\!=\!\min_{i\in[k/2]}\abs{\pi_{n-1}(i\!+\!1)\!-\!\pi_{n-1}(i)}\!=\!k/8$. For $\a\!=\!4$, when $i\!=\!k/2\!-\!2$ and $i\!+\!3\!=\!k/2\!+\!1$, we have $\abs{\pi_n(i\!+\!3)\!-\!\pi_n(i)}\!=\!k/2\!+\!1\!-\!(k/2\!-\!2)\!=\!3$, hence $\text{SP}_{4}\!=\!\min(6,\text{SP}_{3})$.~\IEEEQED
\end{lemma}
\vspace{-0.1in}

%
\subsection{Inliers and Outliers}

%
%
%
\begin{theorem}\vspace{-0.05in}
\label{th:theorem_P_sum_K}
For any permutation $\pi$ that fixes the zero element (i.e., $\pi(0)=0$)
\begin{align}
\label{eq:INL_saw_formula_K_brev}\\[-3.1em]
\#\text{INL}_{\a,\b}
        &= \frac{\a\b}{k} + \sum_{j=0}^{k-1}
            \left[\!\sawfrac{j-\a}{k}\!-\!\sawfrac{j}{k}\!\right]\!\!
            \left[\!\sawfrac{\pi(j)-\b}{k}\!-\!\sawfrac{\pi(j)}{k}\!\right] + K_{\text{INL}}(\a,\b),~\text{where}\\
    K_{\text{INL}}(\a,\b) &=  \hf\!\flrfrac{\pi(\a)-\b}{k} \!+\!\hf\!\flrfrac{\pi(\b)-\a}{k}
                    \!-\!\frac{1}{4}\deltafrac{\pi(\b)-\a}{k}
                    \!-\!\frac{1}{4}\deltafrac{\vphantom{\beta}\a}{k} \!-\!\frac{1}{4}\deltafrac{\b}{k}
                    \!+\!\frac{3}{4} \label{eq:P_f_cong_constant_K}\\[-3.3em]
\end{align}
Also, there exist small positive constants $c_1,c_2$ such that $\left\lceil\! \frac{\a\b}{k}\!-\!c_1\!\right\rceil \!\leq\! \#\text{INL}_{\a,\b} \! \leq \!\flr{\frac{\a\b}{k}\!+\!c_2}$ by evaluating~\eqref{e:INL_definition} without $\flr{\cdot}$.
\end{theorem}
\vspace{-0.1in}

%
%
\begin{corollary}
\label{cor:theorem_P_sum_K}
Specifically, for BRPs, $\#\text{INL}_{\a,\b}$ reduces to\vspace{-0.05in}
\begin{align}
\label{eq:corollary_P_sum_K_brev}\\[-3.1em]
\#\text{INL}_{\a,\b} = \frac{\a\b}{k} + \frac{1}{4k}T(k,\a,\b) + K_{\text{INL}}(\a,\b),\\[-3.2em]
\end{align}
where $T(k,\a,\b)$ is given in~\eqref{e:T_sum_minus}, and $K_{\text{INL}}$ reduces to
\begin{singlespace}\vspace{-0.3in}
\begin{align}
\label{eq:P_f_cong_constant_K_cases}
    K_{\text{INL}}(\a,\b) = \left\{
          \begin{array}{ll}
            \!\!\!\hphantom{-}0,  & ~~\hbox{if $\a=0$ or $\b=0$;} \\[-0.35em]
            \!\!\!\hphantom{-}1/2,  & ~~\hbox{if $\pi_n(\a)=\b\neq 0$;} \\[-0.35em]
            \!\!\!\hphantom{-}3/4,  & ~~\hbox{if $\pi_n(\a)>\b\neq 0,~~ \pi_n(\b)>\a\neq 0$;} \\[-0.35em]
            \!\!\!\hphantom{-}1/4,  & ~~\hbox{if $\pi_n(\a)>\b\neq 0,~~ \pi_n(\b)<\a\neq 0$;} \\[-0.35em]
            \!\!\!\hphantom{-}1/4,  & ~~\hbox{if $\pi_n(\a)<\b\neq 0,~~ \pi_n(\b)>\a\neq 0$;} \\[-0.35em]
            \!\!\!-1/4, & ~~\hbox{if $\pi_n(\a)<\b\neq 0,~~ \pi_n(\b)<\a\neq 0$.}
          \end{array}
        \right.\\[-2.5em]
\end{align}
\end{singlespace}
\noindent Equation~\eqref{eq:corollary_P_sum_K_brev} can be evaluated recursively in $\log_2 k - 1$ steps using~\eqref{e:T_sum_minus}. Note that since $k$ is a power of 2, only integer shift and add operations are needed to evaluate~\eqref{e:T_sum_minus} and~\eqref{eq:corollary_P_sum_K_brev}, assuming the product of the constants $\a\b$ is computed off-line.~\IEEEQED
\end{corollary}
\vspace{-0.05in}

%
%
\begin{example} \label{ex:P}
Let $n=32,k=2^n=2^{32},\a=2^{16}-1,\b=2^{16}+1$. Then $\a\b/k=(2^{32}-1)/2^{32}$ and $K_{\text{INL}}=3/4$ since $\pi_{32}(\a)>\b$ and $\pi_{32}(\b)>\a$. Using~\eqref{e:T_sum_minus}, we have $c^*=\pi_{31}(2^{15})=2^{15}$ and $T(2^{32},2^{16}-1,2^{16}+1)=2T(2^{31},2^{16}-1,2^{15})+2^{33}-2^{18}-2^2$. Next we have $T(2^{31},2^{16}-1,2^{15})=2T(2^{30},2^{16}-1,2^{14})$. These steps are repeated using~\eqref{e:T_sum_minus}, resulting in $T(2^{32},2^{16}-1,2^{16}+1)=4294967300=2^{33}+2^2$. Therefore, using~\eqref{eq:corollary_P_sum_K_brev} we have $\#\text{INL}_{2^{16}-1,2^{16}+1} = (2^{32}-1)/2^{32} + (2^{33}+2^2)/2^{34} + 3/2^2 = 2$.
\end{example}

%
\begin{corollary}
\label{cor:P(j+1)}
For BRPs and $\a\neq0,\b\neq0$, it follows that $\sum_{j=0}^{k-1}\flrfrac{\vphantom{\pi_n(j\!+\!1)}j\!-\!\a}{k}\!\!\flrfrac{\pi_n(j\!+\!1)\!-\!\b}{k} = \#\text{INL}_{\a+1,\b} - 1$.~\IEEEQED
\end{corollary}
\vspace{-0.05in}

%
\begin{theorem}[Probability of Bounded Inliers]
\label{th:prob_a1<=j<a2 and_b1<=Xj<b2}
The probability that $\b_1\! \leq \!X_{j}\! <\! \b_2$ for $\a_1\!\leq\! j\!<\!\a_2$ is
\begin{align*}\\[-3.1em]
    \frac{  \left(\#\text{INL}_{\a_2,\b_2} - \#\text{INL}_{\a_2,\b_1}\right) -
            \left(\#\text{INL}_{\a_1,\b_2} - \#\text{INL}_{\a_1,\b_1}\right)}
         {k},\\[-3.1em]
\end{align*}
where $0\!\leq\! \a_1\! <\!\a_2 \!\leq\! k, ~0\!\leq\! \b_1\! <\!\b_2\! \leq\! k$, and $\#\text{INL}$ is given by~\eqref{eq:corollary_P_sum_K_brev}.
\IEEEproof $\text{INL}_{\a_2,\b_2}$ counts the number of integers $j<\a_2$ such that $\pi_n(j)<\b_2$, while $\text{INL}_{\a_2,\b_1}$ counts those integers such that $\pi_n(j)<\b_1$. Hence the difference counts all $i<\a_2$ such that $\b_1\! \leq \!\pi_{n}(j)\! <\! \b_2$. Similarly for $\#\text{INL}_{\a_1,\b_2} - \#\text{INL}_{\a_1,\b_1}$. Therefore $\left(\#\text{INL}_{\a_2,\b_2} - \#\text{INL}_{\a_2,\b_1}\right) - \left(\#\text{INL}_{\a_1,\b_2} - \#\text{INL}_{\a_1,\b_1}\right)$ counts all $\a_1\!\leq\! j\!<\!\a_2$ such that $\b_1\! \leq \!\pi_{n}(j)\! <\! \b_2$.~\IEEEQED
\end{theorem}
\vspace{-0.05in}

%
%
%
We can similarly count the integers $j\in[k]$ which have \emph{successive inliers}, i.e. those such that $X_j<\a$ and $X_{j+1}<\b$: $\text{SINL}_{\a,\b}\triangleq\{j\in[k] ~|~ \pi(j) < \a, ~\pi(j+1) < \b\}, ~0<\a,\b \leq k$,
which is given by the summation:
\begin{align}\\[-3.3em]
\label{e:points_pi_j_a_pi_j+1_b}
\#\text{SINL}_{\a,\b} = \sum_{j=0}^{k-1}\flrfrac{\pi(j)\!-\!\a}{k}\!\!\flrfrac{\pi(j+1)\!-\!\b}{k}\\[-3.5em]
\end{align}
\vspace{-0.25in}

%
%
\begin{theorem}[Successive Inliers]
\label{th:theorem_S_sum_K}
The number of elements in $\text{SINL}_{\a,\b}$ for $\a,\b\neq0$ is
\begin{align}
\label{eq:theorem_S_sum_K_brev}\\[-3.3em]
\#\text{SINL}_{\a,\b}
    &= \frac{\a\b}{k} + \sum_{j=0}^{k-1}\!
       \left[\!\sawfrac{\pi_{n}(j)\!-\!\a}{k}\!\!-\!\!\sawfrac{\pi_{n}(j)}{k}\!\right]\!\!\!
       \left[\!\sawfrac{\pi_n(j\!+\!1)\!-\!\b}{k}\!\!-\!\!\sawfrac{\pi_n(j\!+\!1)}{k}\!\right] +
       K_{\text{SINL}}(\a,\b)\notag\\[-0.2em]
    &= \frac{\a\b}{k} + \frac{1}{k^2}W(k,\a,\b)+K_{\text{SINL}}(\a,\b)\\[-3.0em]
\end{align}
where $W(k,\a,\b)$ is given in~\eqref{e:W_k_a_b_def},~\eqref{e:W_k_a_b_iterative_formula},\vspace{-0.1in}
\begin{align}
\label{eq:constant_K_S}
    K_{\text{SINL}}(\a,\b)  &= \hf\!\flrfrac{\b'\!-\!\a}{k} \!-\!\frac{1}{4}\deltafrac{\b'\!-\!\a}{k}
                            \!+\!\frac{1}{4}\deltafrac{\a\!+\!1}{k} \!+\! \hf\!\flrfrac{\a'\!-\!\b}{k}
                            \!-\! \hf\!\flrfrac{k/2\!-\!\b}{k} \!+\!\frac{1}{4}\deltafrac{k/2\!-\!\b}{k},\\[-3.1em]
\end{align}
$\a'\!=\!\pi_{n}\!\!\left(\pi_{n}(\a)\!+\!1\right)$, and $\b'\!=\!\pi_{n}\!\!\left(\pi_{n}(\b)\!-\!1\right)$. Moreover, if $\a\leq k/2$ and $\b \leq k/2$, then $\#\text{SINL}_{\a,\b}=0$.
\IEEEproof We expand~\eqref{e:points_pi_j_a_pi_j+1_b} in terms of saw-functions similar to Theorem~\eqref{th:theorem_P_sum_K}, multiply out terms and then simplify the expression using~\eqref{e:property_sawsum_pi_j},~\eqref{e:property_sawsum_pi_j_plus_w}.
Equations~\eqref{eq:theorem_S_sum_K_brev} and~\eqref{eq:constant_K_S} follow. Further, it is easy to
show that $K_{\text{SINL}}(\a,\b)$ evaluates to $-1,-3/4,-1/2,-1/4,0,1/4$ depending on $\a,\b$.
The details of the proof are omitted. Finally, $\#\text{SINL}_{\a,\b}=0$ if both $\a\leq k/2$ and $\b \leq k/2$ since either $j$ or $j+1$ is odd which implies either $\pi_{n}(j)\geq k/2$ or $\pi_{n}(j+1)\geq k/2$, and therefore all $j \notin\text{SINL}_{\a\leq k/2,\b\leq k/2}$.~\IEEEQED
\end{theorem}

%
%
%
%
%
%
\begin{example} \label{ex:S}
Using the numbers from Example~\ref{ex:P}, we have $\a\b/k\!=\!(2^{32}\!-\!1)/2^{32}$. Also
$\pi_{32}(\a)\!=\!\pi_{32}(2^{16}-1)\!=\!2^{16}(2^{16}-1)$, $\a'\!=\!\pi_{32}\!\!\left(2^{16}(2^{16}-1)\!+\!1\right)\!=\!k/2\!+\!2^{16}\!-\!1\!=\!k/2\!+\!\a \!>\! \b$, $\pi_{32}(\b)\!=\!\pi_{32}(2^{16}\!+\!1)\!=\!k/2\!+\!2^{15}$, and $\b'\!=\!\pi_{32}\!\!\left(k/2\!+\!2^{15}\!-\!1\right)\!=\!2^{15}(2^{16}\!-\!2)\!+\!1 \!>\! \a$. Therefore $K_{\text{SINL}}\!=\!1/2\!-\!1/2\!=\!0$.

Next, using~\eqref{e:W_k_a_b_iterative_formula} we have $W(2^{32},2^{16}-1,2^{16}+1)=8W(2^{31},2^{15}-1,2^{15})-(2^{17}-1)k$;
$W(2^{31},2^{15}-1,2^{15})=8W(2^{30},2^{14}-1,2^{14})-2^{15}k/2$; $W(2^{30},2^{14}-1,2^{14})=8W(2^{29},2^{13}-1,2^{13})-2^{14}k/2^2$;
$\cdots$; $W(2^{17},2^{1}-1,2^{1})=8W(2^{16},0,1)-2^{1}k/2^{15} = -2k/2^{15}$. Summing all terms, we get
$W(2^{32},2^{16}-1,2^{16}+1) = -(2^{17}-1)k -k \sum_{i=0}^{14} 2^{14-2i} \times 8^{i+1} = k-k^2$.
Therefore $\#\text{SINL}_{\a,\b}= (k-1)/k + (k-k^2)/k^2 + 0 = 0$.~\IEEEQED
\end{example}

%
\begin{theorem}[Probability of Bounded Successive Inliers]
\label{th:prob_a1<=Xj<a2 and_b1<=Xj+1<b2} The probability that $\a_1\! \leq \!X_{j}\! <\! \a_2$ and $\b_1\! \leq \!X_{j+1}\! <\! \b_2$ is\vspace{-0.08in}
\begin{align*}
    \frac{  \left(\#\text{SINL}_{\a_2,\b_2} - \#\text{SINL}_{\a_2,\b_1}\right) -
            \left(\#\text{SINL}_{\a_1,\b_2} - \#\text{SINL}_{\a_1,\b_1}\right)}
         {k},\\[-3.3em]
\end{align*}
where $0\!\leq\! \a_1\! <\!\a_2 \!\leq\! k, ~0\!\leq\! \b_1\! <\!\b_2\! \leq\! k$, and $\#\text{SINL}$ is given by~\eqref{eq:theorem_S_sum_K_brev}. This result is similar to Theorem~\ref{th:prob_a1<=j<a2 and_b1<=Xj<b2}.~\IEEEQED
\end{theorem}
\vspace{-0.2in}

%
\subsection{Inversions}\vspace{-0.0in}
\begin{lemma}[Inversions]
\label{lem:I_definition}\vspace{-0.1in}
The number of inversions $\#\text{INV}$ is given by\vspace{-0.05in}
\begin{align}
\label{e:lemma_I_definition}
\#\text{INV}(\pi_n)   &= \frac{k(k-1)}{2} - \sum_{\a=0}^{k-1}\sum_{j=0}^{k-1}\flrfrac{j-\a}{k}\!\!\flrfrac{\pi(j)-\pi(\a)}{k}
                = \frac{k^2}{4} - \frac{(n+1)k}{4}\\[-3.2em]
\end{align}
\IEEEproof Using~\eqref{e:INL_definition} then~\eqref{eq:INL_saw_formula_K_brev} in~\eqref{e:I_definition} with $\b=\pi_n(\a)$, we have\vspace{-0.05in}
\begin{align*}\\[-3.1em]
\#\text{INV}(\pi_n)   &= \frac{k(k-1)}{2} \!-\!
        \frac{1}{k}\sum_{\a=0}^{k-1}\a\pi_n(\a) \!-\!
        \sum_{\a=0}^{k-1}\!\sum_{j=0}^{k-1}\!
            \left[\!\!\sawfrac{j-\a}{k}\!\!-\!\!\sawfrac{j}{k}\!\!\right]\!\!\!
            \left[\!\!\sawfrac{\pi_n(j)\!-\!\pi_n(\a)}{k}\!\!-\!\!\sawfrac{\pi_n(j)}{k}\!\!\right] \!-\!
        \sum_{\a=0}^{k-1}\!K_{\text{INL}}\!\left(\a,\pi_n(\a)\right)\\[-0.2em]
        &= \frac{k(k-1)}{2} - \frac{J_1(k)}{k} - U(k) - \sum_{\a=1}^{k-1}\hf\\[-3.3em]
\end{align*}
where $J_1(k),U(k)$ are given in equations~\eqref{e:sum_j1_pi_j_formula} and~\eqref{e:U_sum_minus}, respectively,
and $K_{\text{INL}}\left(\a,\pi_n(\a)\right)=1/2$ from~\eqref{eq:P_f_cong_constant_K_cases} when $\a\neq 0$, and 0 otherwise.
Substituting~\eqref{e:sum_j1_pi_j_formula} and~\eqref{e:U_sum_minus} in the second equation and simplifying terms, equation~\eqref{e:lemma_I_definition} follows.~\IEEEQED
\end{lemma}
\vspace{-0.1in}

%
\subsection{Serial Correlations}\vspace{-0.05in}
A necessary condition for the apparent randomness of $\{X_j\}$ is the small size of the serial correlation statistic\vspace{-0.05in}
\begin{align}\vspace{-0.05in}
\label{e:serial_correlations}
    \theta_p = \frac{\text{Cov}(X_i,X_{i+p})}{\text{Var}(X_i)}
           = \frac{\Ex{\left(X_i-\Ex{X_{i}}\right)\!\!\left(X_{i+p}-\Ex{X_{i+p}}\right)}}
                  {\Ex{\left(X_i-\Ex{X_{i}}\right)^2}},\quad p=1,2,\cdots,k-1,\\[-3.3em]
\end{align}
between $X_i$ and its $p$-th successors $X_{i+p}$, where $\Ex{\cdot}$ is the expectation operator.
$\theta_p$ is called the serial correlation coefficient, a measure of the extent to which $X_{i+p}$ depends on $X_i$.
To compute $\theta_p$, we first determine the variance $\text{Var}(X_i)$: $\Ex{X_i} \!=\! \frac{1}{k}\sum_{j=0}^{k-1}X_j \!=\! \frac{1}{k}\sum_{j=0}^{k-1}j \!=\! \frac{k-1}{2}$, and
$\Ex{X_i^2} \!=\! \frac{1}{k}\sum_{j=0}^{k-1}X_j^2 \!=\! \frac{1}{k}\sum_{j=0}^{k-1}j^2\!=\! \frac{(k-1)(2k-1)}{6}$.
Hence $\text{Var}(X_i) \!=\! \Ex{X_i^2} \!-\! {\Ex{X_i}}^2 \!=\! \frac{k^2-1}{12}$. The only difficult part of~\eqref{e:serial_correlations} is the covariance:

%
%
\begin{theorem}[Covariance]\vspace{-0.05in}
\label{th:theorem_covariance}
\begin{align}
\label{e:theorem_covariance}\\[-3.6em]
 \text{Cov}(X_i,X_{i+p})
    &=\frac{1}{k}C(k,p)\!+\!\frac{1}{4}\! +\!\frac{k}{2}\!\left(\!1 -\frac{3}{2^{v+1}}\!\right)\\[-3.3em]
\end{align}
where $C(k,p)$ is given by~\eqref{e:C_k_p_sum}, and $0\leq v<n$ is the position of the least-significant one-bit in $p_{(2)}$ (starting from 0).
\end{theorem}
\begin{corollary}[Serial correlations for $p=1$]\vspace{-0.06in}
    \label{cor:serial_cor_p=1}
\begin{align*}\\[-3.3em]
    \theta_1 = \frac{\text{Cov}(X_i,X_{i+1})}{\text{Var}(X_i)} = -\frac{5k^2+5k+12}{7k(k+1)}\\[-3.3em]
\end{align*}
\IEEEproof Substitute~\eqref{e:ser_corr_C_k_1} for $C(k,1)$ and $v=0$ in~\eqref{e:theorem_covariance}, then divide by the variance $(k^2-1)/12$.~\IEEEQED
\end{corollary}
A correlation coefficient always lies between $\pm1$. When it is small, it indicates that $X_i$ and $X_{i+p}$ are almost independent. Hence it is desirable to have $\theta_1$ close to zero. Since $\lim_{k\rightarrow\infty}\theta_1 = -5/7$, it follows that BRPs have weak correlation properties.
\vspace{-0.2in}

%
\section{Serially-Pruned Bit-Reversal Interleavers and Minimal Inliers}\label{s:pruned_interl_min_inliers}\vspace{-0.05in}
The permutation inliers problem is applied to study pruned bit-reversal interleavers (BRIs). A BRI maps an $n$-bit integer $x$ into $n$-bit integer $y$ such that $y\!=\!\pi_n(x)$, where $x,y \in[k]$ and $k\!=\!2^n$ is the interleaver size. A \emph{serially-pruned} BRI (PBRI) of size $\a<k$ and pruning length $\b\!<\!k$, with $\a\!<\!\b$, is defined by $\cpi_n\!:\!\mathcal{D}\rightarrow \mathcal{R}, x\!\mapsto\!y\!=\!\ppin{n}{x}\!=\!\pi_n(p(x))$, such that: 1) $\ppin{n}{x}\!<\!\b$, and 2) $p(x)\!\triangleq\!x\!+\!\Delta_x$ is the \emph{serial pruning function} where $\Delta_x$ is the pruning gap of $x$ defined to be the minimum $\Delta\geq\!0$ such that $\#\text{INL}_{x+\Delta,\b} \!=\! x$ (i.e., for $j\!=\!0,\cdots,x\!+\!\Delta_x\!-\!1$, $\pi(j)\!<\!\b$ is satisfied exactly $x$ times). The domain and range of $\cpi$ are $\mathcal{D}\!=\![\a]$ and $\mathcal{R}\!=\!\pi_n\!\left(p\!\left([\a]\right)\right)$. Pruned interleavers are used when blocks of arbitrary lengths (other than powers-of-2) are needed. To interleave a block of size $\b$, a mother interleaver whose size is the smallest power-of-2 that is $\geq \b$ is selected and pruned. Hence, in the following, we assume that $k/2\!<\!\b\!<\!k$.

\begin{figure}[t]
\centering
\subfloat[]{\includegraphics[scale=0.9]{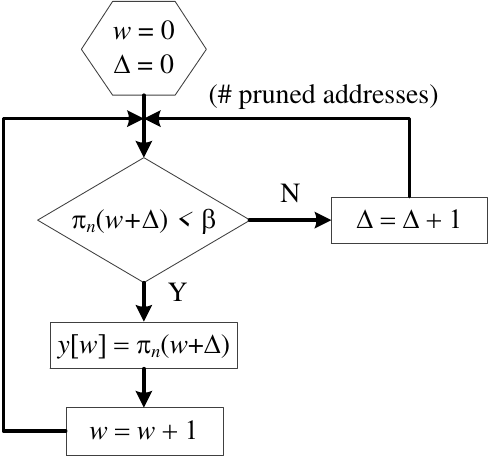}\vspace{-0.1in}\label{f:flowchart}}\qquad\qquad
\subfloat[]{\includegraphics[scale=0.9]{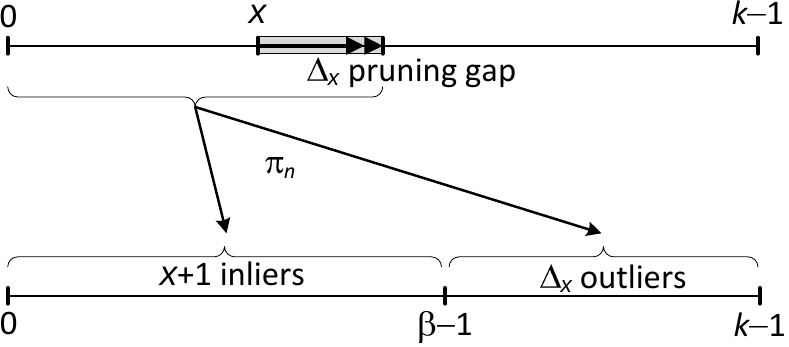}\vspace{-0.1in}\label{f:min_delta}}\vspace{-0.05in}
\caption{(a) Flowchart of the serial pruning algorithm. (b) The smallest interval of addresses $x\!+\!\Delta_x\!+\!1$ that has exactly $x\!+\!1$ inliers with respect to $\b$.}
\label{f:pruning}
\end{figure}

There are several ways to prune addresses from the mother interleaver. One method is to ignore positions beyond $\b\!-\!1$ in the permuted sequence, which we consider in this work (see also~\cite{2002_ferrari,2005_Dinoi_Benedetto_S_random}).
Other methods prune addresses beyond $\b\!-\!1$ in the original sequence, or prune a mixture of addresses from both the
original and permuted sequences~\cite{2005_Dinoi_Benedetto_S_random}. Hence any address that maps to an address $\geq \b$ is dropped and the next consecutive address is tried instead. To determine where an address $x$ is mapped, a \emph{serial} PBRI (S-PBRI) starts from $w\!=\!0$ and maintains the number of invalid mappings $\Delta$ (pruning gap) that have been skipped along the way (see Fig.~\ref{f:flowchart}). If $w\!+\!\Delta$ maps to a valid address (i.e., $\pi_n(w\!+\!\Delta) \!< \!\b$), then $w$ is incremented by 1. If $w\!+\!\Delta$ maps to an invalid address (i.e., $\pi_n(w\!+\!\Delta) \!\geq\! \b$), $\Delta$ is incremented by 1. These steps are repeated until $w$ reaches $x$ and $\pi_n(x\!+\!\Delta)\!<\!\b$, and hence $\Delta_x \!=\! \Delta$. Therefore, $x \!\mapsto \!\ppin{n}{x} \!=\! \pi_n(x\!+\!\Delta_x)$. Algorithm~\ref{a:s-pbri_algorithm} shows the pseudo-code of a generic cascadable S-PBRI with $\mathcal{D}\!=\!\{w_1,w_1\!+\!1,\cdots,w_2\}$, $\a\!=\!w_2\!-\!w_1$, $\Delta_{w_1}$ the pruning gap up to $w_1$, and $\Delta_{w_2}$ up to $w_2$. The parameters $w_1,w_2,\Delta_{w_1}$ are set to $w_1=0,w_2=x,\Delta_{w_1}=0$ to compute $\Delta_x$.
\begin{algorithm}
\caption{Serial PBRI Algorithm: $[\mathbf{y},\Delta_{w_2}]=\text{S-PBRI}(k,w_1,w_2,\b,\Delta_{w_1})$}\label{a:s-pbri_algorithm}
\begin{algorithmic}[0]
\State $w \gets w_1,\Delta \gets \Delta_{w_1}$
\While{$w\leq w_2$}
\If {$\pi_n(w+\Delta)<\b$}
    \State $\mathbf{y}[w] \gets \pi_n(w+\Delta)$
    \State $w \gets w + 1$
\Else
    \State $\Delta \gets \Delta + 1$
\EndIf
\EndWhile
\State  $\Delta_{w_2} \gets \Delta$
\end{algorithmic}
\end{algorithm}

The time complexity to determine $\Delta$ is $\mathcal{O}(k)$. However, using the inliers problem formulation, $\Delta$ is simply the minimum non-negative integer to be added to $\a$ such that $\text{INL}_{\a+\Delta,\b}$ has exactly $\a$ inliers: $ \min \Delta \geq 0 ~\text{such that}~\#\text{INL}_{\a+\Delta,\b} = \a$ (see Fig.~\ref{f:min_delta}). Out of the first $\a$ addresses, there are $\#\text{OUL}_{\a,\b}$ outliers $\geq \b$. Hence $\Delta \geq \#\text{OUL}_{\a,\b}$. Next consider the expanded interval of addresses $\a_1=\a+\#\text{OUL}_{\a,\b}$. This set contains $\#\text{OUL}_{\a_1,\b}$ outliers. Hence again $\Delta \geq \#\text{OUL}_{\a_1,\b}$. This process is repeated by expanding the interval into $\a_2=\a+\#\text{OUL}_{\a_{1},\b}$ and determining the corresponding number of outliers. The process terminates when $\#\text{OUL}_{\a_t,\b}=\#\text{OUL}_{\a_{t-1},\b}$ at some step $t$ when there are no more outliers, and hence $\Delta= \#\text{OUL}_{\a_t,\b}$. This process for computing the minimum number of inliers is implemented in Algorithm~\ref{a:minimal_inliers_algorithm}.\vspace{-0.1in}
\begin{algorithm}
\caption{Minimal Inliers (MI) Algorithm: $\Delta = \text{MI}(k,\a,\b)$}\label{a:minimal_inliers_algorithm}
\begin{algorithmic}[0]
\State $t\gets 0$
\State $\Delta^{(0)}\gets 0$
\Repeat
\State $\Delta^{(t+1)} \gets \#\text{OUL}_{\a+\Delta^{(t)},\b}$
\State $t \gets t + 1$
\Until{$\Delta^{(t)}=\Delta^{(t-1)}$}
\State $\Delta \gets \Delta^{(t)}$
\end{algorithmic}
\end{algorithm}

%
%
\begin{example}
Let $n\!=\!32,k\!=\!2^n\!=\!2^{32},\a\!=\!2^{12},\b\!=\!2^{31}\!+\!10$. Applying the MI algorithm, we have $\Delta^{(1)}=\#\text{OUL}_{2^{12},2^{31}+10}\!=\!2047$ using~\eqref{e:outliers_set_definition},~\eqref{eq:corollary_P_sum_K_brev}. Next we expand $\a$ to $\a+2047$ and recompute $\Delta^{(2)}\!=\!\#\text{OUL}_{2^{12}+2047,2^{31}+10}=3070$. Similarly at step 3 we have $\Delta^{(3)}=\#\text{OUL}_{2^{12}+3070,2^{31}+10}=3582$. The operations are repeated until $t=12$ with $\Delta^{(12)}=\#\text{OUL}_{2^{12}+4093,2^{31}+10}=4093.$
\end{example}

\begin{figure*}[hbtp]
    \centerline{\subfloat[]{\includegraphics[scale=0.9]{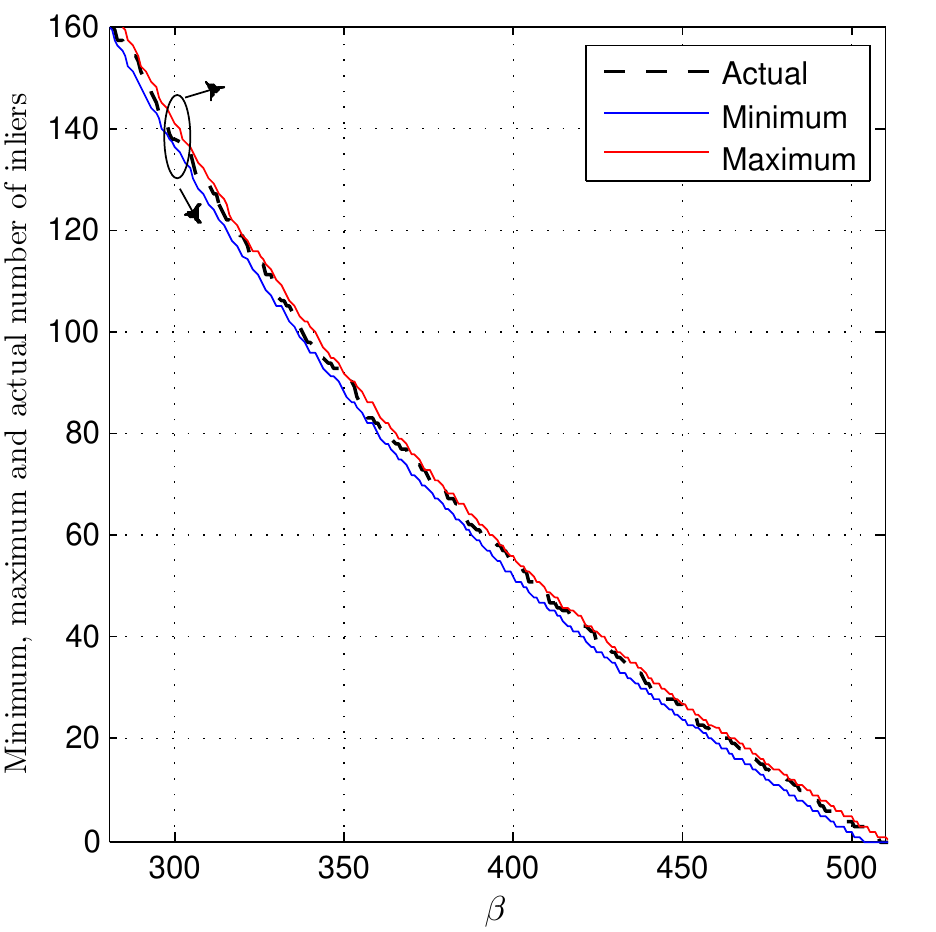}
    \label{f:bounds}}
    \hfil
    \subfloat[]{\includegraphics[scale=0.9]{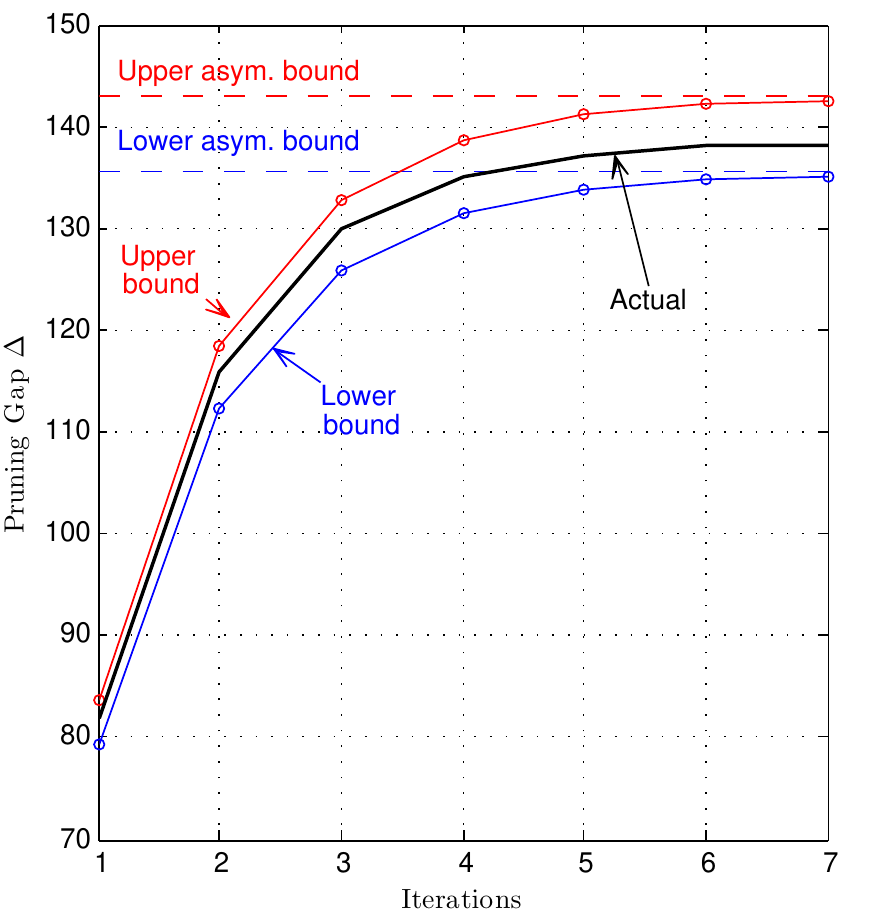}
    \label{f:convergence}}}
    \caption{(a) Lower and upper bounds on the pruning gap $\Delta$ for $k=512,\a=200$, and (b) convergence rate of the MI algorithm for $\b=300$.}
    \label{f:bounds_convergence}
\end{figure*}

The convergence rate of the MI algorithm is $\a\!-\!\b/k$ as shown in Theorem~\ref{th:conv_rate}. The proof is based on deriving exact expressions for tight lower and upper bounds on $\Delta$. Figure~\ref{f:bounds_convergence} plots these bounds for $k\!=\!2^9,\a\!=\!200$, and the convergence rate when $\b\!=\!300$.

%
%
\begin{theorem}[Rate of Convergence]
\label{th:conv_rate}
The minimal inliers algorithm converges at a rate $\mu = 1-\b/k$.
\end{theorem}

Using the MI algorithm a \emph{parallel} PBRI of length $\b$ with a parallelism factor of $p$ over the S-PBRI can be designed by employing $p$ (or $p\!+\!1$ if $\b\!\neq\! 0\! \pmod{p}$) S-PBRIs of size $\flr{\b/p}$ and pruning length $\b$ as shown in Algorithm~\ref{a:p-pbri_algorithm}.
\begin{algorithm}
\caption{Parallel PBRI Algorithm: $\mathbf{y}=\text{P-PBRI}(k,p,\b)$}\label{a:p-pbri_algorithm}
\begin{algorithmic}[0]
\ForAll{$i =0 \to p-1$}
\State $\Delta_{i} \gets \text{MI}(k,i\!\flr{\b/p},\b)$
\State $\left[\mathbf{y}[i\!\flr{\b/p}\!:\!(i\!+\!1)\!\flr{\b/p}\!-\!1],\Delta_{i+1}\right] \gets \text{S-PBRI}(k,i\!\flr{\b/p},(i\!+\!1)\!\flr{\b/p}\!-\!1,\b,\Delta_i)$
\EndFor
\If{$\b \bmod p > 0$}
\State $\Delta_{p} \gets \text{MI}(k,p\!\flr{\b/p},\b)$
\State $\mathbf{y}[p\!\flr{\b/p}:\b\!-\!1] \gets \text{S-PBRI}(k,p\!\flr{\b/p},\b\!-\!1,\b,\Delta_{p-1})$
\EndIf
\end{algorithmic}
\end{algorithm}
\vspace{-0.15in}

%
\section{Extension to 2D Block and Stream Interleavers}\label{s:ext_composite_interleavers}
We extend the discussion in this section to composite interleavers that employ smaller interleavers to construct a larger interleaver, such as 2-dimensional (2D) block and stream interleavers. \vspace{-0.15in}

%
\subsection{2D Block Interleavers}\label{s:2D_interleavers}\vspace{-0.05in}
A 2D block interleaver~\cite{2001_Garello,1999_Eroz_Hammongs_prunable_interleavers} of size $k$ is defined by a permutation $\pi$ composed of two smaller permutations $\s_1$ and $\s_2$ of size $k_1$ and $k_2$, respectively, where $k\!=\!k_1k_2$. Let $x_1\in[k_1]$, $x_2\in[k_2]$, and $x=x_1k_2+x_2\in[k]$. Then $\pi(x)\triangleq\s_2(x_2)k_1+\s_1(x_1)$. Alternatively, we say $(x_1,x_2)\in[k_1]\times[k_2]$ is mapped to $(\s_2(x_2),\s_1(x_1))\in[k_2]\times[k_1]$. This is equivalent to writing the sequence of integers $[k]$ into a $k_1\times k_2$ array row-wise, permuting the entries in each column by $\s_1$ and in each row by $\s_2$, then reading the entries from the array column-wise, and hence the interleaver is referred to as \emph{row-by-column}. The reversal of dimensions in general improves the spread properties of $\pi$. If identical permutations $\s_1$ are applied to all columns and identical $\s_2$ applied to all rows, then the order of applying the permutations does not matter. Otherwise, if $\s_1$ is column-specific, say $\s_{1,x_2}$, and $\s_2$ is row-specific, say $\s_{2,x_1}$, then the order matters. In a \emph{row-first} block interleaver, an entry $(x_1,x_2)$ maps to row $x_1'\!=\!\s_{1,x_2}(x_1)$ then to column $x_2'\!=\!\s_{2,x_1'}(x_2)$, while in a \emph{column-first} interleaver, it maps to column $x_2'\!=\!\s_{2,x_1}(x_2)$ then to row $x_1'\!=\!\s_{1,x_2'}(x_1)$. For simplicity, we assume identical $\s_1$'s and identical $\s_2$'s in the discussion below.

A pruned 2D block interleaver of size $\a\!=\!\a_1k_2\!+\!\a_2\!<\!k$ and pruning length $\b\!=\!\b_1k_1\!+\!\b_2\!<\!k$, with $\a\!<\!\b$, is defined by the map $\ppi{\cdot}:\mathcal{D}\rightarrow \mathcal{R}$ where $\abs{\mathcal{D}}\!=\!\abs{\mathcal{R}}\!=\!\a$ similar to a pruned 1D block interleaver. Here, $\a_1\!=\!\flr{\a/k_2},\a_2\!=\!\a\bmod k_2,\b_1\!=\!\flr{\b/k_1},\b_2\!=\!\b\bmod k_1$, and the integer $x\!=\!x_1k_2\!+\!x_2\!<\! \a_1k_2\!+\!\a_2 \in[k]$ maps to $\ppi{x}\!=\!\pi(y)\!<\!\b_1k_1\!+\!\b_2$, where $y\!=\!x+\Delta_x\!=\!\!y_1k_2\!+\!y_2$, $y_1\!=\!\flr{y/k_2},y_2\!=\!y\bmod k_2$, and $\pi(y)\!=\!\s_2(y_2)k_1\!+\!\s_1(y_1)$ is a 2D permutation. In a pruned 2D block BRI (P2BRI), $\s_1\!=\!\pi_{1,n_1}$ and $\s_2\!=\!\pi_{2,n_2}$ are bit-reversal permutations on $n_1$ and $n_2$ bits, respectively, and $k_1\!=\!2^{n_1},k_2\!=\!2^{n_2}$.

To count the $(\a,\b)$-inliers $\#\text{INL}_{\a,\b}(\s_1,\!\s_2)\!\triangleq\! \#\text{INL}_{\a,\b}(\pi)$ in a pruned block interleaver, we count the number of times $\pi(x)\!=\!\s_2(x_2)k_1\!+\!\s_1(x_1) \!< \!\b_1k_1\!+\!\b_2$ for $x_1k_2\!+\!x_2\!<\! \a_1k_2\!+\!\a_2$. This is satisfied if: 1a) $\s_2(x_2)\!<\!\b_1$, or 1b) $\s_2(x_2)\!=\!\b_1$ and $\s_1(x_1)\!<\!\b_2$, and 2a) $x_1\!<\! \a_1$, or 2b) $x_1\!=\! \a_1$ and $x_2\!<\! \a_2$. Conditions 1a), 2a) are both satisfied $\a_1\!\b_1$ times. Conditions 1a), 2b) simply count the $(\a_2,\b_1)$-inliers for $\s_2$, which is $\#\text{INL}_{\a_2,\b_1}(\s_2)$. Similarly, 1b), 2a) count the $(\a_1,\b_2)$-inliers for $\s_1$, which is $\#\text{INL}_{\a_1,\b_2}(\s_1)$. Finally, 1b), 2b) are satisfied once if $\s_1(\a_1)\!<\!\b_2$ and $\s_2^{-1}(\b_1)\!<\!\a_2$. Adding the results we get:\vspace{-0.10in}
\begin{align}
\label{e:INL_2DPBRI}
\#\text{INL}_{\a,\b}(\s_1,\!\s_2)=\a_1\b_1 + \#\text{INL}_{\a_2,\b_1}(\s_2)+\#\text{INL}_{\a_1,\b_2}(\s_1)+
\left\{
  \begin{array}{ll}
    1, & \hbox{if $\s_1(\a_1)\!<\!\b_2,\s_2^{-1}(\b_1)\!<\!\a_2$;} \\[-0.8em]
    0, & \hbox{otherwise.}
  \end{array}
\right.
\end{align}

%
%
\begin{example} \label{ex:INL_2DPBRI}
Consider a P2BRI with $n_1\!=\!20,n_2\!=\!12,\a\!=\!2^{18}\!-\!99,\b\!=\!2^{31}\!+\!2^{19}\!+\!133$. We have
$k_1\!=\!2^{n_1}\!=\!2^{20}$;$\s_1\!=\!\pi_{20},k_2\!=\!2^{n_2}\!=\!2^{12},\s_2\!=\!\pi_{12},n\!=\!n_1\!+\!n_2\!=\!32,k\!=\!2^n\!=\!2^{32}$; $\a_1\!=\!\flr{\a/k_2}\!=\!63,\a_2\!=\!\a\bmod k_2\!=\!3997$;$\b_1\!=\!\flr{\b/k_1}\!=\!2048,\b_2\!=\!\b\bmod k_1\!=\!524421$. Using~\eqref{eq:corollary_P_sum_K_brev}, we compute $\#\text{INL}_{3997,2048}(\pi_{12})\!=\!1999$, $\#\text{INL}_{63,524421}(\pi_{20})\!=\!33$. Since $\pi_{20}(63)\!=\!1032192\! \nless \! 524421$, conditions 1b), 2b) are not satisfied. Hence $\#\text{INL}_{2^{18}-99,2^{31}+2^{19}+133}(\pi)\!=63\!\times\!2048\!+\!1999\!+\!33\!=\!131056$.~\IEEEQED
\end{example}

The minimal inliers algorithm can be applied to compute the pruning gap of a P2BRI with outliers $\#\text{OUL}_{\a,\b}(\pi)\!=\!\a\!-\!\#\text{INL}_{\a,\b}(\s_1,\!\s_2)$ computed using~\eqref{e:INL_2DPBRI}. A parallel P2BRI can be realized as well using Algorithm~\ref{a:p-pbri_algorithm}. Extensions to multi-dimensional hyper-block pruned interleavers can be similarly defined, but the details are omitted due to lack of space.

%
\subsection{Stream Interleavers}\label{s:stream_interleavers}
In some communication systems (e.g.~\cite{IEEE_802.11n,IEEE_802.20_3GPP2}), a block of information bits is divided into sub-blocks each of which is interleaved independently. Interleaved bits out of each sub-block are treated as streams that are concatenated (or even further interleaved) to form the interleaved bits of the original block. For example, a 2-stream interleaver divides a block of length $k=2^n$ into two sub-blocks of size $k/2$, interleaves sub-block 0 using $\s_0$ and sub-block 1 using $\s_1$, and then combines the outputs bits from both streams in an alternating fashion. The resulting permutation is given by \vspace{-0.1in}
\begin{align}
    \pi(2x)&=2\s_0(x)\\[-0.5em]
    \pi(2x+1)&=2\s_1(x)+1
\end{align}
where $x\!=\!0,1,\cdots,k/2-1$. A 2-stream bit-reversal interleaver uses bit-reversal maps on $n-1$ bits to interleave the sub-blocks, i.e., $\s_0(x)\!=\!\s_1(x)\!=\!\pi_{n-1}(x)$. A pruned 2-stream bit-reversal interleaver is defined similar to a PBRI.

To count its $(\a,\b)$-inliers $\#\text{INL}_{\a,\b}(\s_0,\cdots,\s_{m-1};\omega)$, we simply count the $(\a_j,\b_j)$-inliers of $\s_{\omega(j)}$ for $j\!=\!0,\cdots,m\!-\!1$ and add the results:
\begin{align}
\label{e:INL_2streamPBRI}
\#\text{INL}_{\a,\b}(\s_0,\!\s_1)=\#\text{INL}_{\ceil{\a/2},\ceil{\b/2}}(\s_0)+\#\text{INL}_{\flr{\a/2},\flr{\b/2}}(\s_1).
\end{align}

In fact, the above formula can be generalized to a pruned $m$-stream interleaver employing $m$ generic constituent permutations $\s_0,\cdots,\s_{m-1}$ of size $k/m$, where the $m$ output bits from the $m$ streams are permuted according to some permutation $\omega$ of size $m$. The resulting permutation is given by \vspace{-0.1in}
\begin{align}
\label{e:INL_mstream_genperm}
    \pi(mx+j)&=m\times\s_{\omega(j)}(x)+\omega(j),~j=0,1,\cdots,m-1,\\[-2.75em]
\end{align}
where $x\!=\!0,1,\cdots,k/m-1$. To count its $(\a,\b)$-inliers $\#\text{INL}_{\a,\b}(\s_0,\cdots,\s_{m-1};\omega)$, we simply count the $(\flr{(\a\!-\!j\!-\!1)/m\!+\!1},\flr{(\b\!-\!\omega(j)\!-\!1)/m\!+\!1})$-inliers of $\s_{\omega(j)}$ for $j\!=\!0,
\cdots,m\!-\!1$ and add the results, to obtain
\begin{align}
\label{e:INL_mstream_genperm_inliers}
\#\text{INL}_{\a,\b}(\s_0,\cdots,\s_{m-1};\omega)=\sum_{j=0}^{m-1}\#\text{INL}_{\flr{(\a\!-\!j\!-\!1)/m\!+\!1},\flr{(\b\!-\!\omega(j)\!-\!1)/m\!+\!1}}(\s_{\omega(j)}).
\end{align}

%
\section{Application to Pruned FFT Algorithm and Pruned LTE Turbo Interleavers}\label{s:applications} In this section, we apply the inliers problem to design parallel bit-reversal permuters for pruned FFTs, as well as parallel pruned interleavers for LTE turbo codes.\vspace{-0.15in}

%
\subsection{Pruned FFT Algorithm}\label{s:pruned_FFT}\vspace{-0.05in}
The fast Fourier transform (FFT) is widely used in signal processing and communications
such as digital filtering, spectral analysis, and polyphase filter multicarrier demultiplexing (MCD)~\cite{1987_Holm,1996_He,1980_Sreenivas,2005_Hu}. In some of these FFT applications, there exist cases where the input vector has many zeros or the required outputs may be very sparse compared to the transform length. For example, in digital filtering, one may only require the spectrum corresponding to certain frequency windows of the FFT, or in MCD, only a few carriers out of the overall range of available carriers at any given time are needed. In digital image processing, only part of the images are of interest to certain applications. In these cases, most of the FFT outputs are not required. Several FFT pruning algorithms have been proposed to deal with such cases~\cite{1971_Markel,1979_Sreenivas,1993_Sorensen,2005_Hu,2012_Wang} and avoid redundant computations on zero inputs or for unused outputs. However, most of these algorithms do not consider the cost of pruned bit reversal reordering of the inputs or outputs when performing in-place FFT computations.

\begin{figure}[t]
\centering
\includegraphics[scale=0.9]{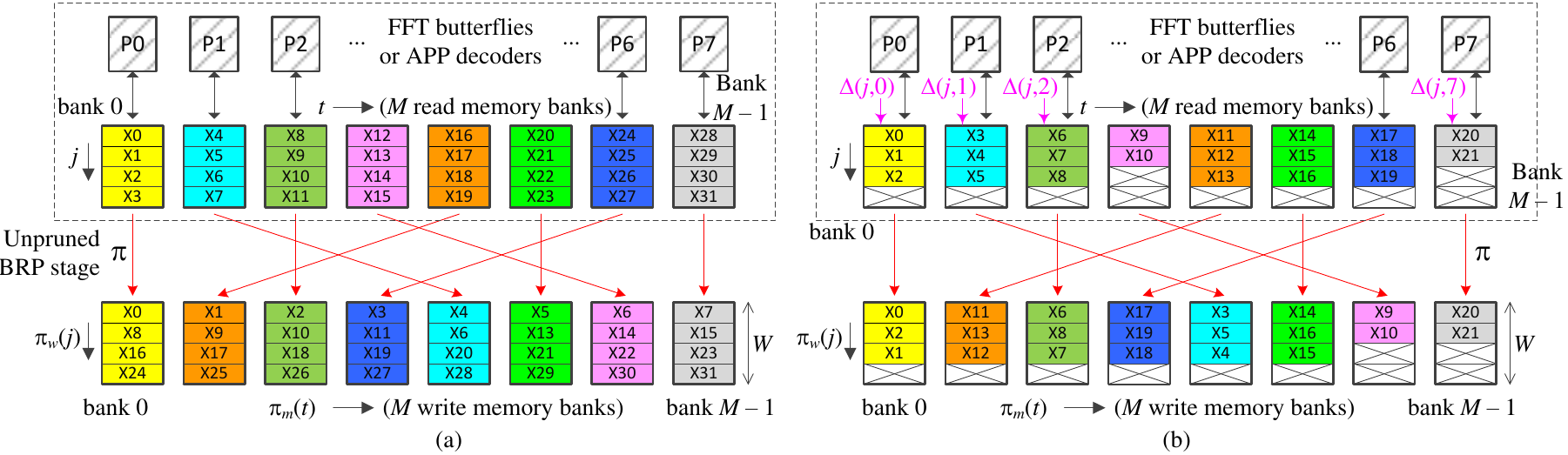}\vspace{-0.1in}
\caption{8-way parallel contention-free mapping for (a) an unpruned, and (b) a pruned FFT bit-reversal mapping with $k\!=\!32,\a\!=\!\b\!=\!22,W\!=\!4,M\!=\!8$.}
\label{f:contention_free}
\end{figure}

For simplicity of exposition, we assume in the following that only a narrow spectrum is of interest, but the resolution within that band has to be very high. Hence, the DFT $x\!\leftrightarrow\!X$ has some $k\!=\!2^n$ input values $x_i$, but fewer than $k$ outputs $X_i$ are needed. We also assume a radix-2 FFT algorithm is employed with in-place FFT computations using a set of butterflies that compute the final outputs in a set of $M$ memory banks in bit-reversed order. A subsequent stage performs BRP re-ordering of the outputs back to natural order. Note that since a BRP is an involution (i.e., $\pi_n\!=\!\pi_n^{-1}$), re-ordering a bit-reversed output is analogous to bit-reversed ordering of the output in natural order. Hence, we assume that the FFT outputs are written in natural order in the output memory banks, and the BRP stage does bit-reversal ordering. Figure~\ref{f:contention_free}a illustrates the BRP stage for the unpruned FFT case, which reads from the FFT memory banks and writes to the input memory banks at the receiving end. We show next that this BRP stage (both unpruned or serially pruned) can be parallelized to match the parallelism degree $M$ of the FFT, eliminating its serial bottleneck on throughput.

A permutation of length $k\!=\!W\!\cdot\!M$ in general is said to be contention-free~\cite{Nimbalker_tcom_2008} with degree $M\!=\!2^m$ if an array of $k$ data elements stored in one set of $M$ \emph{read} memory banks each of size $W\!=\!2^w$ can be permuted and written into another set of $M$ \emph{write} memory banks, such that at each step, $M$ data elements are read in parallel from the $M$ read banks and written in parallel into the $M$ write banks without reading or writing more than one element from/to any bank (see Fig.~\ref{f:contention_free}a). Data is stored sequentially in the read banks such that linear address $i\!=\!j\!+\!tW$ corresponds to location $j$ in bank $t\!=\!\flr{i/W}$, where $0\!\leq\! j\! <\! W$ and $0\!\leq \!t\! <\!M$. To permute any $M$ data entries at linear addresses $j,j\!+\!W,\cdots,j\!+\!(M\!-\!1)W$ in parallel, the contention-free property stipulates that $B\left(\pi(j\!+\!tW);W,M\right)\!\neq\!B\left(\pi(j\!+\!vW);W,M\right)$ for all $0\!\leq\! j\! <\! W$ and $0\!\leq\! t\!\neq\! v\! < \! M$, where the bank addressing function $B$ is either $B(i;W,M)\!\triangleq\!\flrfrac{i}{W}$ or $B(i;W,M)\!\triangleq\! i\!\bmod{M}$. This is a more general condition than~\cite{Nimbalker_tcom_2008}, and effectively uses either the $m$ most or least significant bits (MSBs/LSBs) of $\pi(j\!+\!tW)$ as a permuted bank address.

It is easy to prove that the bit-reversal permutation is contention-free for any $k\!=\!2^n,M\!=\!2^m,W\!=\!2^w$, where $n\!=\!m\!+\!w$ and $m\!<\!n$, using the property $\pi_n(j\!+\!tW)\!=\!M\cdot \pi_w(j)\!+\!\pi_m(t)$. For any pair of distinct windows $t,u$, we have\vspace{-0.15in}
\begin{align*}
    M\cdot \pi_w(j) + \pi_m(t) \neq M\cdot \pi_w(j) + \pi_m(u)  \pmod{M}, \qquad j\!=\!0,1\,\cdots,W\!-\!1.
\end{align*}\\[-3.1em]
Hence the $m$ LSBs designate a permuted bank address. Figure~\ref{f:contention_free}a illustrates the contention-free property of the BRP map for a 32-point unpruned FFT block whose outputs are stored in $M\!=\!8$ read memory banks. The BRP stage permutes the 8 banks in $W\!=\!4$ steps and stores the data in the write memory banks in parallel without any stalls due to address collisions.

An arbitrary pruning of a permutation does not preserve its contention-free property. However, serial pruning does, and a contention-free pruned permuter can be designed as shown in Theorem~\ref{th:contention_free}. First, the serial-pruning map $p(i)\!=\!i\!+\!\Delta_i$ itself is contention free. To show this, take two addresses $i_1\!=\!j\!+\!t_1W$ and $i_2\!=\!j\!+\!t_2W$ that correspond to banks $t_1$ and $t_2\!>t_1$. Then $\flr{(j\!+\!t_1W\!+\!\Delta_{i_1})/W}\!\neq \! \flr{(j\!+\!t_2W\!+\!\Delta_{i_2})/W}$ for any $0\!\leq\! j \!<\! W$ since $p(\cdot)$ is monotonically increasing and hence $\Delta_{i_2}\!\geq\!\Delta_{i_1}$.

%
\begin{theorem}\label{th:contention_free} Any serially-pruned, contention-free permutation (interleaver) remains contention free.\vspace{-0.05in}
\begin{proof} One scenario is to insert zero filler bits in the pruned positions while storing the data sequentially in memory across the banks. This requires comparing $\pi(j)$ with $\b$ serially for every $j$ before writing to memory. Hence the contention-free property applies for the pruned interleaver across all the banks if the mother interleaver is contention-free.

Another scenario is to store the data across the banks without filler bits as shown in Fig.~\ref{f:contention_free}b. To interleave properly, we need to keep track of the inliers that fall within each window. First, since the number of inliers up to window $t$ is
$\Delta_{(t+1)\cdot W}\!=\!\#\text{INL}_{(t+1)\cdot W,\b}(\pi)$, data located between address $\Delta_{t\cdot W}$ and
$\Delta_{(t+1)\cdot W}\!-\!1$ are stored sequentially in bank $t$. We know that addresses $j,j\!+\!W,\cdots,j\!+\!(M\!-\!1)W$ map to distinct windows under $\pi$. Address $j$ in window $t$, which might be pruned, actually corresponds to the unpruned address $j\!+\!tW\!-\!\Delta(j,t)$, where $\Delta(j,t)$ is defined as:\vspace{-0.1in}
\begin{align}\label{eq:contention_Delta}
\Delta(j,t) = \left\{
                \begin{array}{ll}
                  \Delta_{t\cdot W},    & \hbox{if $j\!=\!0$;} \\[-1.1em]
                  \Delta(j-1,t),            & \hbox{if $j\!>\!0,~\pi_n(j+tW)\!<\!\b$;} \\[-1.1em]
                  \Delta(j-1,t)+1,          & \hbox{if $j\!>\!0,~\pi_n(j+tW)\!\geq \!\b$.}
                \end{array}
              \right.
\end{align}\\[-2.0em]
with initial condition $\Delta_0\!=\!0$. Then, for $0\!\leq\! j\! <\! W$ and $0\leq t\neq v\leq M$, we have\vspace{-0.1in}
\begin{align*}
    B\!\left(\cpi(j\!+\!tW\!-\!\Delta(j,t));W,M\right) \!=\! B\!\left(\pi(j\!+\!tW);W,M\right) \neq B\!\left(\pi(j\!+\!vW);W,M\right) \!=\! B\!\left(\cpi(j\!+\!vW\!-\!\Delta(j,v));W,M\right)
\end{align*}\\[-3.0em]
Hence a serially-pruned interleaver is contention-free when the banks are accessed sequentially using a counter from $j\!=\!0,1,\cdots,W\!-\!1$, if the mother interleaver is contention-free.
\end{proof}
\end{theorem}
The pruning gaps in~\eqref{eq:contention_Delta} can be computed efficiently using Algorithm~\ref{a:minimal_inliers_algorithm} together with any scheme to enumerate the inliers depending on the permutation at hand. In Fig.~\ref{f:contention_free}b, the theorem is applied to parallelize the pruned BRP stage of a 32-point FFT algorithm pruned to $\b\!=\!22$ points and permute the its outputs in parallel without contention when accessing the memory banks. Pruned locations are marked as $\boxtimes$. Each read memory bank $t$ is initialized with the appropriate $\Delta(j,t)$ using~\eqref{eq:contention_Delta}, and accessed by a counter $j$ that runs from 0 to $W\!-\!1$. When reading bank $t$ at step $j$, the actual address corresponds to $j\!+\!tW\!-\!\Delta(j,t)$. If $\pi_n(j\!+\!tW)\!<\!\b$, the read is successful. Otherwise, the location is pruned, reading from bank $t$ is stalled and $\Delta(j,t)$ is incremented. The FFT results are written in parallel in pruned BRP order in the write memory banks in 3 steps.

%
\vspace{-0.15in}
\subsection{Pruned LTE Turbo Interleavers}\label{s:pruned_LTE_turbo_interleavers}\vspace{-0.05in}
Serial pruning is also valuable in turbo coding applications because it can accommodate for flexible codeword lengths. In a typical communication system employing adaptive modulation and coding, only a small set of discrete codeword lengths $k$ are supported. Bits are either punctured or filled in to match the nearest supported length. For a pruned interleaver $\cpi$ of length $\b$ to be useful, it is desirable to have the following characteristics: 1) It does not require extra memory to store the pruned indices, 2) pruning preserves the contention-free property~\cite{Nimbalker_tcom_2008,Takeshita_contention_free_2006} of its mother interleaver (if present), and 3) its spread factor~\cite{Dolinar_1995} degrades gracefully with the number of pruned indices $g\!\triangleq\!k\!-\!\b$, and hence the impact on BER performance is limited.

Serial pruning satisfies properties 1 and 2 as shown in Section~\ref{s:pruned_FFT}. The implications are that serially-pruned contention-free interleavers are parallelizable at a low implementation cost using the schemes proposed in this work to enumerate inliers. When coupled with windowing techniques to parallelize the constituent a posteriori probability (APP) decoders (see Fig.~\ref{f:contention_free}b with APP decoders instead of FFT blocks), a turbo decoder can be efficiently parallelized to meet throughput requirements in 4G wireless standards and beyond. We next show that serial pruning also satisfies property 3.

The spread factor of an interleaver is a popular measure of merit for turbo codes~\cite{Dolinar_1995}. The spread measures of $\pi$ and $\cpi$ associated with two indices $i,j$ are $S(i,j)\!=\!\abs{\pi(i)\!-\!\pi(j)}\!+\!\abs{i\!-\!j}$ and $S_p(i,j)\!=\!\abs{\cpi(i)\!-\!\cpi(j)}\!+\!\abs{i\!-\!j}\!=\!\abs{\pi(p(i))\!-\!\pi(p(j))}\!+\!\abs{i\!-\!j}$. The minimum spreads of $\pi$ and $\cpi$ are defined as
$S_{\text{min}}\!\triangleq\!\min_{i,j<k} S(i,j)\!=\!\min_{\a} \text{SP}_{\a}(\pi)$ and $S_{p,\text{min}}\!\triangleq\!\min_{i,j<\b} S_p(i,j)\!=\!\min_{\a} \text{SP}_{\a}(\cpi)$, $i\!\neq\!j$. The following theorem shows that $S_{p,\text{min}}$ remains close to $S_{\text{min}}$ when $g$ is small.
%
\begin{theorem}\label{th:spread} The minimum spread of a serially-pruned interleaver of length $\b$ is at least\vspace{-0.15in}
\begin{align}\label{eq:spread_lb}
    S_{p,\text{min}} \geq \frac{S_{\text{min}}}{\left(1+\g+g/k\right)^t}
\end{align}\\[-2.75em]
where $\g$ is a small positive constant and $t\!=\!-\log(1\!-\!\g\!-\!g/k)/\log(1\!+\!\g\!+\!g/k)$.~\IEEEQED
\end{theorem}
The proof relies on the fact that $S_{p,\text{min}}\!+\!\abs{p(i_0)\!-\!p(j_0)}\!-\!\abs{i_0\!-\!j_0} \!\geq S_{\text{min}}$, where $i_0,j_0$ are such that $\abs{\pi(p(i_0))\!-\!\pi(p(j_0))}\!+\!\abs{i_0\!-\!j_0}\!=\!S_{p,\text{min}}$. The difference $D\!\triangleq\!\abs{p(i_0)\!-\!p(j_0)}\!-\!\abs{i_0\!-\!j_0}$ is upper bounded as $D\!\leq\!p(j_0)\!-\!j_0$, assuming $j_0\!>\!i_0$, since $p(\cdot)$ is a monotonically increasing function. Since $i_0,j_0$ cannot be separated by more than $S_{p,\text{min}}\!-\!1$ positions, we need to find the maximum of $p(j_0)\!-\!j_0$ when $j_0\!=\!S_{p,\text{min}}$. This is equivalent to finding the maximum expansion of an interval of length $S_{p,\text{min}}$ such that it contains \emph{at least} $S_{p,\text{min}}$ inliers. From Theorem~\ref{th:theorem_P_sum_K}, this expansion leads to finding the minimum $t\!\geq 0$ that satisfies $S_{p,\text{min}}(1\!+\!\g\!+\!g/k)^t(1\!-\!\g\!-\!g/k)\!\geq\!S_{p,\text{min}}$, from which~\eqref{eq:spread_lb} follows. For example, the QPP interleaver $\pi(j)\!=\!63j\!+\!128j^2\!\pmod{2048}$ has $S_{\text{min}}\!=\!64$ and $\g\!=\!0.076$. If $g\!=\!20$ positions are pruned, then $S_{\text{p,min}}\!\geq\!58$. In fact, the actual $S_{p,\text{min}}$ is 62.

\begin{figure}[!t]
    \hspace{-0.0in}
    \subfloat[]{\includegraphics[scale=0.7]{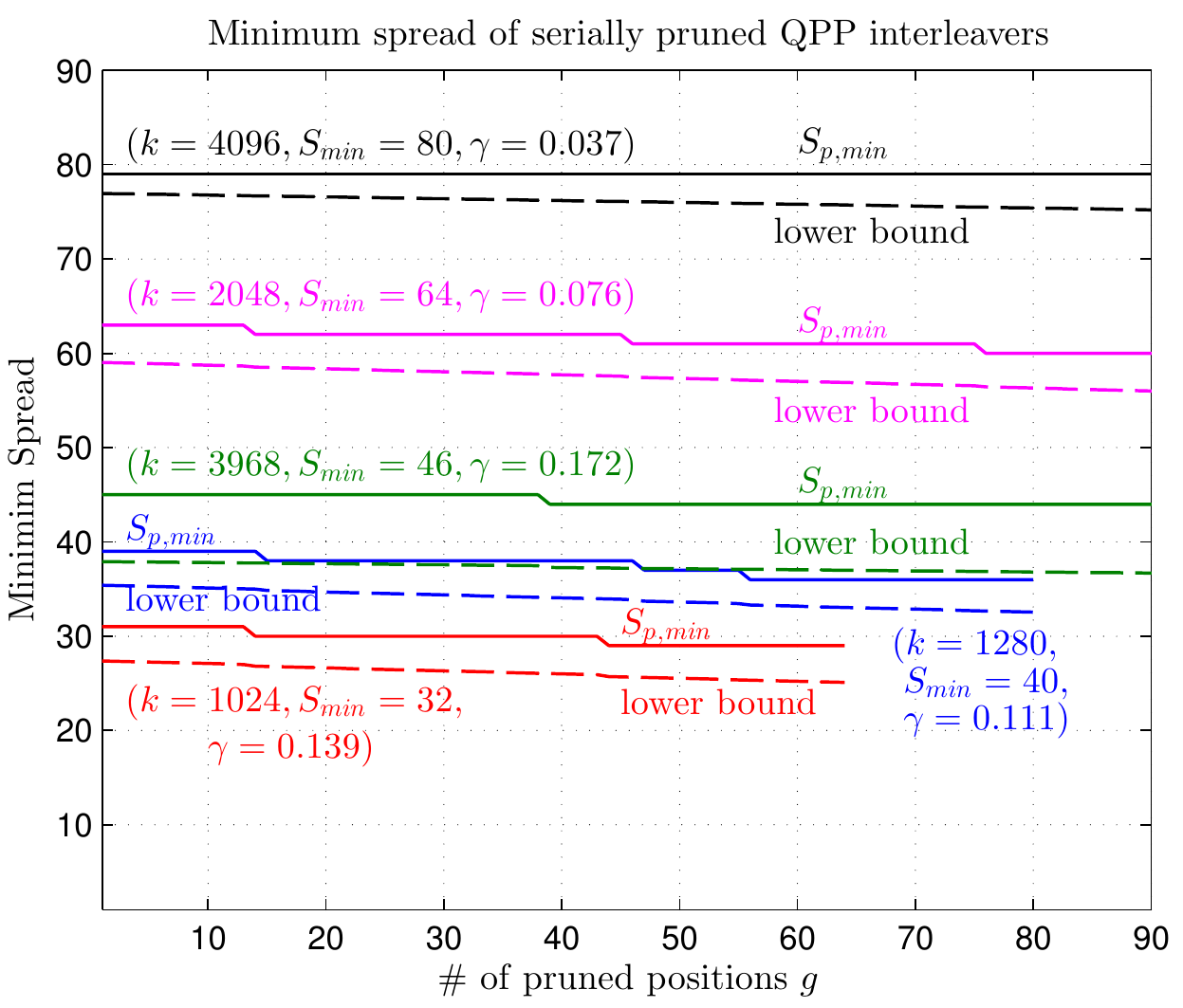}\label{f:spread_plot}}
    \subfloat[]{\includegraphics[scale=0.7]{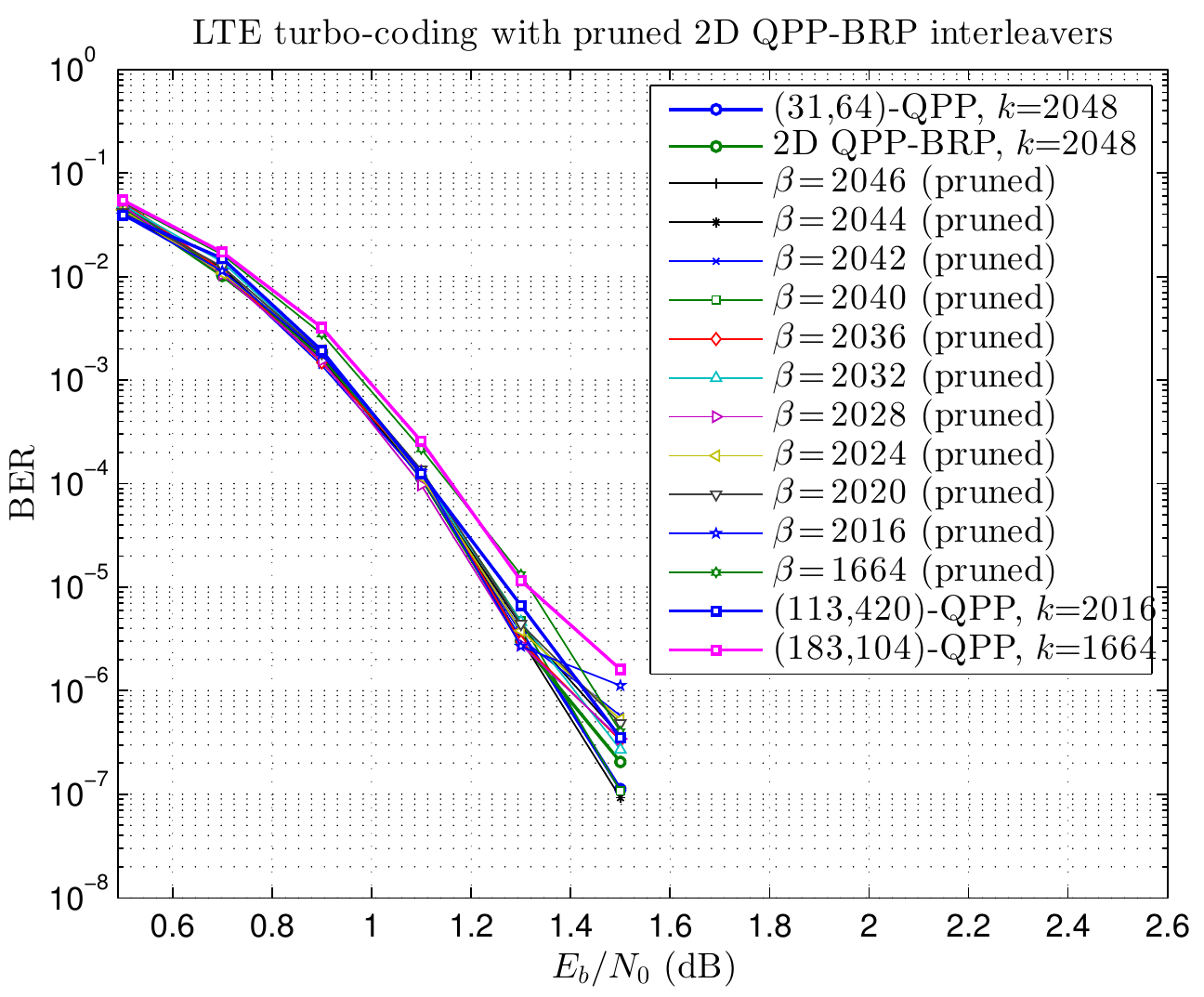}\label{f:BER2K_plot}}\vspace{-0.05in}
    \caption{(a) Minimum spread of pruned QPP interleavers in LTE. (b) BER of LTE turbo codes with pruned 2D QPP-BRP interleavers of length $k\!=\!2048$.}
    \label{f:BER_plots}
\end{figure}

Figure~\ref{f:spread_plot} plots the minimum spread of serially-pruned QPP interleavers as a function of $g$, for several mother QPP interleavers. The lower bound in~\eqref{eq:spread_lb} is plotted as well. The length $k$, minimum spread $S_{\text{min}}$ and constant $\gamma$ of the mother interleavers are shown in brackets. As shown, $S_{p,\text{min}}$ of the pruned interleavers remains very close to $S_{\text{min}}$ when up to $g\!=\!2S_{\text{min}}$ indices are pruned, and the lower bounds predicted by~\eqref{eq:spread_lb} are rather tight.
%
%

To assess the impact of serial pruning on error-correction performance, the BER of 3GPP LTE turbo codes employing serially-pruned 2D QPP-BRP interleavers were simulated over an AWGN channel. The 2D mother interleaver of length $k\!=\!2048$ is a concatenation of a QPP and a BRP defined by $\pi(x)\!=\!\flr{\!\s_2(x)/k_1\!}k_1\!+\!\s_1(x\bmod k_1)$, where $k_1\!=\!16,n_1\!=\!4$, $\s_1(x)\!=\!\pi_4(x)$, $k_2\!=\!512$, and $\s_2(x)\!=\!31x\!+\!64x^2\!\pmod{2048}$. 500,000 frames were simulated assuming BPSK modulation and log-MAP decoding with up to 6 decoding iterations. Figure~\ref{f:BER2K_plot} shows the results using the 2D mother interleaver and eleven serially-pruned interleavers of lengths as indicated in the figure. Also shown for comparison are results for three other 1D QPP interleavers of lengths 2048, 2016 and 1664 that are supported in LTE (the other 9 lengths from 2016 to 2046 are not supported). In almost all cases, the pruned interleavers perform very close to the 2D mother QPP-BRP and 1D QPP interleavers.\vspace{-0.1in}

%
\section{Implementation Aspects and Practical Examples}\label{s:practical examples_impl_aspects}
\begin{figure}[hbtp]
\centering
\includegraphics[scale=0.55]{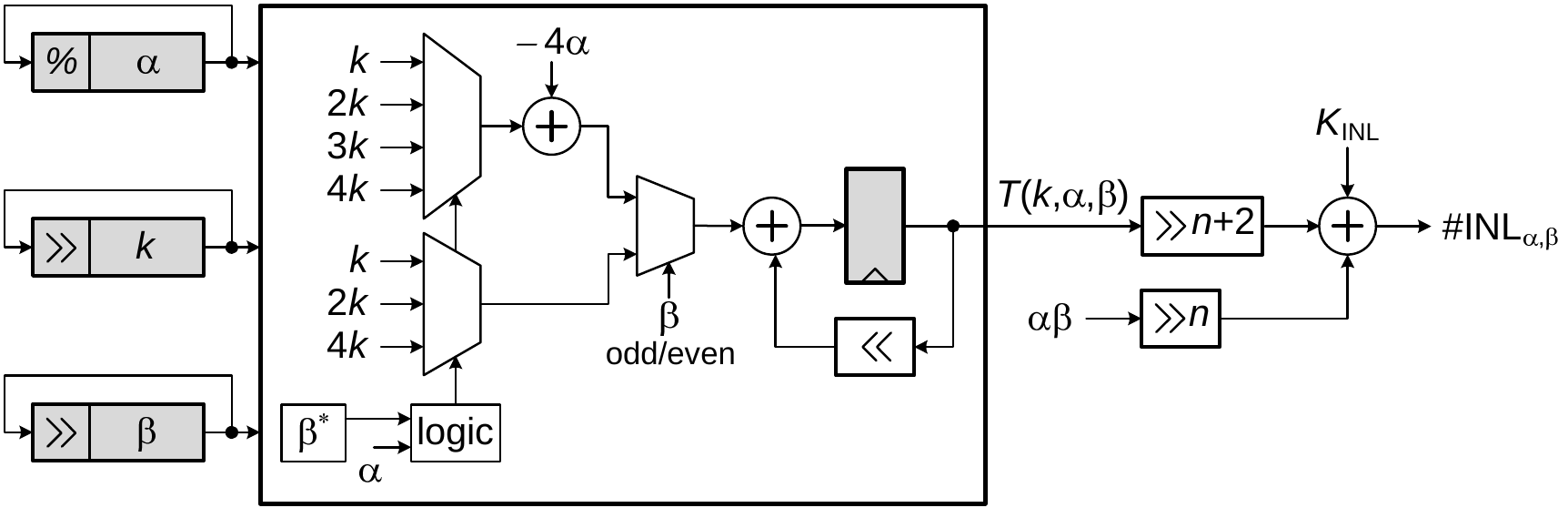}\vspace{-0.1in}
\caption{Architecture for computing $\#\text{INL}_{\a,\b}$ in (\ref{eq:corollary_P_sum_K_brev}) for bit-reversal permutations using the $T(k,\a,\b)$ function in (\ref{e:T_sum_minus}).}
\label{f:brev_inliers}
\end{figure}

\vspace{-0.05in}
Figure~\ref{f:brev_inliers} shows the architecture for computing $\#\text{INL}_{\a,\b}$ in~\eqref{eq:corollary_P_sum_K_brev} for bit-reversal permutations using the $T(k,\a,\b)$ function in~\eqref{e:T_sum_minus} using elementary logic gates. The block is clocked for $n\!-\!1$ clock cycles to produce the result. The three shift registers on the left are initialized with $\a,k,\b$. The register with symbol $\%$ drops out the most significant bit every cycle and stores the resulting contents back in the register, while the registers with symbols $\gg$ perform a left shift by one position every cycle. The block $\b^*$ reverses the bits of $\b/2$ or $(\b\!-\!1)/2$ depending on whether $\b$ is odd or even. The multiplexer logic simply selects what the expression in the $T(k,\a,\b)$ recursion in~\eqref{e:T_sum_minus} evaluates to (see the proof of Lemma~\ref{lemma:T_sum_minus} in the Appendix). The block with symbol $\ll$ multiplies the previous output of the register by 2 to generate $T(k/2,\a,\b/2)$ or $T(k/2,\a,(\b\!-\!1)/2)$. After $n\!-\!1$ clock cycles, the output $T(k,\a,\b)$ is then divided by $4k$ using the $\gg\!n\!+\!2$ block, and then $\a\b/k$ (the block $\gg\!n$ performs division by $k$) and $K_{\text{INL}}$ are added to generate $\#\text{INL}_{\a,\b}$.\vspace{-0.05in}

\begin{figure}[hbtp]
\centering
\includegraphics[scale=0.6]{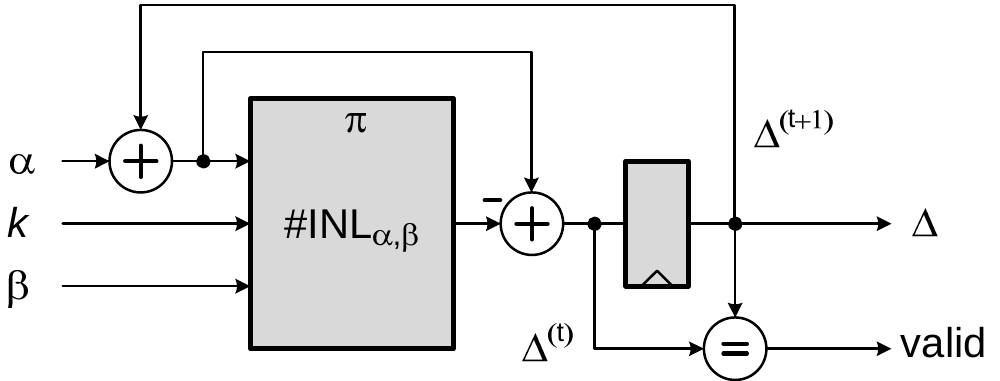}\vspace{-0.1in}
\caption{Architecture of the minimal inliers algorithm in Algorithm~\ref{a:minimal_inliers_algorithm}.}
\label{f:min_inliers}
\end{figure}

Figure~\ref{f:min_inliers} shows the implementation of the minimal inliers algorithm in Algorithm~\ref{a:minimal_inliers_algorithm}. The architecture can be used to compute the minimal inliers of any permutation by using the appropriate block in $\#\text{INL}_{\a,\b}$. For bit-reversal permutation, the block in Fig.~\ref{f:brev_inliers} is used. For linear congruential permutations, the block proposed in~\cite{2009_Mansour_LCSI} can be used. For a generic permutation, a lookup table implementation can be used when the size is small. A parallel pruned interleaver can be realized simply by cascading several minimal inliers blocks according to Algorithm~\ref{a:p-pbri_algorithm}.

%
\subsection{Practical Examples}\label{s:practical examples}
\begin{figure}[hbtp]
\centering
\includegraphics[scale=0.5]{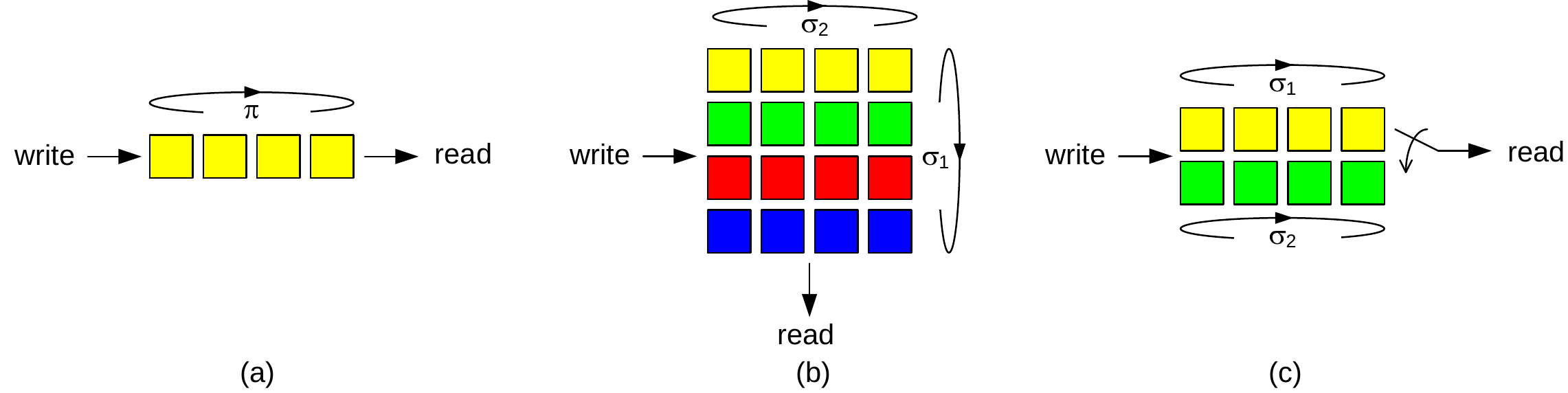}\vspace{-0.1in}
\caption{(a) 1D interleaver, (b) 2D block interleaver, and (c) 2-stream interleaver.}
\label{f:case_studies}
\end{figure}

To demonstrate the performance advantage of the proposed schemes in this paper, several pruned interleavers were constructed and simulated using the proposed pruning algorithms as well as existing serial pruning algorithms in the literature. One dimensional, 2D block, and 2-stream interleavers are considered (see Fig.~\ref{f:case_studies}). For the 1D case, bit-reversal (\textbf{brev}) and linear congruential sequence (\textbf{lcs})~\cite{Knuth_Seminumerical} are considered (refer to Table~\ref{t:interleaver_case_studies}). For the 2D case, four combinations of permutations across the two dimensions are considered: \textbf{brev} across both, \textbf{brev} across the first and reversed \textbf{brev} across the second, \textbf{lcs} across the first and \textbf{brev} across the second, \textbf{lcs} across the first and a quadratic permutation polynomial (\textbf{qpp}) across the second. The \textbf{lcs} permutations $\s_1\!=\!hj\pmod{k_1}$ vary from column to column by changing $h$ (odd). The \textbf{qpp} permutation has size 32 and its inliers are implemented using a look-up table. These interleavers are used in practice for example in~\cite{IEEE_802.11n,IEEE_802.16e,IEEE_802.20_3GPP2,LTE_phy_layer}.

For the 2-stream case, three combinations of permutations across the two streams are considered: \textbf{brev} across the first dimension and reversed \textbf{brev} across the second, \textbf{lcs} across the first and \textbf{brev} across the second, \textbf{lcs} across both dimensions. The parameters of all interleavers are listed in Table~\ref{t:interleaver_case_studies}.
\begin{table}[hbtp]
\singlespace
\footnotesize
\centering
\caption{Parameters of 1D, 2D, and 2-stream interleavers considered.}\vspace{-0.1in}
\begin{tabular}{|r|l|l|}
  \hline
  \textbf{Interleaver}              & \textbf{Permutation} $\pi$         & \textbf{Size} $k\!=\!2^n$  \\\hline\hline
  brev1D                            & $\pi\!=\!\pi_n$               & $n\!=\!10,11,\cdots,24$ \\ \hline
  lcs1D                             & $\pi(j)\!=\!(k/2\!-\!1)j\!\pmod{k}$    & $n\!=\!10,11,\cdots,24$ \\ \hline \hline
  \multicolumn{1}{r|}{\textbf{2-Dimensional}}  & \multicolumn{1}{l|}{$\pi\!=\!\s_2k_1\!+\!\s_1$}  & \multicolumn{1}{l}{$k_1\!=\!2^{n_1},k_2\!=\!2^{n_2},n\!=\!n_1\!+\!n_2$}\\ \hline
  \multirow{2}{*}{brev-brev2D}      & $\s_1\!=\!\pi_{n_1}$          & $n_1\!=\!5,6,\cdots,11$  \\
                                    & $\s_2\!=\!\pi_{n_2}$          & $n_2\!=\!6,7,\cdots,12$  \\ \hline
  \multirow{2}{*}{brev-brevrev2D}   & $\s_1\!=\!\pi_{n_1}$          & $n_1\!=\!6,7,\cdots,12$  \\
                                    & $\s_2\!=\!k_2\!-\!1\!-\!\pi_{n_2}$    & $n_2\!=\!6,7,\cdots,12$  \\ \hline
  \multirow{3}{*}{lcs-brev2D}       & $\s_1(j)\!=\!hj\!\pmod{k_1}$     & $n_1\!=\!5,6,\cdots,18$  \\
                                    &                           & $h\!=\!\text{randpermute}\{1,3,\cdots,(k_1\!-\!1)/2\}$\\
                                    & $\s_2\!=\!\pi_{n_2}$          & $n_2\!=\!6$  \\\hline
  \multirow{3}{*}{lcs-qpp2D}        & $\s_1(j)\!=\!hj\!\pmod{k_1}$     & $n_1\!=\!6,7,\cdots,18$  \\
                                    &                           & $h\!=\!\text{randpermute}\{1,3,\cdots,(k_1\!-\!1)/2\}$\\
                                    & $\s_2(j)\!=\!(k_2/2\!-\!1)j\!+\!2j^2\!\pmod{k_2}$   & $n_2\!=\!5$ \\\hline\hline
  \multicolumn{1}{r|}{\textbf{2-Stream}}  & \multicolumn{1}{l|}{$\pi\!=\!2\s_1  || (2\s_2\!+\!1)$}  & \multicolumn{1}{l}{$k_1\!=\!k_2=2^{n-1}$}\\ \hline
  \multirow{2}{*}{brev-brev2S}      & $\s_1\!=\!\pi_{n-1}$          & $n\!=\!10,11,\cdots,24$  \\
                                    & $\s_2\!=\!k_2\!-\!1\!-\!\pi_{n-1}$          &                       \\ \hline
  \multirow{2}{*}{lcs-brev2S}       & $\s_1(j)\!=\!(k/4\!-\!1)j\!\pmod{k/2}$     & $n\!=\!10,11,\cdots,24$  \\
                                    & $\s_2\!=\!\pi_{n-1}$          &   \\\hline
  \multirow{3}{*}{lcs-lcs2S}        & $\s_1(j)\!=\!(k/4\!-\!1)j\!\pmod{k/2}$   & $n\!=\!10,11,\cdots,24$  \\
                                    & $\s_2(j)\!=\!(k/4\!+\!1)j\!\pmod{k/2}$   &                       \\\hline\hline
\end{tabular}
\label{t:interleaver_case_studies}
\end{table}

\begin{figure*}[!t]
\centerline{\subfloat[1D and 2D]{\includegraphics[width=3.75in]{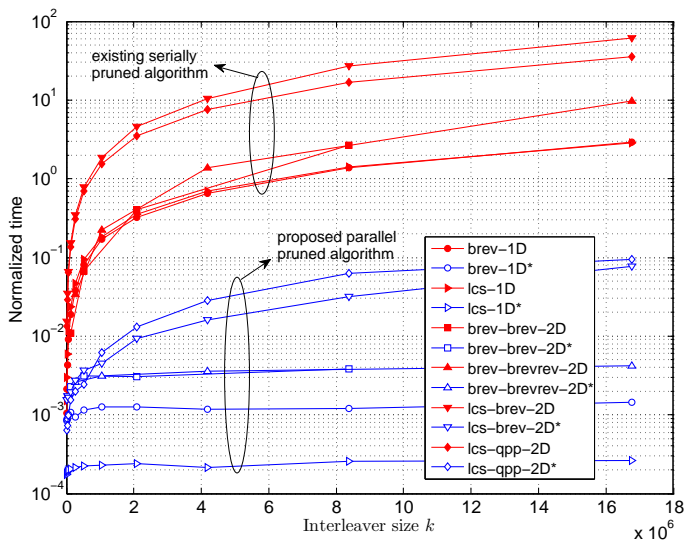}%
\label{f:1D}}
\hfil
\subfloat[2-stream]{\includegraphics[width=3.75in]{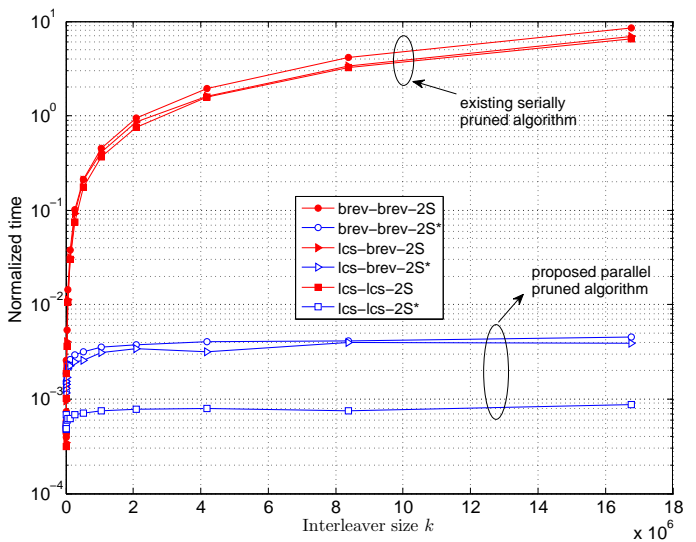}%
\label{f:2D}}}
\caption{Normalized pruned interleaving time as a function of interleaver size for (a) 1D and 2D interleavers, and (b) 2-stream interleavers.}
\label{f:performance_plots}
\end{figure*}

Figure~\ref{f:1D} plots the normalized time of the proposed pruning algorithm for the 1D and 2D pruned interleavers as a function of interleaver size. Also shown are the corresponding normalized times of serially-pruned algorithms. Figure~\ref{f:2D} shows the results for the 2-stream interleavers. The plots demonstrate a significant improvement between 3 to 4 orders of magnitude in pruning time compared to the serial case.

%
\section{Conclusions and Remarks}\label{s:conclusion} A mathematical formulation for analyzing the pruning of bit-reversal permutations has been presented. Pruning a permutation has been cast mathematically in terms of a problem of evaluating sums involving integer floors and saw-tooth functions. Bit-reversal permutations have been characterized in terms of permutation statistics, and have been shown to possess weak correlation properties. Moreover, using a new permutation statistic called permutation inliers that characterizes the pruning gap of BRPs, a computationally efficient algorithm for parallelizing serially-pruned bit-reversal interleavers has been proposed. Extensions to block and stream interleavers have been considered as well. The efficiency of this algorithm in terms of reducing interleaving latency and memory overhead has been demonstrated in the context of LTE turbo codes and pruned FFTs. The importance of this algorithm further lies in that it enables flexible and high speed implementations of PBRIs and other pruned permutations employed in communication standards that support multiple data rates and variable-length codewords.

The work proposed in this paper can be applied to more general block interleavers that involve generic permutations. We are investigating the class of interleavers based on permutation polynomials of general degree $p$~\cite{LTE_phy_layer,2007_Takeshita_alggeom}. Similar to~\eqref{eq:sum_product_saw_fractions}, these permutations require evaluating sums of the form $\dedsuminline{j=0}{k-1}{\vphantom{j^2b}j}{k}{h_0+jh_1+\cdots+j^ph_p}{k}$
with constant coefficients $h_j\in\mathds{Z}$, including the class of second-degree QPP interleavers, for arbitrary $k$. We conjecture that there exist recursive Euclidean-like algorithms to evaluate these sums that are analogous to those used for evaluating sums for linear permutation polynomials based on generalized Dedekind sums $\dedsuminline{j=0}{k-1}{\vphantom{jh}j}{k}{h_0+jh_1}{k}$~\cite{Knuth_Seminumerical}.

%
\vspace{-0.1in}
\appendix
%
%
\vspace{-0.1in}
\section*{Proof of Theorem~\ref{th:sum_j^m_pi_j_formula}}\vspace{-0.35in}
\begin{align*}
    J_m(k)
        = \sum_{j=0}^{k/2-1}\!j^m\pi_n{(j)} + \sum_{j=k/2}^{k-1}\! j^m\pi_n{(j)}
        = 2\sum_{j=0}^{k/2-1}\!j^m\pi_{n-1}{(j)} + \sum_{j=1}^{k/2-1}\! (j+k/2)^m\left(2\pi_{n-1}{(j)}+1\right)
                    +(k/2)^m.\\[-3.1em]
\end{align*}
Applying the Binomial theorem to expand $(j+k/2)^m$, followed by the Bernoulli expansion $\sum_{j=1}^{k/2-1} j^{m-r} \!=\! \frac{1}{m-r+1}\sum_{s=0}^{m-r}(1\!-\!k/2)^{-s}B_s\binom{m-r+1}{s}$,
where $B_s$ are the Bernoulli numbers ($B_0=1,B_1=1/2,B_2=1/6, \text{etc.}$), the result follows.~\IEEEQED

%
\vspace{-0.1in}
\section*{Proof of Lemma~\ref{lem:saw_sums_over_residue_system}}\vspace{-0.1in}
For~\eqref{e:property_sawsum_j}, we have $\sum_{j=0}^{k-1}\!\sawfracinline{j}{k} \!=\! \sum_{j=0}^{k-1}\left(\frac{j}{k}\!-\!\flrfrac{j}{k} \!- \!\hf \!+\!\hf\deltafrac{j}{k}\right)\!=\! \frac{k(k-1)}{2k} - 0 - \frac{k}{2} + \hf \!=\!0$. For~\eqref{e:property_sawsum_j_plus_w}, let $w\!=\!n\!+\!\theta$ for integer $n$ and real $0\!<\!\theta\!<\!1$. Then $\sum_{j=0}^{k-1}\sawfracinline{j+w}{k} \!=\! \sum_{j=0}^{k-1}\left[\!\saww{\frac{j+n}{k}} \!+\! \frac{\theta}{k}
\!-\!\hf\deltafrac{j+n}{k}\!\right] \!= \!0 + \frac{\theta}{k}\cdot k - \hf \!=\! \theta - \flr{\theta} - \hf + \hf\delta\!\left(\theta\right) \!=\! \saw{\theta}$,
where the second to last equality follows since $0\!<\!\theta\!<\!1$. For~\eqref{e:property_sawsum_pi_j}, $\sum_{j=0}^{k-1}\!\sawfracinline{\pi(j)}{k}\!=\!\sum_{j=0}^{k-1}\!\sawfracinline{\vphantom{\pi(j)}j}{k}\!=\!0$ since the two sum the same elements but in a different order. For~\eqref{e:property_sawsum_jh}, let $g\!=\!\gcd{(h,k)}, h'\!=\!h/g,k'\!=\!k/g$, then we have $\sum_{j=0}^{k-1}\!\!\sawfracinline{jh}{k}
    \!=\! \sum_{i=0}^{g-1}\sum_{j=0}^{k'-1}\!\!\sawfracinline{(ik'+j)h}{k}
    \!=\! \sum_{i=0}^{g-1}\sum_{j=0}^{k'-1}\!\!\sawfracinline{jh'}{k'}$ using~\eqref{e:saw_property_x_plus_integer}. Since $jh'\bmod k'$ is a permutation on $[k']$, then $\sum_{j=0}^{k'-1}\!\!\sawfracinline{jh'}{k'}\!=\!0$, and hence the sum is 0.
Finally, the proof of~\eqref{e:property_sawsum_pi_j_plus_w} is similar to~\eqref{e:property_sawsum_j_plus_w} by noting that $(\pi(j)\!+\!n)\bmod k$ is a permutation and that $\sum_{j=0}^{k-1}\deltafrac{\pi(j)\!+\!n}{k}\!=\!1$.~\IEEEQED

%
\vspace{-0.1in}
\section*{Proof of Lemma~\ref{lem:saw_sums_over_half_residue_system}}\vspace{-0.05in}
If $b\!\geq\! k$, then $\sawfracinline{\vphantom{\pi_n(j)}j\!-\!b}{k}\!=\!\sawfracinline{(\vphantom{\pi_n(j)}j\!-\!b)\bmod k}{k}$ and $\sawfracinline{\pi_n(j)\pm b}{k}\!=\!\sawfracinline{(\pi_n(j)\pm b)\bmod k}{k}$ using property~\eqref{e:saw_property_x_plus_integer}. Hence we assume $0\!\leq\! b\!<\!k$. If $0\!\leq\! b\!<\!k/2$, then using~\eqref{e:saw_function} $4\sum_{j=0}^{k/2-1}\!\!\sawfracinline{j-b}{k} \!=\!
        4\sum_{j=0}^{k/2-1}\frac{j-b}{k}
        \!-\! 4\sum_{j=0}^{b-1}\flrfrac{j-b}{k}
        \!-\! 4\sum_{j=b}^{k/2-1}\flrfrac{j-b}{k}
        \!-\! 2k\sum_{j=0}^{k/2-1}\!\!1
        \!+\! 2\sum_{j=0}^{k/2-1}\!\!\deltafrac{j-b}{k}$,
which reduces to $-k/2 \!+\! 2b \!+\! 1$ since $\flrfrac{j-b}{k}\!=\!-1$ in the second sum on the right and $0$ in the third sum, while $\deltafrac{j-b}{k}\!=\!1$ only once when $j\!=\!b$ in the last sum. On the other hand, if $b\!\geq\! k/2$, then $4\!\sum_{j=0}^{k/2-1}\!\!\sawfracinline{j-b}{k}\!=\!  4\!\sum_{j=0}^{k/2-1}\!\frac{j-b}{k}
               \!-\!4\!\sum_{j=0}^{k/2-1}\!\flrfrac{j-b}{k}
               \!-\!2\!\sum_{j=0}^{k/2-1}\!\!1
               \!+\!2\!\sum_{j=0}^{k/2-1}\!\!\deltafrac{j-b}{k}$,
which simplifies to $3k/2 \!-\! 2b \!-\! 1$ since $\flrfrac{j-b}{k}\!=\!-1$ and $\deltafrac{j-b}{k}\!=\!0$. For~\eqref{e:saw_sum_pi_+b_2}, first note the following useful property that relates $\pi_n$ on $n$ bits to $\pi_{n-1}$ on $n-1$ bits:\vspace{-0.1in}
\begin{align}
\label{e:property_pi_n_k_2_k}
    \pi_n(j) {=} \left\{\!
               \begin{array}{ll}
                 \!\!\!2\pi_{n-1}(j), & \hbox{$j=0,1,\cdots,k/2-1$;} \vspace{-0.1in}\\
                 \!\!\!2\pi_{n-1}(j)+1, & \hbox{$j=k/2,\cdots,k-1$.}
               \end{array}
             \right.\\[-3.1em]
\end{align}
Then $\sum_{j=0}^{k/2-1}\!\!\sawfracinline{\pi_n(j)\pm b}{k} = \sum_{j=0}^{k/2-1}\!\!\sawfracinline{2\pi_{n-1}(j)\pm b}{k} =\saw{\pm b/2}$ which equals zero using~\eqref{e:saw_property_n_over_2}.~\IEEEQED


%
\vspace{-0.15in}
\section*{Proof of Lemma~\ref{lem:saw_sum_j2_over_residue_system}}\vspace{-0.1in}
We first show that the sum $Q(k)\!=\!\sum_{j=0}^{k-1}\!\sawfracinline{j^2}{k}$ in~\eqref{e:saw_sum_j2} satisfies the recursion $Q(k)\!=\!2Q(k/4)\!-\!\frac{3}{2}$ for $k\!\geq8$
by counting the number of quadratic residues modulo $2^n$, which are integers of the form $q\!=\!j^2\!\pmod{2^n}$ and $\gcd(q,2^n)\!=\!1$. A well-known result from number theory is that these residue classes are the \emph{odd} integers in $[k]$ of the form $8r\!+\!1$, where $r\!=\!0,1,\cdots,k/8\!-\!1$. It follows that since there are a total of $2^{n-1}$ odd integers in $[k]$, and odd integers can only be congruent to either $1,3,5,7\!\pmod{8}$ (equally distributed), the number of quadratic residues is $2^{n-1}/4\!=\!2^{n-3}$. Moreover, if the odd integer $j$ maps to the residue $q$ modulo $2^n$, then so do the integers $-j\bmod 2^n,(k/2\!-\!j)\bmod 2^n,(k/2\!+\!j)\bmod 2^n$. Hence, $\mathop{\sum_{\stackrel{j=0}{j~\text{odd}}}^{k-1}}\!\sawfracinline{j^2}{k}
        \!=\!4\sum_{r=0}^{k/8-1}\!\sawfracinline{\vphantom{j^2}8r\!+\!1}{k}\!=\!4\sum_{r=0}^{k/8-1}\!\left[\frac{8r\!+\!1}{k}\!-\!\hf\right]
        \!=\!-\frac{3}{2}$
which is independent of $k$. Therefore, $Q(k) \!=\! \mathop{\sum_{\stackrel{j=0}{j~\text{even}}}^{k-1}}\!\sawfracinline{j^2}{k} \!+\!
            \mathop{\sum_{\stackrel{j=0}{j~\text{odd}}}^{k-1}}\!\sawfracinline{j^2}{k}
         \!=\! \sum_{j=0}^{k/2-1}\!\sawfracinline{(2j)^2}{k}\!-\!\frac{3}{2}
         \!=\! 2\sum_{j=0}^{k/4-1}\!\sawfracinline{j^2}{k/4}\!-\!\frac{3}{2}$
which proves $Q(k)\!=\!2Q(k/4)\!-\!\frac{3}{2}$, with initial conditions $Q(2)\!=\!0,Q(4)\!=\!-1/2$. We can rewrite this recursion in a different form for $k\!\geq\! 4$: $Q(k) \!=\!Q(k/2) \!-\! \frac{\sqrt{k}}{4}$ if $\log_2(k)$ is even, and $Q(k) \!=\!Q(k/2)\! -\! \frac{\sqrt{k/2}}{2}$ if $\log_2(k)$ is odd,
with $Q(2)\!=\!0$. Solving this recursion, equation~\eqref{e:saw_sum_j2} follows. Moreover, substituting~\eqref{e:saw_function} in~\eqref{e:saw_sum_j2}, and noting that $\deltafrac{j^2}{k}\!=\!1$ when $j\!=\!m\sqrt{k},m=0,1,\cdots,\sqrt{k}-1$, if $\log_2(k)$ is even, and when $j\!=\!2m\sqrt{k/2},m=0,1,\cdots,\sqrt{k/2}-1$, if $\log_2(k)$ is odd, equation~\eqref{e:floor_sum_j2} follows.~\IEEEQED

\vspace{-0.1in}
\section*{Proof of Lemma~\ref{lem:prod_saw_fractions}}\vspace{-0.1in}
Applying~\eqref{e:property_pi_n_k_2_k}, we can split $R(k)$ in~\eqref{e:R_k_definition} as follows:
\begin{align}\\[-3.1em]
    R(k)&=  4k\!\sum_{j=0}^{k/2-1}\!\!\sawfrac{j}{k}\!\!\!\sawfrac{2\pi_{n-1}(j)}{k} +
            4k\!\sum_{j=0}^{k/2-1}\!\!\sawfrac{j+k/2}{k}\!\!\!\sawfrac{2\pi_{n-1}(j)+1}{k}\label{e:S_k_sum_split}\\[-0.2em]
        &=  4k\!\sum_{j=0}^{k/2-1}\!\!\sawfrac{j}{k}\!\!\!\sawfrac{2\pi_{n-1}(j)}{k} +
            4k\!\sum_{j=0}^{k/2-1}\!\!\left[\!\sawfrac{2j}{k}\!-\!\sawfrac{j}{k} \!\right]\!\!\!
                \left[\! \sawfrac{2\pi_{n-1}(j)}{k} +\frac{1}{k}
                       -\hf\deltafrac{2\pi_{n-1}(j)}{k}\!\right]\\[-3.1em]
            \label{e:S_k_sum_split_simplified}
\end{align}
where in~\eqref{e:S_k_sum_split_simplified}, property~\eqref{e:saw_property_x_plus_half} and eq.~\eqref{eq:lemma_2x+c_over_k} with $c=0$ are applied. After simplification, we obtain $ R(k)\!=\!2R(k/2) \!+\! k/2\!-\!1$,
after applying~\eqref{e:saw_sum_k+b_2} with $b\!=\!0$. Solving the recurrence with initial condition $R(1)\!=\!0$ yields $R(k) \!=\! \sum_{i=1}^{n} 2^{n-i}\!\left( \frac{2^i}{2}\!-\!1 \right)\!=\!\frac{nk}{2} \!- \!k \!+\! 1$.

\vspace{-0.1in}
%
\section*{Proof of Lemma~\ref{lemma:S_sum}}\vspace{-0.1in} Using~\eqref{e:property_pi_n_k_2_k} we can split $S(k,b,c)$ similar to~\eqref{e:S_k_sum_split}.
If $c$ is odd, set $c^*=\pi_{n-1}^{-1}\left( (k-c-1)/2 \right)=\pi_{n-1}\left( (c-1)/2 \right)$. Then using~\eqref{eq:lemma_2x+c_over_k},
$S(k,b,c)$ reduces to $ S(k,b,c)=-4\sum_{j=0}^{k/2-1}\!\!\sawfracinline{j-b}{k} + 2k\sawfracinline{c^*-b}{k} +2S(k/2,b,(c-1)/2)$.
On the other hand, if $c$ is even, set $c^*=\pi_{n-1}^{-1}\left(c/2 \right)=\pi_{n-1}\left( c/2 \right)$. Then using~\eqref{eq:lemma_2x+c_over_k}, $S(k,b,c)$
reduces to $S(k,b,c)=-4\!\sum_{j=0}^{k/2-1}\!\!\sawfracinline{j-b}{k} \!-\!2k\!\saww{\!\frac{c^*-b}{k}\!+\!\hf\!} \!+\! 2S(k/2,b,c/2)$.
Evaluating the first sum in both cases using~\eqref{e:saw_sum_k+b_2}, equations~\eqref{e:S_sum} and~\eqref{w:K_S} follow.~\IEEEQED

\vspace{-0.1in}
\section*{Proof of Lemma~\ref{lemma:T_sum_minus}}\vspace{-0.1in} From~\eqref{e:T_definition}, it is obvious that if $c=0$ or $b=0$, then $T(k,b,c)=0$. Hence in the following we assume $c\neq 0, b\neq 0$.  Using~\eqref{e:property_pi_n_k_2_k} we can split $T(k,b,c)$ similar to~\eqref{e:S_k_sum_split}, then apply~\eqref{eq:lemma_2x+c_over_k} with $c=0$ to obtain\vspace{-0.1in}
\begin{align*}
T(k,b,c) &= 4k\sum_{j=0}^{k/2-1}\!\left[\!\sawfrac{j-b}{k}\!\!-\!\!\sawfrac{j}{k}\!  \right]\!\!
            \left[\!\sawfrac{2\pi_{n-1}(j)-c}{k}\!\!-\!\!\sawfrac{2\pi_{n-1}(j)-c+1}{k}\!
                  -\!\left\{\! -\frac{1}{k} + \hf\deltafrac{2\pi_{n-1}(j)}{k} \!\right\} \!\right]\\[-0.2em]
         &\quad + 4k\sum_{j=0}^{k/2-1}\!
            \left[\!\sawfrac{2(j-b)}{k}\!\!-\!\! \sawfrac{2j}{k}\!\right]\!\!
            \left[\!\sawfrac{2\pi_{n-1}(j)-c+1}{k} \!-\!
                \left\{ \!\sawfrac{2\pi_{n-1}(j)}{k} \!+\!\frac{1}{k}\!-\!\hf\deltafrac{2\pi_{n-1}(j)}{k}\!\right\}
            \!\right]\\[-3.1em]
\end{align*}
Denote by $T_1$ the first sum, and by $T_2$ the second. \textbf{Case 1}: When $c$ is odd, applying~\eqref{eq:lemma_2x+c_over_k} again, $T_1$ and $T_2$ reduce to $T_1=2k\!\left[\!\sawfracinline{c^*-b}{k}\! - \!\sawfracinline{c^*}{k}\! +\!\sawfracinline{\vphantom{c^*}b}{k} \!\right]$ and $T_2=2T(k/2,b,(c-1)/2) - 2k\!\sawfracinline{\vphantom{c^*}2b}{k}$,
where $c^*\!=\!\pi_{n-1}((c\!-\!1)/2)$. Adding the two and using~\eqref{e:saw_property_x_plus_half}, we obtain $T(k,b,c) \!=\! 2T(k/2,b,(c-1)/2) \!+\!
                2k\!\left[\! \sawfracinline{c^*-b}{k} \!-\! \sawfracinline{c^*}{k}\!
                -\!\saww{\!\frac{\vphantom{c^*}b}{k}\!+\!\hf\!}
                \!\right]$.
Simplifying the saw functions using~\eqref{e:saw_function}, the first equation in~\eqref{e:T_sum_minus} follows. Moreover, it is easy to show that these saw functions evaluate to either $-4b,-4b+k,-4b+2k,-4b+3k,-4b+4k$ depending on $c^*$ and $b$. \textbf{Case 2}: When $c$ is even, $T_1$ still simplifies as shown above but with $c^*=\pi_{n-1}(c/2)$, while $T_2$ becomes $T_2\!=\!-2k\!\left[\!\sawfracinline{2(c^*-b)}{k}\!-\!\sawfracinline{2c^*}{k}\!+\!\sawfracinline{\vphantom{c^*}2b}{k} \!\right] \!+\! 2T(k/2,b,c/2)$.
Hence $T(k,b,c) \!=\! 2T(k/2,b,c/2) \!-\!2k\!\left[\!\saww{\!\frac{c^*-b}{k}\!+\!\hf\!} \!-\! \saww{\!\frac{c^*}{k}\!+\!\hf\!} \!+\!
                \saww{\!\frac{\vphantom{c^*}b}{k}\!+\!\hf\!} \!\right]$.
Again, simplifying the saw functions, the second equation in~\eqref{e:T_sum_minus} follows. Moreover, it is easy to show that these saw functions evaluate to either $0,k,2k$ depending on $c^*$ and $b$.~\IEEEQED
\vspace{-0.05in}

%
\section*{Proof of Lemma~\ref{lemma:U_sum_minus}}\vspace{-0.1in} We split $U(k)$ similar to~\eqref{e:S_k_sum_split} with respect to both $j$ and $b$, and apply~\eqref{eq:lemma_2x+c_over_k}\vspace{-0.05in}
\begin{align}\\[-3.1em]
    U(k) &= \sum_{b=0}^{k/2-1}\!\!\left\{\!
                \sum_{j=0}^{k/2-1}\!\!\left[\!\sawfrac{j-b}{k}\!-\!\sawfrac{j}{k}  \!\right]\!\!
            \left[\!\sawfrac{2\pi_{n-1}(j)-\pi_n(b)}{k}\!\!-\!\!\sawfrac{2\pi_{n-1}(j)-\pi_n(b)+1}{k}
                  \!-\!\left\{\! -\frac{1}{k} + \hf\deltafrac{2\pi_{n-1}(j)}{k} \!\right\} \!\right]\right.\notag\\[-0.2em]
         &\hphantom{=} +\left.  \sum_{j=0}^{k/2-1}\!\!
            \left[\!\sawfrac{2(j-b)}{k}\!-\! \sawfrac{2j}{k}\!\right]\!\!
            \left[\!\sawfrac{2\pi_{n-1}(j)-\pi_n(b)+1}{k} \!-\!
                \left\{\! \sawfrac{2\pi_{n-1}(j)}{k} +\frac{1}{k}-\hf\deltafrac{2\pi_{n-1}(j)}{k}\!\right\}
            \!\right]\right\}  \notag\\[-0.2em]
        &\hphantom{=}+\sum_{b=k/2}^{k-1}\!\!\left\{\!
                \sum_{j=0}^{k/2-1}\!\!\left[\!\sawfrac{j-b}{k}\!-\!\sawfrac{j}{k}  \!\right]\!\!
            \left[\!\sawfrac{2\pi_{n-1}(j)-\pi_n(b)}{k}\!\!-\!\!\sawfrac{2\pi_{n-1}(j)-\pi_n(b)+1}{k}
                  \!-\!\left\{\! -\frac{1}{k} + \hf\deltafrac{2\pi_{n-1}(j)}{k} \!\right\} \!\right]\right.\notag\\[-0.2em]
        &\hphantom{=}\left. + \sum_{j=0}^{k/2-1}\!\!
            \left[\!\sawfrac{2(j-b)}{k}\!-\! \sawfrac{2j}{k}\!\right]\!\!
            \left[\!\sawfrac{2\pi_{n-1}(j)-\pi_n(b)+1}{k} \!-\!
                \left\{\! \sawfrac{2\pi_{n-1}(j)}{k} +\frac{1}{k}-\hf\deltafrac{2\pi_{n-1}(j)}{k}\!\right\}
            \!\right]\right\}\notag\\[-0.2em]
        &= \sum_{b=0}^{k/2-1}\!\!\left\{ U_1 + U_2 \right\} + \sum_{b=k/2}^{k-1}\!\!\left\{ U_3 + U_4 \right\}
        \label{e:U_sum_minus_split}\\[-3.1em]
\end{align}
For $b\!=\!0,\cdots,k/2-1$, then $\pi_n(b)$ is even. Therefore $c_{\text{even}}^* \!\triangleq\! \pi_{n-1}\!\left( \pi_n(b)/2 \right)\!=\! b$. On the other hand, for $b\!=\!k/2,\cdots,k-1$, then $\pi_n(b)$ is odd. Therefore $c_{\text{odd}}^* \!\triangleq\! \pi_{n-1}\!\left( (\pi_n(b)\!-\!1)/2 \right)\!=\! b \!-\! k/2$. Applying~\eqref{eq:lemma_2x+c_over_k} when $\pi_n(b)$ is even for $U_1$ and $U_2$, we get $U_1\!=\!0$ and $U_2\!=\!\sum_{j=0}^{k/2-1}\!\left[\!\sawfracinline{j-b}{k/2}\!\!- \!\!\sawfracinline{j}{k/2}\!\right]\!\!\!
            \left[\!\sawfracinline{\pi_{n-1}(j)-\pi_n(b)/2}{k/2} \!\!-\!\! \sawfracinline{\pi_{n-1}(j)}{k/2} \!\right]$.
Next, applying~\eqref{eq:lemma_2x+c_over_k} when $\pi_n(b)$ odd for $U_3$, and rearranging terms in $U_4$, we get $U_3\!=\!\frac{1}{4} \!-\! \frac{1}{4}\deltafrac{b-k/2}{k}$ and $U_4\!=\!\sum_{j=0}^{k/2-1}\!\!
            \left[\!\sawfracinline{j-b}{k/2}\!\!-\!\! \sawfracinline{j}{k/2}\!\right]\!\!
            \left[\!\sawfracinline{\pi_{n-1}(j)-(\pi_n(b)-1)/2}{k/2} \!\!-\!\!\sawfracinline{\pi_{n-1}(j)}{k/2}\!\right]
            \!-\! \hf\!\sawfracinline{\vphantom{\pi_{n-1}(j)}2b}{k}$.
After substituting for $U_1,U_2,U_3,U_4$ in~\eqref{e:U_sum_minus_split}, applying~\eqref{e:property_pi_n_k_2_k} on $\pi_n{(b)}$, and simplifying terms, $U(k)$ reduces to $U(k)\!=\!2U(k/2) \!+\! \frac{k}{8} \!- \!\frac{1}{4}$.
Solving the recurrence similar to Lemma~\ref{lem:prod_saw_fractions} with initial condition $U(1)\!=\!0$, the result in~\eqref{e:U_sum_minus} follows.~\IEEEQED
\vspace{-0.05in}

%
\vspace{-0.1in}
\section*{Proof of Lemma~\ref{lemma:C_k_p_sum}}\vspace{-0.05in}
When $p=0$, we have $C(k,0) = k^2\sum_{j=0}^{k-1}\!\sawfracinline{\pi_{n}(j)}{k}\!\!\!\sawfracinline{\pi_n(j)}{k}
          =k^2\sum_{j=0}^{k-1}\!\sawfracinline{\vphantom{\pi_{n}(j)}j}{k}\!\!\!\sawfracinline{\vphantom{\pi_{n}(j)}j}{k}
          = \frac{k(k-1)(k-2)}{12}$,
where the second equality follows the first since the two sum the same elements but in a different order. When $p=k/2$, we split $C(k,k/2)$ and apply~\eqref{e:property_pi_n_k_2_k} to obtain $C(k,k/2)=8(k/2)^2\sum_{j=0}^{k/2-1}\!\!\sawfracinline{\pi_{n-1}(j)}{k/2}\!\!\!\sawfracinline{\pi_{n-1}(j)}{k/2}=8C(k/2,0)=\frac{k(k-2)(k-4)}{12}$.
When $1\!\leq \!p\!<\!k/2$, we split $C(k,p)$ and apply~\eqref{e:property_pi_n_k_2_k} to $\pi_n(j)$ and $\pi_n(j+p)$. For simplicity, let $q\!=\!j\!+\!p$. Then\vspace{-0.05in}
\begin{align}
    \hspace{-0.25in}C(k,p)
        &=  k^2\!\sum_{j=0}^{k/2-1}\!\!\!\sawfrac{2\pi_{n-1}(j)}{k}\!\!\!\sawfrac{2\pi_{n-1}(q)}{k}+
            k^2\!\!\!\sum_{j=k/2-p}^{k/2-1}\!\!\!\!\sawfrac{2\pi_{n-1}(j)}{k}\!\!\!
            \left[\!\sawfrac{2\pi_{n-1}(q)+1}{k}\!\!-\!\!
                \sawfrac{2\pi_{n-1}(q)}{k}\!\right]\label{e:ser_corr_C_k_p_step1}\\[-0.2em]
        &\hphantom{=} +
            k^2\!\sum_{j=0}^{k/2-1}\!\!\!\sawfrac{2\pi_{n-1}(j)\!+\!1}{k}\!\!\!\sawfrac{2\pi_{n-1}(q)\!+\!1}{k}\!+\!
            k^2\!\!\!\sum_{j=k/2-p}^{k/2-1}\!\!\!\!\!\sawfrac{2\pi_{n-1}(j)\!+\!1}{k}\!\!\!
            \left[\!\!\sawfrac{2\pi_{n-1}(q)}{k}\!\!-\!\!
                \sawfrac{2\pi_{n-1}(q)\!+\!1}{k}\!\!\right]\label{e:ser_corr_C_k_p_step2}\\[-3.1em]
\end{align}
In~\eqref{e:ser_corr_C_k_p_step1}, the terms of the first sum when $\pi_n(j\!+\!p)\!=\!2\pi_{n-1}(j\!+\!p)$ are subtracted and replaced by $\pi_n(j\!+\!p)\!=\!2\pi_{n-1}(j\!+\!p)\!+\!1$ for $j\!=\!k/2\!-\!p,\cdots,k/2\!-\!1$ in the second sum. Similarly in~\eqref{e:ser_corr_C_k_p_step2}. Combining~\eqref{e:ser_corr_C_k_p_step1}-\eqref{e:ser_corr_C_k_p_step2}, applying~\eqref{eq:lemma_2x+c_over_k} four times, multiplying out terms, and simplifying the resulting expressions, $C(k,p)$ reduces to
$C(k,p) =   8C(k/2,p) + k^2/2 - p - k\left[ \pi_{n-1}(k/2-p) + \pi_{n-1}(p) \right]$.
Let $v$ denote the position of the least significant one bit in the binary representation of $p$. Then\vspace{-0.1in}
\begin{align}
\label{e:pi_v_binary_rep}
\pi_{n-1}(k/2-p) + \pi_{n-1}(p) = 2^{n-v-1} + 2^{n-v-2} - 1 
                                = 3k/2^{v+2} - 1\\[-3.2em]
\end{align}
Substituting back in $C(k,p)$ we get $C(k,p) =   8C(k/2,p) +\left(1 - \frac{3}{2^{v+1}}\right)\frac{k^2}{2} +k - p$. Finally, when $k/2<p<k-1$, following a similar derivation as above, we obtain $C(k,p) =   8C(k/2,p) +\left(1 - \frac{3}{2^{v+1}}\right)\frac{k^2}{2} + p$.~\IEEEQED
\vspace{-0.05in}

%
\section*{Proof of Lemma~\ref{lem:V_k_a_b}} Assume $a,b$ are both even. The proof for other cases is similar. Splitting the summation, applying~\eqref{e:property_pi_n_k_2_k} to $\pi_n(j)$ and $\pi_n(j+1)$, then adjusting missing terms for $j=k/2-1$ and $j=k-1$, we obtain
\begin{align*}\\[-3.1em]
    V(k,a,b)
        \!&=\! k^2\!\sum_{j=0}^{k/2-1}\!\!\sawfrac{2\pi_{n-1}(j)-a}{k}\!\!\!\sawfrac{2\pi_{n-1}(j+1)-b}{k}
            \!+\! k^2\!\left[\!\sawfrac{a+1}{k} \!-\! \sawfrac{a+2}{k} \!\right]\!\!\!
                    \left[\!\sawfrac{b}{k} \!-\! \sawfrac{b-1}{k} \!\right]\\[-0.2em]
        &\quad \!+\! k^2\!\sum_{j=0}^{k/2-1}\!
            \left[\!
                \sawfrac{2\pi_{n\!-\!1}(j)\!-\!a}{k}\!+\!\frac{1}{k}\!-\!\hf\deltafrac{2\pi_{n\!-\!1}(j)\!-\!a}{k}
            \!\right]\!\!\!
            \left[
                \!\sawfrac{2\pi_{n\!-\!1}(j\!+\!1)\!-\!b}{k}\!+\!\frac{1}{k}\!-\!\hf\deltafrac{2\pi_{n\!-\!1}(j\!+\!1)\!-\!b}{k}
            \!\right],\\[-3.1em]
\end{align*}
where Lemma~\eqref{lemma:lemma_2x+c_over_k} is applied twice. Next, multiplying out terms and using property~\eqref{e:property_sawsum_pi_j_plus_w}, the above expression simplifies to
\begin{align*}\\[-3.1em]
V(k,a,b)
        \!&=\!  2k^2\!\sum_{j=0}^{k/2-1}\!\!
                \sawfrac{2\pi_{n\!-\!1}(j)\!-\!a}{k}\!\!\!\sawfrac{2\pi_{n\!-\!1}(j\!+\!1)\!-\!b}{k}
            \!+\! k^2\!\left[\!\sawfrac{a+1}{k} \!-\! \sawfrac{a+2}{k} \!\right]\!\!\!
                   \left[\!\sawfrac{b}{k} \!-\! \sawfrac{b-1}{k} \!\right]\\[-0.2em]
        \!&\quad - \frac{k^2}{2}\!\!\sawfrac{2\pi_{n-1}(\pi_{n-1}(b/2)\!-\!1)\!-\!a}{k}
               \!-\!\frac{k^2}{2}\sawfrac{2\pi_{n-1}(\pi_{n-1}(a/2)\!+\!1)\!-\!b}{k}
               \!-\! \frac{k}{2}
                \!+\!\frac{k^2}{4}\deltafrac{2\pi_{n-1}(\pi_{n-1}(a/2)\!+\!1)\!-\!b}{k}\\[-3.1em]
\end{align*}
which is equivalent to~\eqref{e:V_k_a_b} with $a'=a,b'=b,a'' = 2\pi_{n-1}(\pi_{n-1}(a/2)\!+\!1)\!-\!b, b'' = 2\pi_{n-1}(\pi_{n-1}(b/2)\!-\!1)\!-\!a$, and $e=1$.~\IEEEQED
\vspace{-0.05in}

%
%
\section*{Proof of Lemma~\ref{lem:W_k_a_b}} Assume $a,b$ are both even. The proof for other cases is similar. Proceeding similar to Lemma~\ref{lem:V_k_a_b}, we obtain
\begin{align*}\\[-3.1em]\hspace{-0.175in}
    W(k,a,b) &= 8W(k/2,a/2,b/2) + \frac{k^2}{4}\left[\deltafrac{2\pi_{n\!-\!1}(\pi_{n\!-\!1}(a/2)\!+\!1)\!-\!b}{k}
                    \!-\! \deltafrac{2\pi_{n\!-\!1}(\pi_{n\!-\!1}(a/2)\!+\!1)}{k}
                    \!-\! \deltafrac{k/2\!-\!b}{k}\!\right]\\[-0.2em]
         &\quad-\frac{k^2}{2}
            \left[\!\sawfrac{2\pi_{n\!-\!1}(\pi_{n\!-\!1}(b/2)\!-\!1)\!-\!a}{k}\!\!-\!\!\sawfrac{2\pi_{n\!-\!1}(\pi_{n\!-\!1}(b/2)\!-\!1)}{k}
            \!\! + \!\!\sawfrac{2\pi_{n\!-\!1}(\pi_{n\!-\!1}(a/2)\!+\!1)\!-\!b}{k}\!\!
                   -\!\!\sawfrac{2\pi_{n\!-\!1}(\pi_{n\!-\!1}(a/2)\!+\!1)}{k}
            \!\right]\\[-0.2em]
          &\quad \!+ \!\frac{k^2}{2}\left[\!\sawfrac{k/2\!-\!b}{k}\!\!-\!\!\sawfrac{a\!+\!2}{k}
                        \!+\!\frac{2}{k}\!-\!\hf\!\right]
                 \!+\! k^2\!\left[\!\sawfrac{a+2}{k}\!\!-\!\!\sawfrac{a+1}{k}\!\!-\!\!\frac{1}{k} \!\right]\!\!\!
                    \left[\!\sawfrac{b-1}{k}\!\!-\!\!\sawfrac{b}{k}\!+\!\frac{1}{k}\!-\!\frac{1}{2} \!\right]\\[-3.1em]
\end{align*}
Simplifying the saw-fractions using~\eqref{e:saw_function}, and noting that
$\deltafrac{2\pi_{n\!-\!1}(\pi_{n\!-\!1}(a/2)\!+\!1)}{k}= \deltafrac{a\!+\!2}{k}$ since both give 1 when $a=k\!-\!2$, expression~\eqref{e:W_k_a_b_iterative_formula} follows with $e_a=e_b=0,e=1,a'=a,b'=b,a''\!=\! 2\pi_{n\!-\!1}\!\left(\pi_{n\!-\!1}\left(a/2\right)\!+\!1\right), b'' \!=\! 2\pi_{n\!-\!1}\!\left(\pi_{n\!-\!1}\left(b/2\right)\!-\!1\right)$.~\IEEEQED
\vspace{-0.05in}

%
\section*{Proof of Lemma~\ref{lem:sum_of_fixed_points}} Consider the $n$-bit binary representation of an integer $0\!\leq\! j\!<\!k$. When $n$ is even, we have $F_1(k)\!=\! \sum_{j=0}^{2^{n/2}-1}(j \!+\! 2^{n/2}\!\times\! \pi_{n/2}(j)) \!= \!\sum_{j=0}^{2^{n/2}-1}(j \!+ \!2^{n/2}\!\times\! j)$, where the second equality follows since the summation runs over all the integers from 0 to $2^{n/2-1}$. When $n$ is odd, we have $F_1(k)\!=\! \sum_{i=0}^{2^{(n-1)/2}-1}(j \!+\! 2^{(n+1)/2}\!\times\! \pi_{(n-1)/2}(j)) \!+\! \sum_{j=0}^{2^{(n-1)/2}-1}(j \!+ \! 2^{(n+1)/2}\!\times\! \pi_{(n-1)/2}(j) \!+\! 2^{(n-1)/2})$. Simplifying both expressions,~\eqref{eq:sum_of_fixed_points} follows. Similarly for $F_2(k)$, when $n$ is even, we have $F_2(k)\!=\! \sum_{j=0}^{2^{n/2}-1}(j \!+\! 2^{n/2}\!\times\! \pi_{n/2}(j))^2 \!= \! (1\!+\!k)\sum_{j=0}^{2^{n/2}-1}j^2  \!+\! 2\sqrt{k}\sum_{j=0}^{2^{n/2}-1}j\pi_{n/2}(j)$. When $n$ is odd, we have $F_2(k)\!= \! \sum_{j=0}^{2^{(n-1)/2}-1}(j \!+\! 2^{(n+1)/2}\!\times\! \pi_{(n-1)/2}(j))^2 \!+\! \sum_{j=0}^{2^{(n-1)/2}-1}(j \!+\! 2^{(n+1)/2}\!\times\! \pi_{(n-1)/2}(j) \!+\! 2^{(n-1)/2})^2 \!=\! 2(1\!+\!2k)\sum_{j=0}^{2^{(n-1)/2}-1}j^2 \!+\! 4\sqrt{2k}\sum_{j=0}^{2^{(n-1)/2}-1}j\pi_{(n-1)/2}(j) \!+\!(k/2)\sqrt{k/2}\! +\! 2\sqrt{k/2}(1+\sqrt{2k})\sum_{j=0}^{2^{(n-1)/2}-1}j$. Simplifying both expressions and using~\eqref{e:sum_j1_pi_j_formula},~\eqref{eq:sum_of_squared_fixed_points} follows.~\IEEEQED
\vspace{-0.05in}

%
\section*{Proof of Lemma~\ref{lem:sum of excedances}}
The first equality follows because the floor functions are $-1$ when $\pi_n(j)\!>\!j$ and 0 otherwise. Assume $n$ is even and consider the binary representation of the integers. Patterns that lead to excedances are $P_{01}\!=\!0\!\!\times\!\!\!\times\!\!\!\times\!\!\!\times\!1$, $P_{00}\!=\!0\!\!\times\!\!\!\times\!\!\!\times\!\!\!\times\!0$, $P_{11}\!=\!1\!\!\times\!\!\!\times\!\!\!\times\!\!\!\times\!1$, where in $P_{00}$ and $P_{11}$, the middle patterns are excedances. For example, $P_{00}\!=\!000110$ and $P_{11}\!=\!101011$ are excedances. The sum of all integers with binary pattern $P_{01}$ is $\sum_{01}\!=\!2\times2^{n-2}(2^{n-2}-1)/2 + 2^{n-2}\!=\!k^2/16$, with pattern $P_{00}$ is $\sum_{00}\!=\! 2E_1(k/4)$, and with pattern $P_{11}$ is $\sum_{11}= (k/2+1)\times\#\text{EXC}(\pi_{n-2})+2E_1(k/4)$, where $\#\text{EXC}(\pi_{n-2})$ is the excedance number in Lemma~\eqref{lem:prob_j_greater_Xj} for $\pi_{n-2}$. Collecting terms, we get the recursion $E_1(k) \!=\! 4E_1(k/4) \!+\! k\left(k-\sqrt{k}+1\right)/8 \!-\! \sqrt{k}/4$ for $k\geq 4$,
with initial condition $E_1(1)\!=\!0$. Similarly, when $n$ is odd we obtain $E_1(k) \!=\! 4E_1(k/4)\! +\! k\left(k\!-\!\sqrt{2k}+1\right)/8 \!-\! \sqrt{2k}/4$ for $k\geq 8$,
with initial condition $E_1(2)\!=\!0$. Solving both recursions,~\eqref{e:sum of excedances} follows.~\IEEEQED
\vspace{-0.05in}

%
%
%
\section*{Proof of Theorem~\ref{th:theorem_P_sum_K}} First write~\eqref{e:INL_definition} as $\#\text{INL}_{\a,\b}  = \sum_{j=0}^{k-1}\!
            \left(\!\flrfrac{\vphantom{\pi_n(j)}j-\a}{k} \!-\! \flrfrac{\vphantom{\pi_n(j)}j}{k}\! \right)\!\!
            \left( \! \flrfrac{\pi_n(j)-\b}{k} \!-\! \flrfrac{\pi(j)}{k} \!\right)$,
then replace the floor functions with $\saw{\cdot}$.
Multiplying out terms in the summation and using~\eqref{e:property_sawsum_pi_j},~\eqref{e:property_sawsum_pi_j_plus_w} we obtain $\#\text{INL}_{\a,\b} = \frac{\a\b}{k}  + \dedsuminline{j=0}{k-1}{\vphantom{\pi(j)}j-\a}{k}{\pi(j)-\b}{k}-\dedsuminline{j=0}{k-1}{\vphantom{\pi(j)}j-\a}{k}{\pi(j)}{k}  - \dedsuminline{j=0}{k-1}{\vphantom{\pi(j)}j}{k}{\pi(j)-\b}{k}+ \dedsuminline{j=0}{k-1}{\vphantom{\pi(j)}j}{k}{\pi(j)}{k} + K_{\text{INL}}(\a,\b)$, where
\begin{align}\\[-3.1em]\hspace{-0.35in}
K_{\text{INL}}(\a,\b) \!&\triangleq\! - \hf\!\!\sawfrac{\!\b'\!-\!\a\!}{k}\! \!-\!\!\hf\!\!\sawfrac{\!\vphantom{\beta}\a\!}{k} \!\!+\!\!
                \hf\!\!\sawfrac{\!\b'}{k} \!\!- \!\!\hf\!\!\sawfrac{\!\pi(\a) \!-\!\b\!}{k} \!\!+\!\!
                \hf\!\!\sawfrac{\!\pi(\a)\!}{k} \!\!+\!\! \frac{1}{4}\deltafrac{\!\pi(\a)\! -\! \b\!}{k}
                \!\!-\!\!\frac{1}{4}\deltafrac{\!\pi(\a)\!}{k}
                \!\!-\!\! \hf\!\!\sawfrac{\!\b\!}{k} \!\!- \!\! \frac{1}{4}\deltafrac{\!-\b\!}{k} \!\!+\!\!
                \frac{1}{4}\label{eq:P_f_cong_constant_K_full}\\[-3.1em]
\end{align}
and $\b'\!=\!\pi^{-1}(\b)$. Equation~\eqref{eq:P_f_cong_constant_K_full} can be further simplified by expanding $\saw{\cdot}$ in terms of floor functions, resulting in~\eqref{eq:P_f_cong_constant_K}. The condition $\pi(0)=0$ slightly simplifies the expression for the constant $K_{\text{INL}}$ but does not make the result less general.~\IEEEQED
\vspace{-0.05in}

%
\section*{Proof of Corollary~\ref{cor:P(j+1)}}
We have $\sum_{j=0}^{k-1}\flrfrac{j-\a}{k}\!\!\flrfrac{\pi_n(j+1)-\b}{k}
        = \sum_{j=1}^{k}\flrfrac{j-(\a+1)}{k}\!\!\flrfrac{\pi_n(j)-\b}{k}$ after change of variables. The $k$th term of the second sum is 0 since $\a\neq 0$. Adding and subtracting the 0th term, $\flrfrac{-\b}{k}=-1$ (since $\b\neq 0$), the result follows.~\IEEEQED
\vspace{-0.05in}

%
%
\section*{Proof of Theorem~\ref{th:theorem_covariance}}\vspace{-0.35in}
\begin{align*}
    \text{Cov}(X_i,X_{i+p})
    = \frac{1}{k}\sum_{j=0}^{k-1}\left(\pi_n(j)-\Ex{X_{j}}\right)\!\!
        \left(\pi_n(j+p)-\Ex{X_{j+p}}\right)
    = k\sum_{j=0}^{k-1}\!\left[\!\frac{\pi_n(j)}{k}\!-\!\hf\!+\!\frac{1}{2k}\!\right]\!\!\!
        \left[\!\frac{\pi_n(j+p)}{k}\!-\!\hf\!+\!\frac{1}{2k}\!\right]\\[-3.3em]
\end{align*}
for $0\!<\!p\!<\!k$. Writing the summand terms using $\saw{\cdot}$, multiplying out terms, and using $C(k,p)$ in~\eqref{e:C_k_p_sum}, the above sum reduces to $\text{Cov}(X_i,X_{i+p})\!=\!\frac{1}{k}C(k,p)\!-\!\frac{1}{4} \!+\!\frac{k}{2} \!-\!\frac{1}{2}\left[\pi_n(k\!-\!p) \!+\! \pi_n(p)\right]$.
Let $0\!\leq\! v<n$ be the position of the least-significant one-bit in the binary representation of $p$ (starting from 0). Substituting $\pi_n(k\!-\!p) \!+\! \pi_n(p)\!=\!3k/2^{v+1}\!-\!1$ similar to~\eqref{e:pi_v_binary_rep}, eq.~\eqref{e:theorem_covariance} follows.~\IEEEQED
\vspace{-0.05in}

%
\section*{Proof of Theorem~\ref{th:conv_rate}} We first derive bounds on $T(k,\a,\b)$ in~\eqref{e:T_definition}. Table~\ref{t:min_max_values_T} lists the first few terms of the minimum ($T_{\text{min}}(k)$) and maximum values ($T_{\text{max}}(k)$) of $T(k,\a,\b)$ empirically. It is easy to show by induction that $T_{\text{min}}(k)$ and $T_{\text{max}}(k)$ satisfy the recursions $T_{\text{min}}(k) \!=\! 2T_{\text{min}}(k/2) \!- \! ((2\!+\!\sqrt{2})\!\pm\!(2\!-\!\sqrt{2}))\sqrt{k} \!+\! 4$ and $T_{\text{max}}(k) \!=\! 2T_{\text{max}}(k/2) \!+\! 4(k\!\mp\!4)/3$
for $k\!>\!2$ with initial conditions $T_{\text{min}}(2)\!=\!2$ and $T_{\text{max}}(2)\!=\!2$, where $\pm$ and $\mp$ represent cases when $n$ is even/odd. Solving the recursions, we obtain the bound:\vspace{-0.1in}
\begin{align}
\label{eq:Tmin_max_bound}
    -3k - 4 + ((4+3\sqrt{2})\pm(4-3\sqrt{2}))\sqrt{k}  ~\leq~  T(k,\a,\b) \leq (12n-11)k/9 \mp 16/9 \\[-3.1em]
\end{align}
\begin{table}[hbtp]
\singlespace
\footnotesize
\centering
\caption{Minimum and maximum values of $T(k,\a,\b)$.}\vspace{-0.1in}
\begin{tabular}{|>{$}c<{$}|>{$}c<{$}|>{$}c<{$}|}
  \hline
  k & T_{\text{min}} & T_{\text{max}} \\\hline\hline
  4     & 0 & 4 \\ \hline
  8     & -4 & 24 \\ \hline
  16    & -20 & 64 \\ \hline
  32    & -52 & 176 \\ \hline
  64    & -132 & 432 \\ \hline
  128   & -292 & 1040 \\ \hline
  256   & -644 & 2416 \\ \hline
  512   & -1348 & 5520 \\ \hline
  1024  & -2820 & 12400 \\\hline
\end{tabular}
\label{t:min_max_values_T}
\end{table}
Let $\Delta\!^{(t)}$ be the minimum integer added to $\a$ in Algorithm~\ref{a:minimal_inliers_algorithm} at iteration $t$. Then at iteration $t\!+\!1$, $\Delta\!^{(t+1)} \!=\! \#\text{OUL}_{\a\!+\!\Delta\!^{(t)},\b} \!=\! (\a\!+\!\Delta\!^{(t)})\!-\!\#\text{INL}_{\a\!+\!\Delta\!^{(t)},\b} \!= \! (\a\!+\!\Delta\!^{(t)})\!-\!(\a\!+\!\Delta\!^{(t)})\b/k\!-\!T(k,\a\!+\!\Delta\!^{(t)},\b)/4k \!-\! K_{\text{INL}}$ from~\eqref{e:outliers_set_definition},~\eqref{eq:corollary_P_sum_K_brev}. Substituting the maximum and minimum values from~\eqref{eq:Tmin_max_bound} in this equation, and using the maximum and minimum values of $K_{\text{INL}}$ in~\eqref{eq:P_f_cong_constant_K_cases}, we obtain\vspace{-0.1in}
\begin{align}
\label{eq:delta_t+1_bound}
    (\a+\Delta\!^{(t)})(1-\b/k) + W_l  ~\leq~  \Delta\!^{(t+1)} \leq (\a+\Delta\!^{(t)})(1-\b/k) + W_u,\\[-3.1em]
\end{align}
where $W_l\!=\!(11\!-\!12n)/36 \!\pm\! 4/9k \!-\!3/4$ and $W_u\!=\!1\!-\!((1\!+\!3\sqrt{2}/4)\!\pm\!(1\!-\!3\sqrt{2}/4))/\sqrt{k}\!+\!1/k$. To determine the convergence rate, we study the convergence of the bounds in~\eqref{eq:delta_t+1_bound}. The solution of the lower-bound recursion $\Delta_l\!^{(t+1)} \!=\!(\a+\Delta_{l}\!^{(t)})(1\!-\!\b/k)\! + \!W_l$ is the sum of a geometric series $\Delta_{l}\!^{(t)} \!=\! \a\!\left(\!\left(1\!-\!\left(1\!-\!\b/k\right)^{t+1}\right)k / \b\!-\!1\right) +\left(\!1\!-\!\left(\!1\!-\!\b/k\right)^{t}\right)W_l k / \b$ which converges to $\Delta_l^* \!\triangleq\! \lim_{t\rightarrow\infty}\Delta_{l}\!^{(t)}  \!=\! \a(k/\b-1)+W_lk/\b$ at a rate $\abs{\Delta_l\!^{(t+1)} \!-\! \Delta_l^*}/\abs{\Delta_{l}\!^{(t)} - \Delta_l^*} = 1\!-\!\b/k$. Similar equations hold for the upper bound recursion $\Delta_{u}\!^{(t)}$ and $\Delta_u^*$ with all subscripts $l$ replaced by $u$. Hence $\Delta^{(t)}$ converges at a rate $1\!-\!\b/k$.~\IEEEQED

\vspace{-0.05in}

%
\begin{lemma}
\label{lemma:lemma_2x+c_over_k}
Let $k$ be even and $-k< c<k$. Then for $x=0,\cdots,k/2-1$, we have\vspace{-0.1in}
\begin{align}
\label{eq:lemma_2x+c_over_k}
\Lambda\triangleq\sawfrac{2x+c}{k} \!\! -\!\! \sawfrac{2x+c+1}{k} =
    \left\{
      \begin{array}{ll}
        -\frac{1}{k} + \hf\deltafrac{2x+c+1}{k}, & \hbox{$c$ odd;} \vspace{-0.1in}\\
        -\frac{1}{k} + \hf\deltafrac{2x+c}{k}, & \hbox{$c$ even.}
      \end{array}
    \right.\\[-3.1em]
\end{align}
%
\IEEEproof First write $\Lambda\!=\!-\frac{1}{k}\!-\!\flrfrac{2x+c}{k}\!+\!\flrfrac{2x+c+1}{k}
\!+\!\hf\deltafrac{2x+c}{k}\!-\!\hf\deltafrac{2x+c+1}{k}$. \textbf{Case $c$ odd}: Then $2x\!+\!c\!\neq\! 0$ (odd) and $2x\!+\!c\!+\!1\!\neq\! 0$ (even). If $2x\!+\!c\!+\!1\!<\!k$ or $2x\!+\!c\!+\!1\!>\!k$, then $\flrfrac{2x+c}{k}\!=\!\flrfrac{2x+c+1}{k}$, so $\Lambda\!=\!-1/k$. Otherwise if $2x\!+\!c\!+\!1\!=\!k$, then $\Lambda\!=\!-1/k\!+\!1/2$. Therefore, $\Lambda\!=\!-\frac{1}{k} \!+\! \hf\deltafrac{2x+c+1}{k}$. \textbf{Case 2 $c$ even}: Then $2x\!+\!c\!+\!1\!\neq\! 0$. If $2x\!+\!c\!=\! 0$ or $2x\!+\!c\!=\!k$, then $\Lambda\!=\!-1/k\!+\!1/2$. Otherwise, if $2x\!+\!c\!<\!k$ and $2x\!+\!c\!\neq \!0$, or $2x\!+\!c\!>\!k$, then $\flrfrac{2x+c}{k}\!=\!\flrfrac{2x+c+1}{k}$, so $\Lambda\!=\!-1/k$. Therefore, $\Lambda\!=\!-\frac{1}{k} \!+\! \hf\deltafrac{2x+c}{k}$.~\IEEEQED
\end{lemma}

\ifCLASSOPTIONcaptionsoff
  \newpage
\fi



%
\vspace{-0.25in}
\bibliographystyle{IEEEbib}
\singlespace
\bibliography{InterleaversBibFile}

\begin{thebibliography}{10}

\bibitem{Ramsey_1970}
J.~Ramsey,
\newblock ``Realization of optimum interleavers,''
\newblock vol. 16, no. 3, pp. 338--345, May 1970.

\bibitem{Forney_1971}
G.~D. Forney, Jr.,
\newblock ``Burst-correcting codes for the classic bursty channel,''
\newblock vol. 19, no. 5, pp. 772--781, Oct. 1971.

\bibitem{1993_Berrou_turbo_codes}
C.~Berrou, A.~Glavieux, and P.~Thitimajshima,
\newblock ``Near {S}hannon limit error-correcting coding and decoding: Turbo
  codes,''
\newblock in {\em Proc. IEEE Conf. on Commun. (ICC)}, Geneva, Switzerland, May
  1993, vol.~2, pp. 1064--1070.

\bibitem{tanner}
R.~Tanner,
\newblock ``A recursive approach to low complexity codes,''
\newblock vol. 27, pp. 533--547, Sep. 1981.

\bibitem{gallager}
R.~Gallager,
\newblock {\em Low-Density Parity-Check Codes},
\newblock MIT Press, Cambridge, MA, 1963.

\bibitem{Zehavi_1992}
E.~Zehavi,
\newblock ``8-{PSK} trellis codes for a {Rayleigh} channel,''
\newblock vol. 40, no. 5, pp. 873--884, May 1992.

\bibitem{1990_Bingham}
J.~Bingham,
\newblock ``Multicarrier modulation for data transmission: {An} idea whose time
  has come,''
\newblock vol. 28, no. 5, pp. 5--14, May 1990.

\bibitem{Parsons_2009}
A.~Parsons,
\newblock ``The symmetric group in data permutation, with applications to
  high-bandwidth pipelined {FFT} architectures,''
\newblock vol. 16, no. 6, pp. 477--480, Jun. 2009.

\bibitem{CooleyTukey_1965}
J.~Cooley and J.~Tukey,
\newblock ``An algorithm for the machine calculation of complex {Fourier}
  series,''
\newblock {\em Math. of Comp.}, vol. 19, no. 90, pp. 297--301, 1965.

\bibitem{Burrus_1988}
C.~Burrus,
\newblock ``Unscrambling for fast {DFT} algorithms,''
\newblock vol. 36, no. 7, pp. 1086--1087, Jul. 1988.

\bibitem{Skodras_1991}
A.~Skodras and A.~Constantinides,
\newblock ``Efficient input-reordering algorithms for fast {DCT},''
\newblock {\em IEE Electron. Lett.}, vol. 27, no. 21, pp. 1973--1975, Oct.
  1991.

\bibitem{1987_Evans}
D.~Evans,
\newblock ``An improved digit-reversal permutation algorithm for the fast
  {Fourier} and {Hartley} transforms,''
\newblock vol. 35, no. 8, pp. 1120--1125, Aug. 1987.

\bibitem{1999_Kim}
K.~Kim,
\newblock ``Shuffle memory system,''
\newblock in {\em 13th International Parallel Processing Symposium / 10th
  Symposium on Parallel and Distributed Processing (IPPS / SPDP '99), 12-16
  April 1999, San Juan, Puerto Rico, Proceedings}. Apr. 1999, pp. 268--272,
  IEEE Computer Society.

\bibitem{1999_Portnoff}
M.~Portnoff,
\newblock ``An efficient parallel-processing method for transposing large
  matrices in place,''
\newblock vol. 8, no. 9, pp. 1265--1275, Sep. 1999.

\bibitem{1991_Verbauwhede}
I.~Verbauwhede et~al.,
\newblock ``In-place memory management of algebraic algorithms on application
  specific {ICs},''
\newblock {\em Journal of VLSI Signal Proc.}, vol. 3, pp. 193--200, 1991.

\bibitem{1999_Chang}
G.~Chang, F.~Hwang, and L.-D. Tong,
\newblock ``Characterizing bit permutation networks,''
\newblock {\em Networks, John Wiley and Sons}, vol. 33, no. 4, pp. 261--267,
  1999.

\bibitem{2004_Rivest}
R.~Lee et~al.,
\newblock ``On permutation operations in cipher design,''
\newblock in {\em Proc. IEEE Conf. on Inf. Technol.: Coding and Computing
  (ITCC)}, Los Alamitos, CA, USA, 2004, vol.~2, pp. 569--577.

\bibitem{2001_Garello}
R.~Garello, G.~Montorsi, S.~Benedetto, and G.~Cancellieri,
\newblock ``Interleaver properties and their applications to the trellis
  complexity analysis of turbo codes,''
\newblock vol. 49, no. 5, pp. 793--807, May 2001.

\bibitem{LTE_phy_layer}
``Evolved universal terrestrial radio access {(E-UTRA)}: Multiplexing and
  channel coding,''
\newblock 3GPP TS 36.212, 3rd Generation Partnership Project (3GPP), Sep. 2008.

\bibitem{IEEE_802.20_3GPP2}
``{IEEE} standard for local and metropolitan area networks --- part 20: Air
  interface for mobile broadband wireless access systems supporting vehicular
  mobility,''
\newblock 802.20, IEEE, Piscataway, NJ, 2008.

\bibitem{IEEE_802.16e}
``{IEEE} standard for local and metropolitan area networks --- part 16: Air
  interface for broadband wireless access systems,''
\newblock 802.16, IEEE, Piscataway, NJ, 2009.

\bibitem{Takeshita_2005}
J.~Sun and O.~Takeshita,
\newblock ``Interleavers for turbo codes using permutation polynomials over
  integer rings,''
\newblock vol. 51, no. 1, pp. 101--119, Jan. 2005.

\bibitem{2001_Crozier}
S.~Crozier and P.~Guinand,
\newblock ``High-performance low-memory interleaver banks for turbo-codes,''
\newblock in {\em Proc. IEEE Veh. Tech. Conf. (VTC)}, Newark, New Jersey, USA,
  Oct. 2001, vol.~4, pp. 2394--2398.

\bibitem{DVB-T2_ETSI_302_755}
ETSI Std. EN 302~755 v1.2.1,
\newblock ``Frame structure channel coding and modulation for a second
  generation digital terrestrial television broadcasting system {(DVB-T2)},''
\newblock ETSI, 2011.

\bibitem{Knuth_Seminumerical}
D.~Knuth,
\newblock {\em The Art of Computer Programming}, vol.~2,
\newblock Addison-Wesley, Reading, MA, 3rd edition, 1998.

\bibitem{2004_Berrou}
C.~Berrou, Y.~Saouter, C.~Douillard, S.~Kerouedan, and M.~Jezequel,
\newblock ``Designing good permutations for turbo codes: towards a single
  model,''
\newblock in {\em Proc. IEEE Conf. on Commun. (ICC)}, Paris, France, Jun. 2004,
  vol.~1, pp. 341--345.

\bibitem{Nimbalker_2008}
A.~Nimbalker, Y.~Blankenship, B.~Classon, and T.~K. Blankenship,
\newblock ``{ARP} and {QPP} interleavers for {LTE} turbo coding,''
\newblock in {\em Proc. IEEE Wireless Commun. and Netw. Conf. (WCNC)}, Las
  Vegas, USA, Apr. 2008, pp. 1032--1037.

\bibitem{1999_Eroz_Hammongs_prunable_interleavers}
M.~Eroz and A.~R. Hammons, Jr.,
\newblock ``On the design of prunable interleavers for turbo codes,''
\newblock in {\em Proc. IEEE Veh. Tech. Conf. (VTC)}, Houston, Texas, USA, May
  1999, vol.~2, pp. 1669--1673.

\bibitem{2002_ferrari}
M.~Ferrari, F.~Scalise, and S.~Bellini,
\newblock ``Prunable {S}-random interleavers,''
\newblock in {\em Proc. IEEE Conf. on Commun. (ICC)}, New York City, New York,
  USA, Apr. 2002, vol.~3, pp. 1711--1715.

\bibitem{2005_Dinoi_Benedetto_S_random}
L.~Dinoi and S.~Benedetto,
\newblock ``Design of fast-prunable {S-random} interleavers,''
\newblock vol. 4, no. 5, pp. 2540--2548, Sep. 2005.

\bibitem{2009_Mansour_LCSI}
M.~M. Mansour,
\newblock ``Parallel lookahead algorithms for pruned interleavers,''
\newblock vol. 57, no. 11, pp. 3188--3194, Nov. 2009.

\bibitem{1971_Dieter}
U.~Dieter and J.~Ahrens,
\newblock ``An exact determination of serial correlations of pseudo-random
  numbers,''
\newblock {\em Numerische Math.}, vol. 17, pp. 101--123, 1971.

\bibitem{1974_Polge}
R.~Polge, B.~Bhagavan, and J.~Carswell,
\newblock ``Fast computational algorithms for bit reversal,''
\newblock vol. C-23, no. 1, pp. 1--9, Jan. 1974.

\bibitem{1988_Rodriguez}
J.~Rodriguez,
\newblock ``An improved bit-reversal algorithm for the fast {Fourier}
  transform,''
\newblock in {\em Proc. IEEE Conf. on Acoustics, Speech, and Signal Process.
  (ICASSP)}, New York City, New York, USA, Apr. 1988, vol.~3, pp. 1407--1410.

\bibitem{1989_Elster}
A.~Elster,
\newblock ``Fast bit-reversal algorithms,''
\newblock in {\em Proc. IEEE Conf. on Acoustics, Speech, and Signal Process.
  (ICASSP)}, Glasgow, Scotland, May 1989, vol.~2, pp. 1099--1102.

\bibitem{1991_Biswas}
A.~Biswas,
\newblock ``Bit reversal in {FFT} from matrix viewpoint,''
\newblock vol. 39, no. 6, pp. 1415 --1418, Jun. 1991.

\bibitem{1991_Yong}
A.~Yong,
\newblock ``A better {FFT} bit-reversal algorithm without tables,''
\newblock vol. 39, no. 10, pp. 2365--2367, Oct. 1991.

\bibitem{1992_Orchard}
M.~Orchard,
\newblock ``Fast bit-reversal algorithms based on index representations in
  $\text{GF}(\!2^b\!)$,''
\newblock vol. 40, no. 4, pp. 1004--1008, Apr. 1992.

\bibitem{1992_Jeong}
J.~Jeong and W.~Williams,
\newblock ``A unified fast recursive algorithm for data shuffling in various
  orders,''
\newblock vol. 40, no. 5, pp. 1091--1095, May 1992.

\bibitem{1995_Rius}
J.~Rius and R.~De~Porrata-Doria,
\newblock ``New {FFT} bit-reversal algorithm,''
\newblock vol. 43, no. 4, pp. 991--994, Apr. 1995.

\bibitem{2001_Drouiche}
K.~Drouiche,
\newblock ``A new efficient computational algorithm for bit reversal mapping,''
\newblock vol. 49, no. 1, pp. 251--254, Jan. 2001.

\bibitem{2004_Prado}
J.~Prado,
\newblock ``A new fast bit-reversal permutation algorithm based on a
  symmetry,''
\newblock vol. 11, no. 12, pp. 933--936, Dec. 2004.

\bibitem{2007_Pei}
S.-C. Pei and K.-W. Chang,
\newblock ``Efficient bit and digital reversal algorithm using vector
  calculation,''
\newblock vol. 55, no. 3, pp. 1173--1175, Mar. 2007.

\bibitem{2009_Mansour_PBRI}
M.~M. Mansour,
\newblock ``A parallel pruned bit-reversal interleaver,''
\newblock vol. 17, no. 8, pp. 1147--1151, Aug. 2009.

\bibitem{1997_Clarke_euler-mahonian}
R.~Clarke, E.~Steingr\'imsson, and J.~Zeng,
\newblock ``New {Euler-Mahonian} permutation statistics,''
\newblock {\em Advances in Applied Mathematics}, vol. 18, pp. 237--270, 1997.

\bibitem{1960_MacMahon_books}
P.~MacMahon,
\newblock {\em Combinatory Analysis}, vol. 1-2,
\newblock Cambridge University Press, Cambridge, 1915,
\newblock (Reprinted by Chelsea, New York, 1955).

\bibitem{1913_MacMahon}
P.~MacMahon,
\newblock ``The indices of permutations and the derivation therefrom of
  functions of a single variable associated with the permutations of any
  assemblage of objects,''
\newblock {\em American Journal of Mathematics}, vol. 35, no. 3, pp. 281--322,
  1913.

\bibitem{1995_Divsalar}
D.~Divsalar and F.~Pollara,
\newblock ``Multiple turbo codes,''
\newblock in {\em Proc. IEEE Military Commun. Conf. (MILCOM)}, San Diego,
  California, USA, Nov. 1995, vol.~1, pp. 279--285.

\bibitem{IEEE_802.11n}
``{IEEE} standard for local and metropolitan area networks --- part 11:
  Wireless {LAN} medium access control {(MAC)} and physical layer {(PHY)}
  specifications: Enhancements for higher throughput,''
\newblock 802.11n, IEEE, Piscataway, NJ, 2009.

\bibitem{1987_Holm}
S.~Holm,
\newblock ``{FFT} pruning applied to time domain interpolation and peak
  localization,''
\newblock vol. 35, no. 12, pp. 1776--1778, Dec. 1987.

\bibitem{1996_He}
S.~He and M.~Torkelson,
\newblock ``Computing partial {DFT} for comb spectrum evaluation,''
\newblock vol. 3, no. 6, pp. 173--175, Jun. 1996.

\bibitem{1980_Sreenivas}
T.~Sreenivas and P.~Rao,
\newblock ``High-resolution narrow-band spectra by {FFT} pruning,''
\newblock vol. 28, no. 2, pp. 254--257, Apr. 1980.

\bibitem{2005_Hu}
Zhong Hu and Honghui Wan,
\newblock ``A novel generic fast {Fourier} transform pruning technique and
  complexity analysis,''
\newblock vol. 53, no. 1, pp. 274--282, Jan. 2005.

\bibitem{1971_Markel}
J.~Markel,
\newblock ``{FFT} pruning,''
\newblock vol. 19, no. 4, pp. 305--311, Dec. 1971.

\bibitem{1979_Sreenivas}
T.~Sreenivas and P.~Rao,
\newblock ``{FFT} algorithm for both input and output pruning,''
\newblock vol. 27, no. 3, pp. 291--292, Jun. 1979.

\bibitem{1993_Sorensen}
H.V. Sorensen and C.S. Burrus,
\newblock ``Efficient computation of the {DFT} with only a subset of input or
  output points,''
\newblock vol. 41, no. 3, pp. 1184--1200, Mar. 1993.

\bibitem{2012_Wang}
Linkai Wang, Xiaofang Zhou, G.E. Sobelman, and Ran Liu,
\newblock ``Generic mixed-radix {FFT} pruning,''
\newblock vol. 19, no. 3, pp. 167--170, Mar. 2012.

\bibitem{Nimbalker_tcom_2008}
A.~Nimbalker, T.~K. Blankenship, B.~Classon, T.~E. Fuja, and D.~J. Costello,
  Jr.,
\newblock ``Contention-free interleavers for high-throughput turbo decoding,''
\newblock vol. 56, no. 8, pp. 1258--1267, Aug. 2008.

\bibitem{Takeshita_contention_free_2006}
O.~Takeshita,
\newblock ``On maximum contention-free interleavers and permutation polynomials
  over integer rings,''
\newblock vol. 52, no. 3, pp. 1249--1253, Mar. 2006.

\bibitem{Dolinar_1995}
S.~Dolinar and D.~Divsalar,
\newblock ``Weight distributions for turbo codes using random and nonrandom
  permutations,''
\newblock JPL TDA Progress Report 42-122, Aug. 1995.

\bibitem{2007_Takeshita_alggeom}
O.Y. Takeshita,
\newblock ``Permutation polynomial interleavers: An algebraic-geometric
  perspective,''
\newblock vol. 53, no. 6, pp. 2116--2132, Jun. 2007.

\end{thebibliography}
\begin{biography}[{\includegraphics[scale=1]{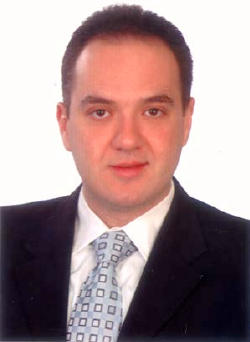}}]{Mohammad M. Mansour} received his B.E. degree with distinction in 1996 and his M.E. degree in 1998 both in computer and communications engineering from the American University of Beirut (AUB), Beirut, Lebanon. In August 2002, Mohammad received his M.S. degree in mathematics from the University of Illinois at Urbana-Champaign (UIUC), Urbana, Illinois, USA. Mohammad also received his Ph.D. in electrical engineering in May 2003 from UIUC.

He is currently an Associate Professor of Electrical and Computer Engineering with the ECE department at AUB, Beirut, Lebanon. From December 2006 to August 2008, he was on research leave with QUALCOMM Flarion Technologies in Bridgewater, New Jersey, USA, where he worked on modem design and implementation for 3GPP-LTE, 3GPP-UMB, and peer-to-peer wireless networking PHY layer standards. From 1998 to 2003, he was a research assistant at the Coordinated Science Laboratory (CSL) at UIUC. During the summer of 2000, he worked at National Semiconductor Corp., San Francisco, CA, with the wireless research group. In 1997 he was a research assistant at the ECE department at AUB, and in 1996 he was a teaching assistant at the same department. His research interests are VLSI design and implementation for embedded signal processing and wireless communications systems, coding theory and its applications, digital signal processing systems and general purpose computing systems.

Prof. Mansour is a member of the Design and Implementation of Signal Processing Systems Technical Committee of the IEEE Signal Processing Society, and a Senior Member of the IEEE. He has been serving as an Associate Editor for IEEE \textsc{Transactions on Circuits and Systems II} since April 2008, Associate Editor for \textsc{IEEE Transactions on VLSI Systems} since January 2011, and Associate Editor for \textsc{IEEE Signal Processing Letters} since January 2012. He served as the Technical Co-Chair of the IEEE Workshop on Signal Processing Systems (SiPS 2011), and as a member of the technical program committee of various international conferences. He is the recipient of the PHI Kappa PHI Honor Society Award twice in 2000 and 2001, and the recipient of the Hewlett Foundation Fellowship Award in March 2006. He joined the faculty at AUB in October 2003.
\end{biography}

%
%
%




\end{document}